\documentclass[12pt,aps,pre,preprint,showpacs,showkeys]{revtex4}

\usepackage{graphicx}
\usepackage{subfigure}
\usepackage{amsmath}
\usepackage{bm}
\usepackage{amssymb}
\usepackage{xcolor}

\newcommand{\Jnn}{J^{\mathrm{nn}}}
\newcommand{\Jfn}{J^{\mathrm{fn}}}

\begin{document}

\title{Spatial data modeling by means of Gibbs Markov random fields based on a generalized planar rotator model}
\author{Milan \v{Z}ukovi\v{c}}
\email{milan.zukovic@upjs.sk}
\affiliation{Department of Theoretical Physics and Astrophysics, Institute of Physics, Faculty of Science, Pavol Jozef \v{S}af\'arik University in Ko\v{s}ice, Park Angelinum 9, 041 54 Ko\v{s}ice, Slovak Republic}
\author{Dionissios T. Hristopulos}
\affiliation{School of Electrical and Computer Engineering, Technical University of Crete, Chania 73100, Greece}

\date{\today}

\begin{abstract}
We introduce a Gibbs Markov random field for spatial data on Cartesian grids which is based on the generalized planar rotator (GPR) model. The GPR model generalizes the recently proposed modified planar rotator (MPR) model by including in the Hamiltonian additional terms that  better capture  realistic features of spatial data, such as smoothness, non-Gaussianity, and geometric anisotropy. In particular, the GPR model includes up to infinite number of higher-order harmonics with exponentially vanishing interaction strength, directional dependence of the bilinear interaction term between nearest grid neighbors, longer-distance neighbor interactions, and two types of an external bias field. Hence, in contrast with the single-parameter MPR model,  the GPR model features five additional parameters: the  number $n$  of higher-order terms and the parameter $\alpha$ controlling their decay rate, the  exchange anisotropy parameter $\Jnn$, the further-neighbor interaction coupling $\Jfn$, and the external field (bias)  parameters $K$ (or $K'$). We present numerical tests on various synthetic data which demonstrate the effects of the respective terms on the model's prediction performance and we discuss these results in connection with the data properties. 
\end{abstract}

\keywords{Generalized planar rotator, spatial prediction, non-Gaussian distribution, conditional simulation, geometric anisotropy}

\maketitle

\section{Introduction}

Technological advances in environmental data collection, such as  remote sensing techniques, have prompted development of new data processing methods. Such processing includes filling of gaps that may arise due to various reasons~\citep{Kadlec17,Lehman04,Bechle13,Cole11,Sun17,Yoo10} and need to be filled to avoid the adverse missing-data impact on statistical estimates of means and trends~\citep{Sickles07}. Considering the fact that such data are typically massive and are collected with high frequency, the new methods should be computationally efficient and able to operate on data with general distribution as much as possible automatically or just with minimal user intervention. Traditional spatial interpolation methods, such as kriging~\citep{wack03} do not comply with these requirements, albeit, several modifications~\citep{cres08,furr06,kauf08,zhong16,marco18,ingram08} and parallelized schemes ~\citep{cheng13,guti14,hu15,pesq11,strz12,misr20} have been proposed primarily in effort to alleviate their computational burden.

A fundamentally different approach to the geostatistical problem has adopted tools from statistical physics and proposed to model spatial correlations by means of short-range interactions between Boltzmann-Gibbs random field variables~\citep{dth03,dthsel07}, instead of subjective and computationally much more intensive geostatistical approach based on the empirical variogram. These so-called \emph{Spartan spatial random field models} have been shown to be computationally efficient and applicable to both gridded and scattered Gaussian data. 

Spatial data on regular grids are often modeled by means of Gaussian Markov random fields (MRF)~\citep{Rue05}, which are based on the principles of conditional independence and the imposition of spatial correlations via local interactions. Much less attention has been paid to using non-Gaussian, such as Gibbs MRF based on spin models from statistical physics. Nevertheless, there have been some attempts to apply several of these models, such as the binary Ising, q-state Potts and clock, and continuous planar rotator models, to image restoration~\citep{nishi99,wong00,inoue01,ino-carl01,tadaki01,saika02} and geostatistical~\citep{mz-dth09a,mz-dth09b,mz-dth18,zuko21} problems. 

Recently, we have introduced a spatial prediction method based on the planar rotator model suitably modified to account for spatial correlations that are typical in geophysical and environmental data sets~\citep{mz-dth18}. In thermodynamic equilibrium, this modified planar rotator (MPR) model was shown to display a flexible type of short-range correlations controlled by the reduced temperature. In the proposed MPR prediction method the reduced temperature is the only model parameter that can be efficiently estimated by means of an ergodic specific energy matching principle. The spatial prediction of missing data is based on performing conditional Monte Carlo (MC) simulations and taking the mean of the respective conditional distribution at the target site given the incomplete measurements. In spite of its simplicity (it involves only one parameter), in comparison with some established prediction methods, the MPR method was shown to be competitive in terms of the prediction performance and due to sparse precision matrix structure, which allowed vectorization and parallelization on GPU~\citep{zuko20}, as well as a flexible hybrid algorithm~\citep{mz-dth20} also computationally very efficient.

In the present investigation we explore possibilities of enhancement of the MPR method’s flexibility by its generalization to include additional parameters that would better capture some realistic spatial features, such as data smoothness, non-Gaussianity or geometric anisotropy. Possible extensions in this direction include the generalization of the MPR Hamiltonian to obtain a generalized planar rotator (GPR) model, which incorporates higher-order couplings, an exchange interaction anisotropy, an interaction beyond nearest neighbors and an external “bias” field. We discuss effects of the respective extensions on the prediction performance in context with the character of the data.

\section{GPR Gibbs Markov random field}
\label{sec:gpr}

The MPR model, introduced in Ref.~\citep{mz-dth18}, is defined by means of the following Hamiltonian
\begin{equation}
\label{Hamiltonian_MPR}
{\mathcal H}=-J\sum_{\langle i,j \rangle}{\mathcal H}_{i,j},
\end{equation}
where ${\mathcal H}_{i,j}=\cos\phi_{i,j}$ is the pairwise potential and $\phi_{i,j}=q(\phi_{i}-\phi_{j})$ is an  angle between the $i$th and $j$th spins modified by the factor $q \leq 1/2$~\footnote{In Ref.~\citep{mz-dth18} as well as in this study we arbitrarily set the value of the modification factor to $q=1/2$.}, $J>0$ is the \emph{exchange interaction parameter}, and $\langle i,j \rangle$ denotes the sum over nearest neighbor spins on the grid. The prediction algorithm based on conditional Monte Carlo simulation of the MPR model is in detail described in Ref.~\citep{mz-dth18}. In the following we generalize the above MPR to GPR model, which includes several additional parameters with the goal of increasing of its flexibility. The GPR model Hamiltonian reads as follows
\begin{eqnarray}
\label{eq:Hamiltonian_GPR}
{\mathcal H}(n,\alpha,\Jnn,\Jfn,K)&=&-\Jnn_{x}\sum_{\langle i,j \rangle_x}{\mathcal H}_{i,j}(n,\alpha)
-\Jnn_{y}\sum_{\langle i,j \rangle_y}{\mathcal H}_{i,j}(n,\alpha)\\ \nonumber
&&-\Jfn\sum_{\langle\langle i,j \rangle\rangle}{\mathcal H}_{i,j}(n,\alpha)-K\sum_{i}B_i,
\end{eqnarray}
where the sums $\langle i,j \rangle_x$, $\langle i,j \rangle_y$, $\langle\langle i,j \rangle\rangle$, and $\sum_{i}$ denote summations over pairs of nearest-neighbor (nn) spins along the $x$- and $y$-axes, further-neighbor (fn) spins and all spins on the grid, respectively. The interaction parameters involve the anisotropic nn interactions $\Jnn_x=1-\Jnn$ in the $x$-direction and $\Jnn_y=\Jnn$ in the $y$-direction, where $\Jnn \in [0,1]$ is the \emph{anisotropy parameter}, and  $\Jfn$ is the isotropic fn interaction. ${\mathcal H}_{i,j}(n,\alpha)$ is a generalized potential function obtained by inclusion of $n$ higher-order couplings with exponentially vanishing strength controlled by the decay rate $\alpha$ (see Section~\ref{ssec:hoc}). The last term in Eq.~\eqref{eq:Hamiltonian_GPR} includes  $B_i=f(\phi_{i}-h_{i})$, which is a function of the distance between the spin $\phi_{i}$~\footnote{$\phi_{i}$ is actually the spin angle, however, with the spin length fixed to unity it fully characterizes the spin state.} and the external bias field $h_i$ at the $i$th site, and the parameter $K$ that controls the degree of alignment of the spins with the external field. The effects of the respective Hamiltonian terms on the model's predictive performance are elaborated below.

\subsection{Higher-order couplings}
\label{ssec:hoc}
First generalization involves inclusion of higher-order harmonics up to infinite order. We note that the coarse-grained Hamiltonian~(\ref{Hamiltonian_MPR}) only includes leading orders in a
Fourier expansion of a general microscopic spin-spin interaction ${\mathcal H}(\phi_i-\phi_j)$. The higher-order terms are often neglected, nevertheless, on many occasions they have turned out to play an important role in modeling of magnetic systems (see, e.g. Refs~\citep{lee85,carp89,shi11,pode11,cano14,cano16,zuko17,zuko18}). In particular, we consider the pairwise potential in the form
\begin{equation}
\label{Potential}
{\mathcal H}_{i,j}=\sum_{k=1}^{n}J_{k}\cos^{k}\phi_{i,j},
\end{equation}
where the constants $J_k$ represent weights of the respective (higher-order) terms in the summation. We note that for $n=1$ the potential reduces to that of the standard MPR model~(\ref{Hamiltonian_MPR}). If we assume that their intensity decays exponentially, i.e., $J_k=\alpha^{-k}$, where $\alpha>1$, then the potential ~(\ref{Potential}) can be for a finite $n$ expressed in a closed form
\begin{equation}
\label{Potential_cf_pfin}
{\mathcal H}_{i,j}(n,\alpha)=J(n,\alpha)\frac{\cos\phi_{i,j}\Big[1-\Big(\frac{\cos\phi_{i,j}}{\alpha}\Big)^{n}\Big]}{\cos\phi_{i,j}-\alpha},
\end{equation}
and for $n \to \infty$ it reduces to 
\begin{equation}
\label{Potential_cf_pinf}
{\mathcal H}_{i,j}(\alpha)=J(\alpha)\frac{\cos\phi_{i,j}}{\cos\phi_{i,j}-\alpha},
\end{equation}
where $J(n,\alpha)=(\alpha-1)/(1-\alpha^{-n})$ and $J(\alpha)=\alpha-1$ are exchange interaction parameters chosen to normalize the weights $J_{k}$ (scaling them so they add up to 1). 

\begin{figure}[t!]
\centering
\subfigure{\includegraphics[scale=0.55,clip]{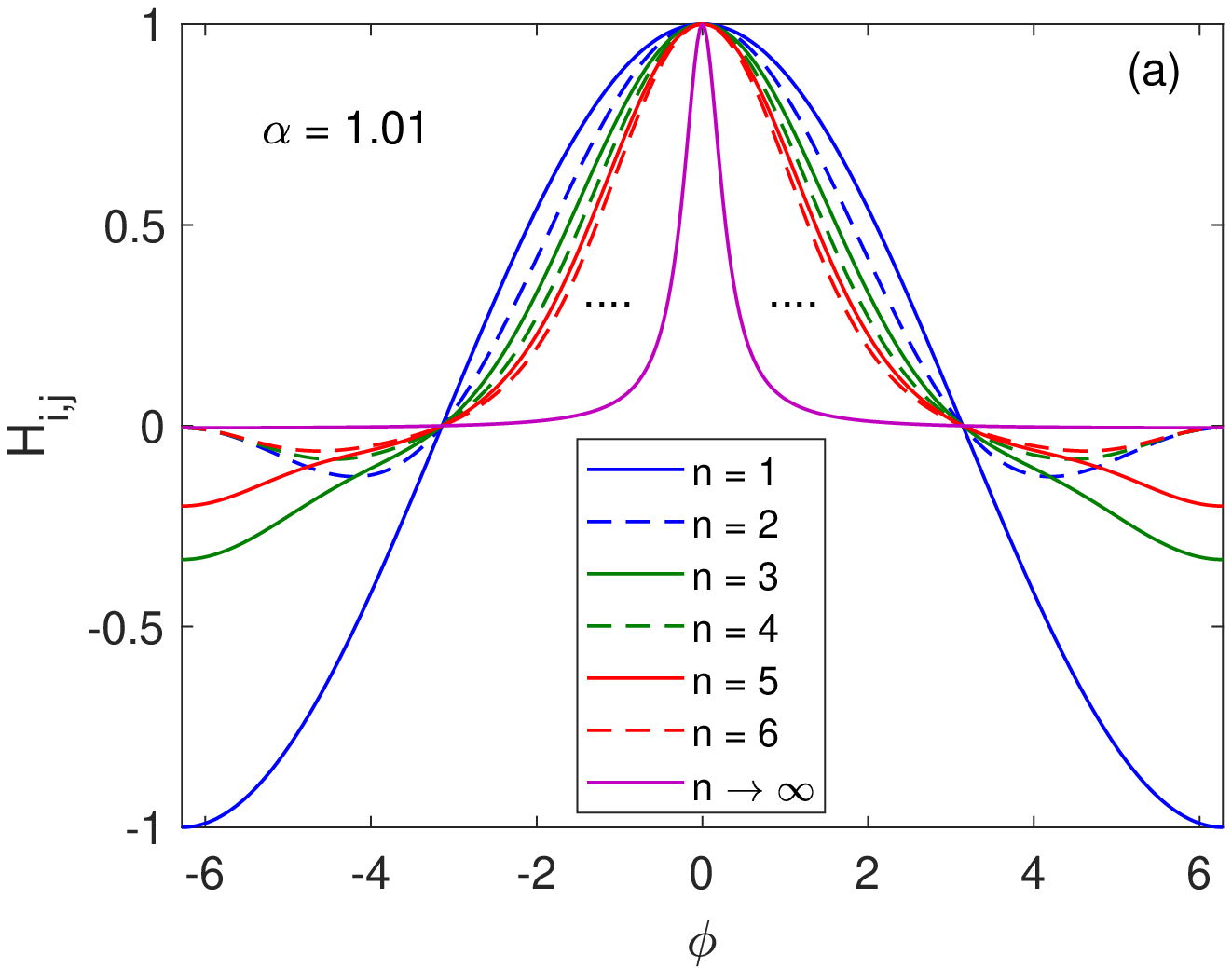}\label{fig:H_ij_phi-n}}
\subfigure{\includegraphics[scale=0.55,clip]{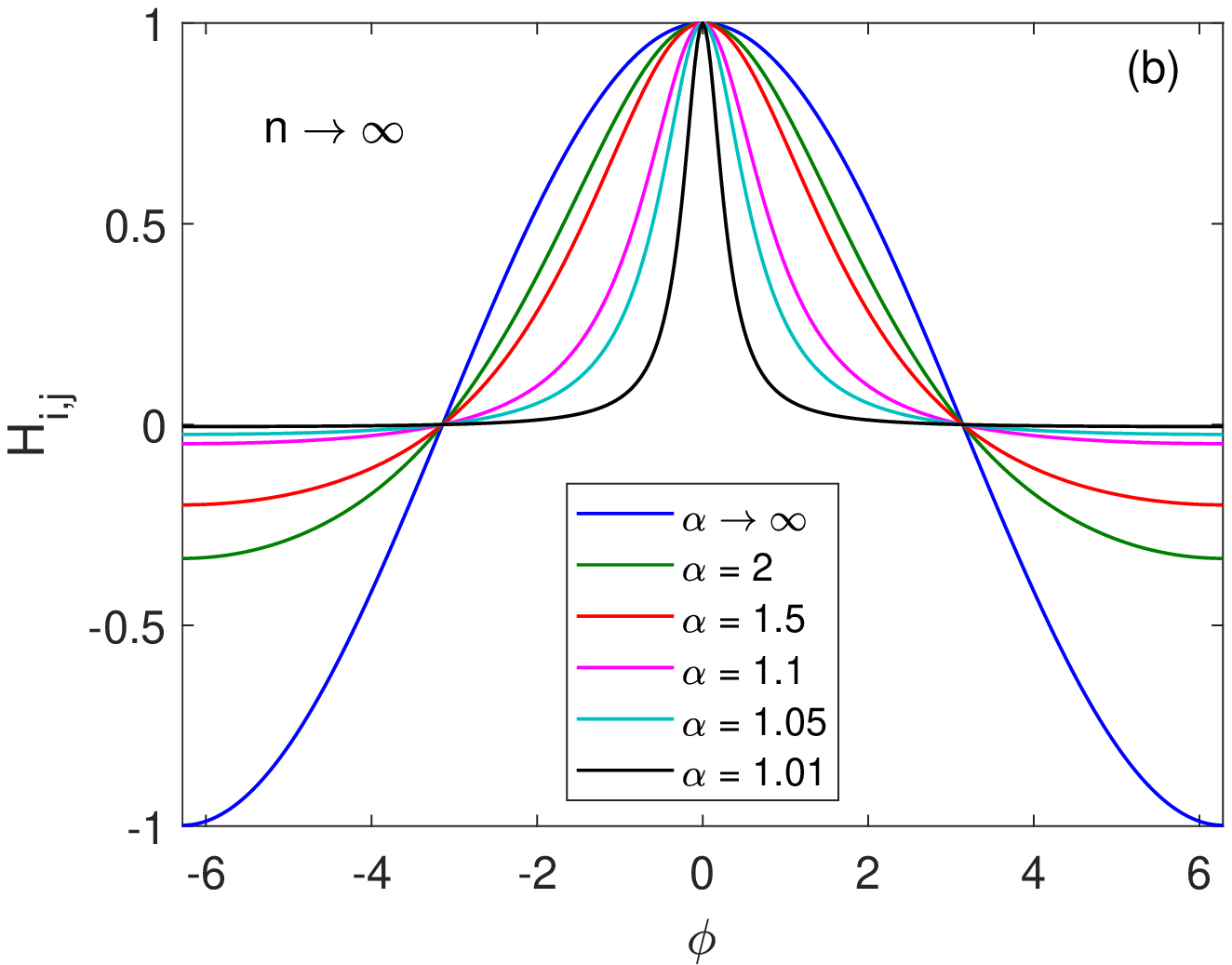}\label{fig:H_ij_phi-alp}}
\caption{Potential functions for (a) different values of the parameter $n$ in the limit of $\alpha \gtrsim 1$ and (b) different values of $\alpha$ in the limit of $n \to \infty$. In (a) the solid (dashed) curves correspond to odd (even) values of $n$.}
\label{fig:potential}
\end{figure}

In Ref.~\citep{mz-dth18} we have shown that the MPR model has the advantage over the Gaussian model with respect to filling gaps in skewed, non-Gaussian spatial data. It was due to the fact that the shape of the MPR potential function translates in higher probability for larger spin angle contrasts, i.e., larger differences between neighboring values of the spin angles. The latter are more likely to occur in skewed data with a heavier than normal right tail (e.g., following the lognormal distribution) and thus the MPR model is more suitable for modeling of such kind of data. In the present extension of the MPR model the role of the parameters $n$ and $\alpha$, which come from inclusion of higher-order couplings, is to control nonlinearity of the potential well. In Fig.~\ref{fig:potential} we show the shape of the potential function for different values of the parameter $n$ in the limit of $\alpha \gtrsim 1$ (Fig.~\ref{fig:H_ij_phi-n}) and for different values of $\alpha$ in the limit of $n \to \infty$ (Fig.~\ref{fig:H_ij_phi-alp}). Both parameters affect the width of the bell-shaped center of the function as well as the character of the wings. In particular both increasing $n$ and decreasing $\alpha$ decrease the width of the center and increase the values in the wings from $-1$ (for $n=1$ or $\alpha \gg 1$) to $0$ (for $n\gg1$ or $\alpha \gtrsim 1$). It is worth noticing that the wing shape depends on whether $n$ is odd or even. Namely, even $n$ encourage more states with larger differences between neighboring values of the spin angles, which may be beneficial in case of some types of non-Gaussian distributions.

\subsection{Anisotropic and further-neighbor interactions}
\label{ssec:anis}
\begin{figure}[t!]
\centering
\includegraphics[scale=0.55,clip]{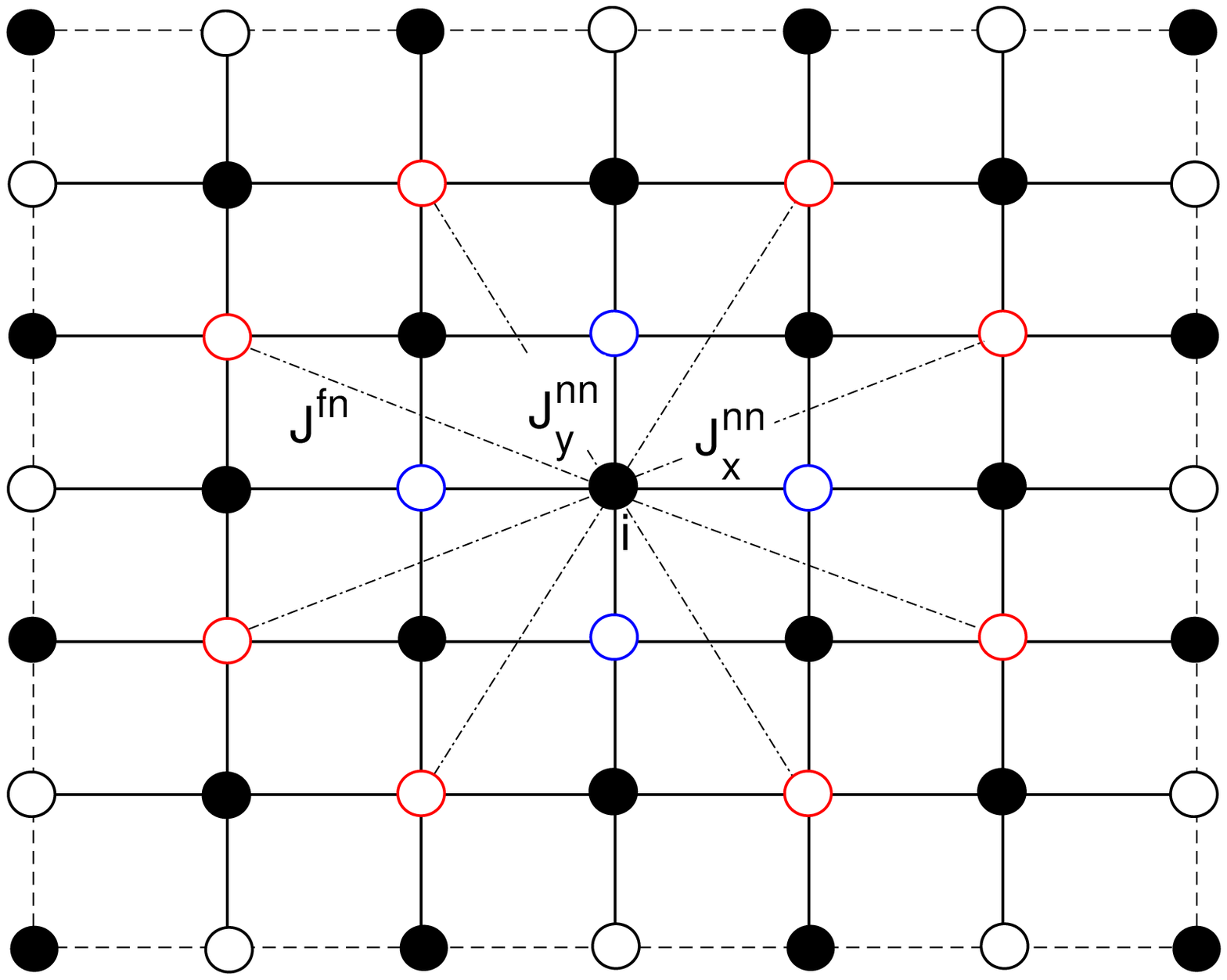}
\caption{Schematic view of the grid, partitioned into two interpenetrating subgrids (filled and empty circles), with the considered interactions. The central site $i$ interacts with its nearest neighbors (blue empty circles) along the x-axis (y-axis) through the interaction $\Jnn_x$ ($\Jnn_y$) and further-neighbors (red empty circles) through the interaction $\Jfn$.}
\label{fig:schematic}
\end{figure}

A schematic view of the grid showing all the interactions considered in the GPR model is presented in Fig.~\ref{fig:schematic}. In order to keep the vectorization and thus as high efficiency of the algorithm as in the original MPR method, we consider splitting the entire grid into two interpenetrating subgrids and restricting the interactions to  pairs of nodes that belong to different subgrids. By applying this so called checkerboard algorithm each Monte Carlo sweep can be completed in just two steps regardless of the grid size. This restriction allows to consider the nearest neighbor (nn) interactions, including their directional distinction to account for geometric anisotropy, and the further-neighbor (fn) interactions, which are are chosen as the fourth nearest considering the entire grid but only the second nearest considering solely the second subgrid (empty circles) with which the central spin is allowed to interact. 

Many spatial prediction methods are based on the assumption of geometric isotropy of the sample, i.e., the assumption that their properties do not depend on the spatial direction. Nevertheless, most spatial data exhibit a certain degree of geometric anisotropy and its neglecting may give rise to prediction errors. Its presence means that the correlation function does not depend only on the the lag vector but also on its direction. Consequently, the correlation lengths along different directions are different and the correlation isolevel
contours have elliptical shapes. The geometric anisotropy in a two-dimensional (2D) sampling coordinate system with $x$ and $y$ axes is fully characterized by the anisotropy ratio $R$ and the orientation angle $\Theta$. $R$ is defined as the ratio of the correlation lengths along the principal axes of anisotropy, which represent the semi-axes of the elliptical isolevel contours. $\Theta$ is the rotation angle between the principal axes of anisotropy and the coordinate system axes.

In the present GPR model the geometric anisotropy can be at least partially captured by introducing directional dependence in the exchange interaction parameters. Assuming that the principal axes of anisotropy coincide with the coordinate system axes (i.e., $\Theta=0$), in the Hamiltonian~(\ref{eq:Hamiltonian_GPR}) it is implemented by distinguishing the nn interactions $\Jnn_x$ and $\Jnn_y$ between nearest-neighbor spins along $x$ and $y$ axes, respectively. This assumption is somewhat restrictive but we note that, in principle, directional dependence can also be introduced to the fn interactions by considering different parameters $\Jfn_d$ in four different diagonal directions $d$, which would further increase the model's ability to capture geometric anisotropy in different directions. Nevertheless, for simplicity in the following we keep the fn interactions isotropic.

By inclusion pairwise couplings going beyond nearest neighbor spins one can naturally expect to impose a better control of correlations at more distant lags. As their importance generally diminishes with distance, it would make sense to start adding to the nearest second, third, and gradually fourth neighbor interactions. However, for the reasons described above we restricted our considerations to only the fourth nearest neighbors, as shown in Fig.~\ref{fig:schematic}. On the other hand, the advantage is that there are eight fn neighbors, compared to only four in cases of the first, second, and third nearest neighbors. Furthermore, they lie in the directions complementary to those corresponding the nn spins, which might be beneficial in the presence of geometric anisotropy.

\subsection{External bias field}
\label{ssec:bias}
Finally, we consider the effect of the external bias field $h_i$ as well as the form of the function $f$ and the coupling parameter $K$. Generally, they are expected to control the simulated data distribution. However, considering the limited ability of the MPR model parameter to appropriately capture the data smoothness, we would like them to also control the degree of smoothness of the spatial variation. The term $-KB_i=-Kf(\phi_i-h_i)$ can be viewed as a cost function at the site $i$ measuring the distance between the simulated spin value $\phi_i$ and the bias field value $h_i$. Its minimal value corresponds to the best match of the two and the parameter $K>0$ serves as a weight of the external bias field term with respect to the other terms in the Hamiltonian. Here we have two choices to be made: the bias field $h_i$ and the form of the function $f$. 

In order to control the smoothness of the spatial configuration, the bias field $h_i$ should be a smooth approximation of the true values which can also be efficiently calculated. Based on our previous tests (see Refs.~\citep{mz-dth13b,mz-dth18}) the bias field obtained by means of bicubic (BC) sample interpolation using the the Matlab\circledR built-in function \verb+griddata+ satisfies these criteria. We considered several suitable forms of the function $f$, such as the square function $y=-(\phi_i-h_i)^2$, commonly used in the optimization problem, $y=-|\phi_i-h_i|$,  $y=-\sqrt{|\phi_i-h_i|}$, $y=-\log(|\phi_i-h_i|)$, as well as the form inspired by the presently used MPR potential function $y=\cos[(\phi_i-h_i)/2]$. By comparing different choices of $f$ we found considerable differences in the values of the parameter $K$, which yield optimal validation measures, but no significant differences in the prediction performance at those parameter values. Therefore, in this paper we opted for presenting the results obtained by the form $y=\cos[(\phi_i-h_i)/2]$.

\begin{figure}[t!]
\centering
\subfigure{\includegraphics[scale=0.55,clip]{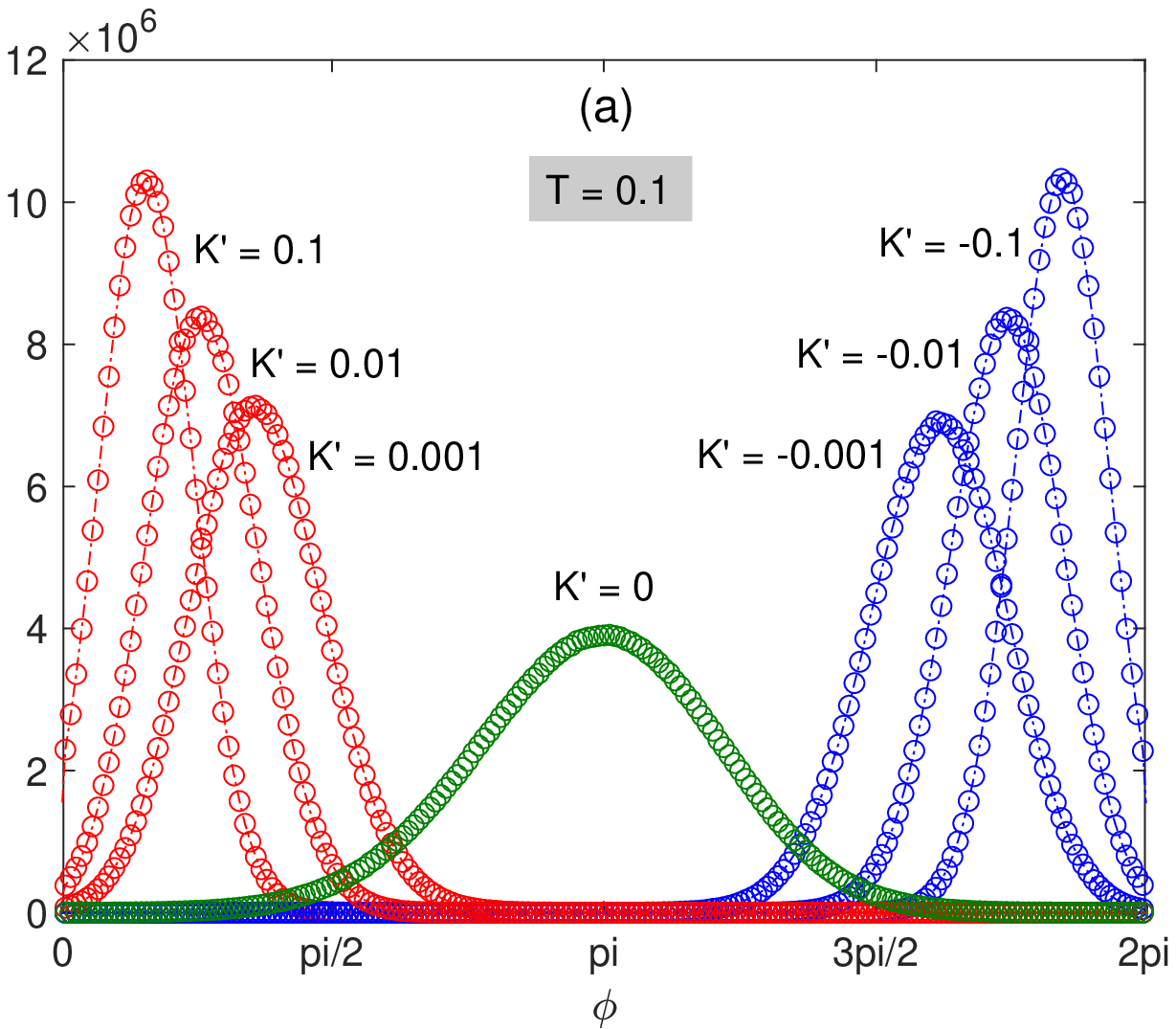}\label{fig:phi-hist_T0_1}}
\subfigure{\includegraphics[scale=0.55,clip]{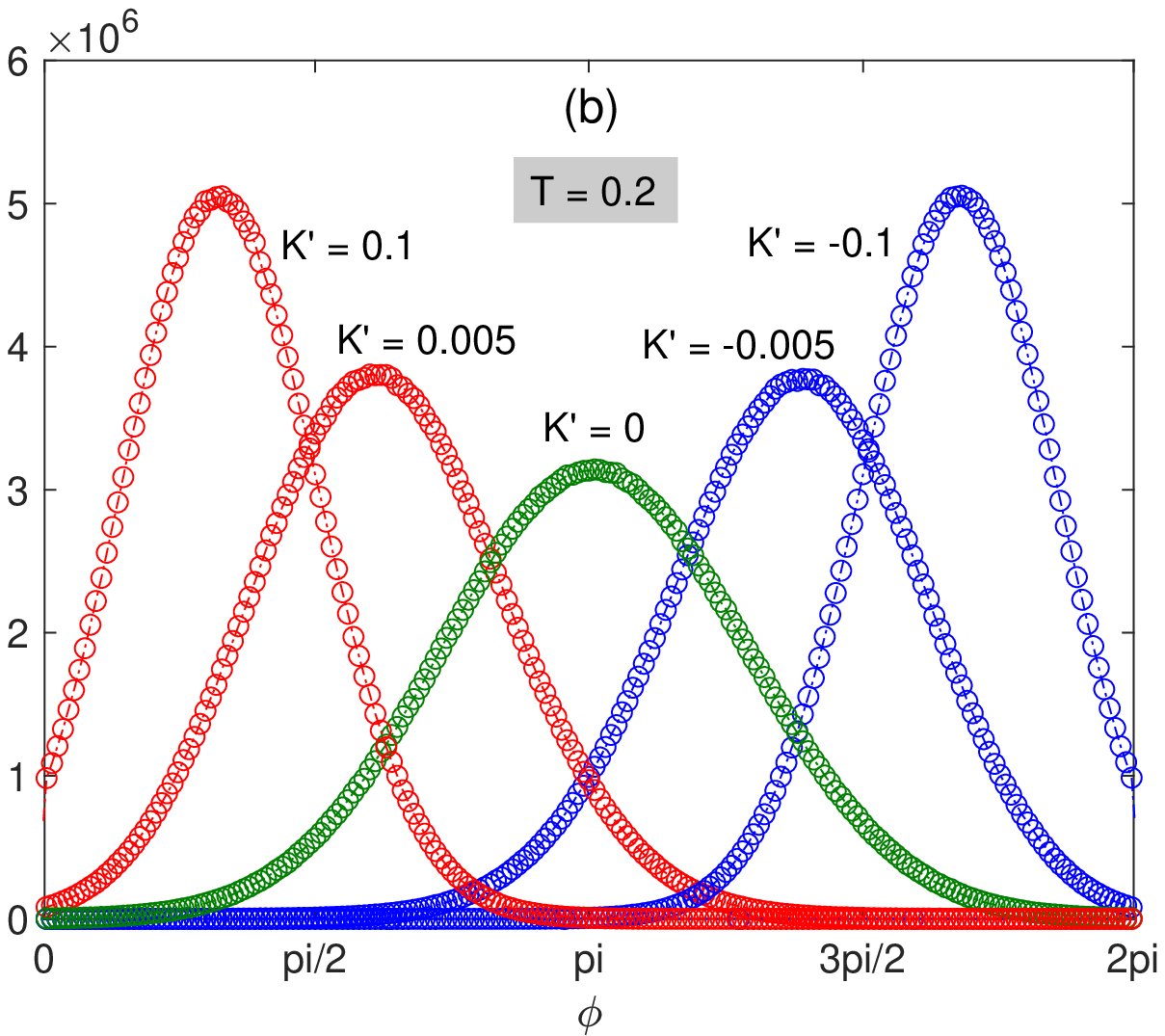}\label{fig:phi-hist_T0_2}}
\caption{Effect of the ``magnetic field'' parameter $K'$ on spin angle distributions obtained from unconditional simulations at (a) $T=0.1$ and (b) $T=0.2$.}
\label{fig:hist-field}
\end{figure}

If the focus is laid on controlling distribution rather than smoothness, then the bias field can be set identically to $h_i = 0$ and the function $f:y=\cos(\phi_i/2)$ can be viewed as ``magnetization'' controlled by an uniform external ``magnetic field'' $K$. To distinguish it from the above case, hereafter, we will refer to the ``magnetic field'' parameter as $K'$ and its value can be both positive and negative. In particular, $K'>0$ will encourage smaller and $K'<0$ larger spin angles. Some typical spin distributions at different fields and temperatures, obtained from unconditional MC simulations, are plotted in Fig.~\ref{fig:hist-field}. Apparently, while the temperature controls the width of the distribution, the field controls its skewness. Thus, one can expect that the external ``magnetic field'' parameter $K'$ in the GPR model can be particularly useful in better reconstruction of non-Gaussian highly skewed distributions.

\subsection{GPR Gibbs Markov random field and model parameters}
\label{ssec:gpr}
The GPR Gibbs Markov random field is defined by means of the Boltzmann-Gibbs distribution
\begin{equation}
 \label{eq:bg-pdf}
 F =\frac{1}{{\mathcal Z}} \exp(-{\mathcal H}/k_{B}T),
\end{equation}
where the energy ${\mathcal H}$ is defined in (\ref{eq:Hamiltonian_GPR}), the normalization constant ${{\mathcal Z}}$ is the partition function, $k_B$ is the Boltzmann constant (hereafter set to one), and $T$ is the temperature. 

We note that the original MPR model had only one parameter (temperature), which was estimated by a so called specific energy matching method. Subsequently, missing values were obtained from conditional Monte Carlo simulations by taking the mean of the conditional distributions at the respective sites given the sample data. On the other hand, the GPR model involves up to six parameters and, compared to the MPR model, finding their optimal values efficiently will become a more involved task and it is left for future considerations. In the present study we instead focus on the exploration of the parameter space with the goal to study the effect of each individual parameter added to the original MPR model on the prediction performance. In particular, we evaluate various prediction validation measures with varying parameters. Typically, we let vary the temperature - the MPR model parameter - and one more parameter, while all the remaining parameters are kept fixed. Thus, the validation measures are presented in the $T-x$ parameter planes, with $x=n,\alpha,\Jnn,\Jfn,$ and $K$ or $K'$.

\section{GPR Prediction and Validation Measures}
\label{sec:vali-design}

Let $G\subseteq \mathbb{Z}^{2}$ be a 2D rectangular grid $G$ of size $N_{G}=L \times L$ with the grid sites denoted as $\mathbf{r}_{i}=(x_i, y_i) \in {\mathbb{R}}^2$, where $i=1, \ldots, N_{G}$ and $\mathbb{R}$ is the set of real numbers. We assume that the data represent a realization of the random field $Z(\mathbf{r};\omega)$ sampled on   $G_{s} \subset G$, where $G_{s}= \{\mathbf{r}_{i}\}_{i=1}^{N}$ and  $N<N_{G}$. The values of the data set are denoted by $Z_{s}=\{z_{i} \in {\mathbb{R}} \}_{i=1}^{N}$. The set of prediction points is denoted by $G_{p}=\{\mathbf{r}_{p}\, \}_{p=1}^{P} $ such that $G_{s} \cup G_{p} = G$, $G_{s} \cap G_{p} = \emptyset$, and $P+N = N_{G}$. The set of the random field values at the prediction sites will be denoted by $Z_{p}$.

The GPR prediction method is based on the GPR Gibbs-Markov random field defined in (\ref{eq:bg-pdf}). The original data are first transformed to continuously-valued ``spin'' variables by mapping from the original space to the spin angle space $[0, 2\pi]$ using the linear transformation

\begin{equation}
\label{map}
Z_{s} \mapsto \Phi_{s} = \frac{2\pi(Z_{s}- z_{s,\min})}{(z_{s,\max} - z_{s,\min})},
\end{equation}
where $z_{s,\min}$ and $z_{s,\max}$ are the minimum and maximum sample values and $\Phi_{s}=\{\phi_{i}\}_{i=1}^{N}$ and $\phi_{i} \in [0,2\pi]$, for $i=1, \ldots, N$. Then spatial correlations, typical in geophysical and environmental data sets, are captured via short-range interactions between the spins of the GPR model with the energy functional (\ref{eq:Hamiltonian_GPR}). The spatial prediction at missing data sites is based on performing conditional MC simulations and taking the mean of the respective conditional distribution in thermodynamic equilibrium. Even the simpler MPR model, controlled by only one parameter - temperature, has been shown to display a rather flexible correlation structure. The corresponding MPR prediction method is designed to operate automatically with high computational efficiency and its detailed algorithm is described in Ref.~\citep{mz-dth18}. The present GPR prediction method follows practically the same algorithm but instead of the MPR it uses a more complex GPR energy functional (\ref{eq:Hamiltonian_GPR}) with additional parameters, which are expected to further increase the method's flexibility and prediction performance.

We employ several validation measures for performance evaluation. Let $Z(\mathbf{r}_p)$ be the true value at $\mathbf{r}_p$ and $\hat{Z}(\mathbf{r}_p)$ its estimated value. The estimation error is defined as $\epsilon(\mathbf{r}_p)= Z(\mathbf{r}_p) - \hat{Z}(\mathbf{r}_p)$. The following validation measures are then defined:

\noindent average absolute error
\begin{equation}
{\rm AAE} = (1/P)\sum_{\mathbf{r}_{p} \in G_{p}}|\epsilon(\mathbf{r}_p)|,
\end{equation}
average relative error
\begin{equation}
{\rm ARE}=(1/P)\sum_{\mathbf{r}_{p} \in G_{p}}\epsilon(\mathbf{r}_p)/Z(\mathbf{r}_p),
\end{equation}
average absolute relative error
\begin{equation}
{\rm AARE} =(1/P)\sum_{\mathbf{r}_{p} \in G_{p}}|\epsilon(\mathbf{r}_p)|/Z(\mathbf{r}_p),
\end{equation}
and root average squared error
\begin{equation}
{\rm RASE} =\sqrt{\frac{1}{P}\sum_{\mathbf{r}_{p} \in G_{p}}\,\epsilon^2(\mathbf{r}_p)}.
\end{equation}
For each complete data set we generate $S$ different sample configurations with missing data and calculate the above validation measures. Averaging over all the sample configurations provide global statistics, denoted by MAAE, MARE, MAARE, and MRASE, where the letter ``M'' denotes the configuration mean.

\section{Results}
\label{sec:results}

\subsection{Data}
\label{ssec:data}
The prediction performance is tested on spatially correlated synthetic data, simulated on the square grid $G$ of the size $N_{G}=L \times L$, using the spectral method of mode superposition~\citep{drum87,dth_2020}. Due to the extensive number of simulations performed for $S$ sampling configurations over a multi-dimensional parameter space,  the grid side size is fixed to the relatively small value $L=64$. The data represent field realizations generated from the joint  Gaussian, $Z \sim N(m=5,\sigma=2)$, and lognormal, $\log Z \sim N(m=5,\sigma=2)$, probability distributions. The spatial continuity of the realizations is imposed by means of  a flexible Whittle-Mat\'{e}rn (WM) covariance model given by
\begin{equation}
\label{mate} G_{\rm Z} (\mathbf{u})=
\frac{{2}^{1-\nu}\, \sigma^{2}}{\Gamma(\nu)} \rho^{\nu}K_{\nu}(\rho).
\end{equation}
In Eq.~\eqref{mate}  the parameter $\sigma^2$ is the variance of the fluctuations, $\nu$ is the smoothness index (higher values of $\nu$ correspond to smoother fields), and $K_{\nu}(\cdot)$ is the modified Bessel function of the second kind of order $\nu$. The normalized lag distance is given by $\rho=\sqrt{u^{2}_{1}/\xi_{1}^{2} + u^{2}_{2}/\xi_{2}^{2}}$, where $\mathbf{u}=(u_{1},u_{2})$ is the lag vector between two grid points $\mathbf{r}_{i}$ and $\mathbf{r}_{j}$, while $\xi_{1}$ and $\xi_{2}$ are correlation lengths in the horizontal and vertical directions, respectively. This definition of $\rho$ allows for geometric (elliptical) anisotropy with principal axes aligned with the coordinate system.  Thus,  $\xi_{1} \neq \xi_{2}$ leads to anisotropic data, while $\xi_{1} = \xi_{2}$ implies isotropy. 

In the cases studied below we consider fixed $\xi_{2}=2$ with  $\xi_{1}=2$  for isotropic and $\xi_{1}=4$ for anisotropic data sets. We also use two values of $\nu = 0.5$ and $\nu = 2.5$, corresponding to relatively rough and smooth spatial processes, respectively. Incomplete samples  ${Z}(G_{s})$ of size $N=N_{G} - \lfloor (p/100\%)\,N_{G} \rfloor$ are generated  by removing (i) randomly $P=\lfloor (p/100\%)\,N_{G} \rfloor$ points or (ii) a randomly selected solid square block of side length $L_B$. The removed (simulated missing) values are set aside to be used as the \emph{validation set}. We generate $S=100$ different sampling configurations for selected degrees of thinning ($p=33\%$ and $66\%$) and block size ($L_B=20$). The predictions at the removed (validation) points are calculated and compared with the true values.

\subsection{Effect of anisotropic interaction}
\label{ssec:anis_eff}

\begin{figure}[t!]
\centering \vspace{-10mm}
\subfigure{\includegraphics[scale=0.43,clip]{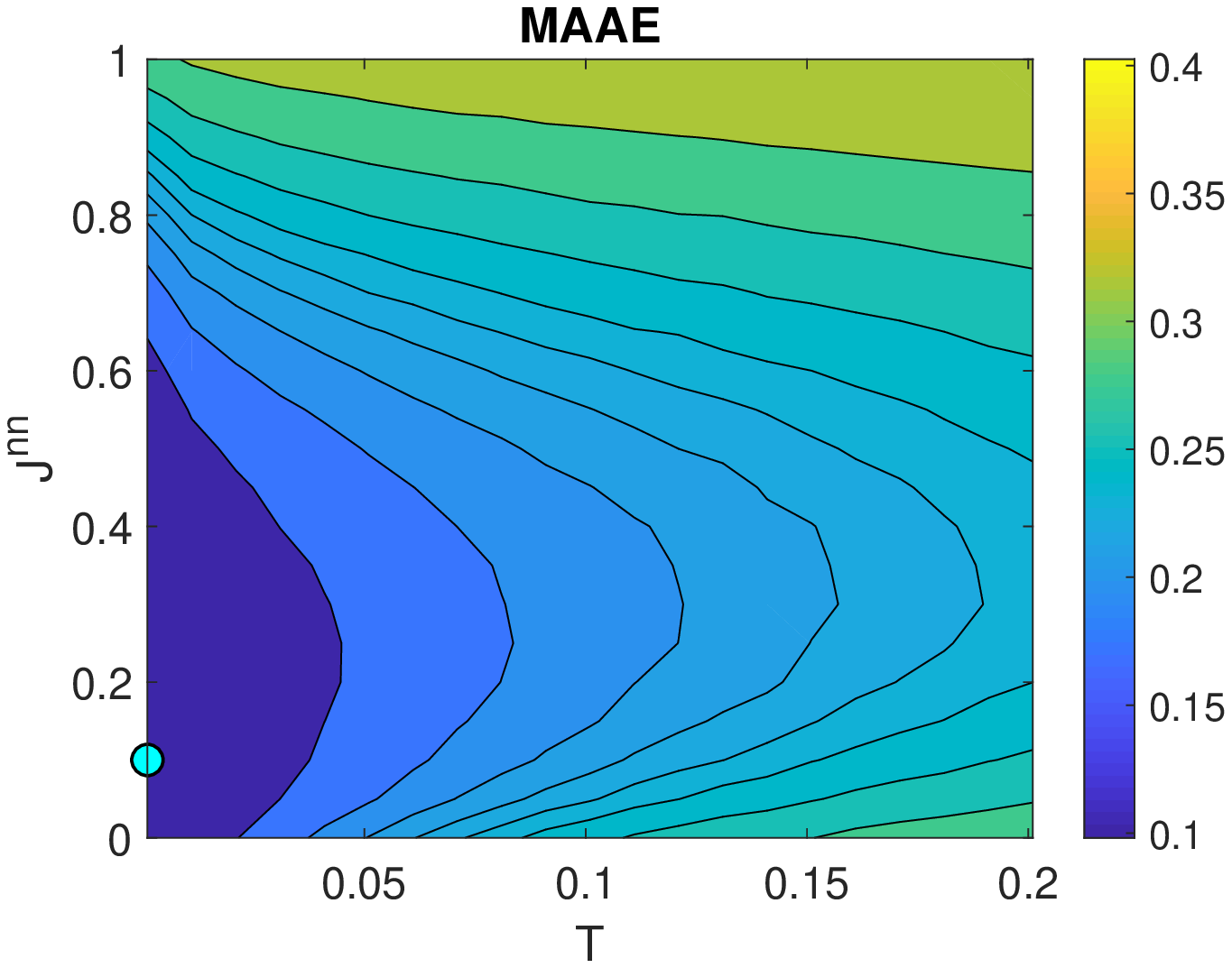}\label{fig:ani_b2_p033_maae}} 
\subfigure{\includegraphics[scale=0.43,clip]{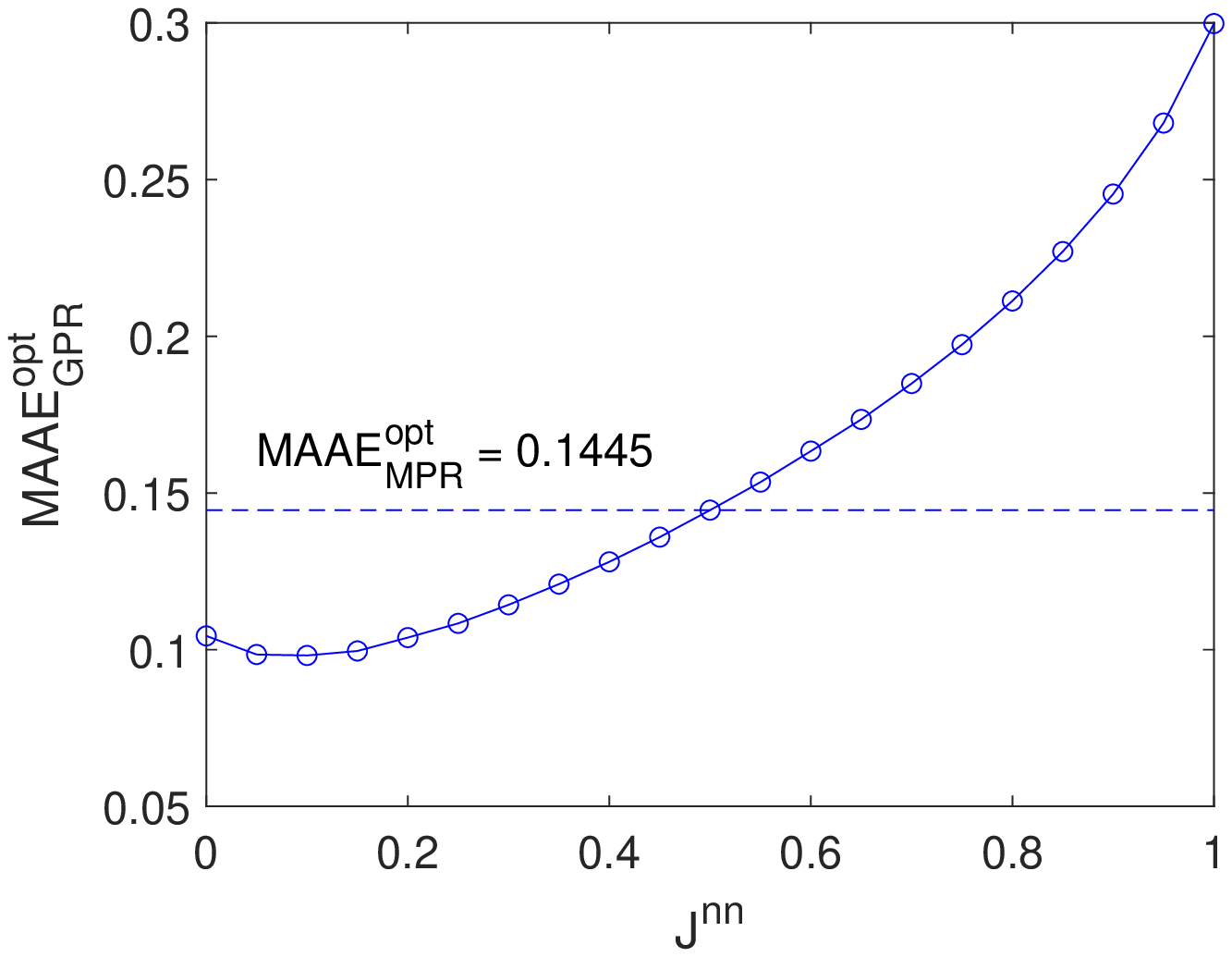}\label{fig:ani_b2_p033_maae_opt}}\\ \vspace{-3mm}
\subfigure{\includegraphics[scale=0.43,clip]{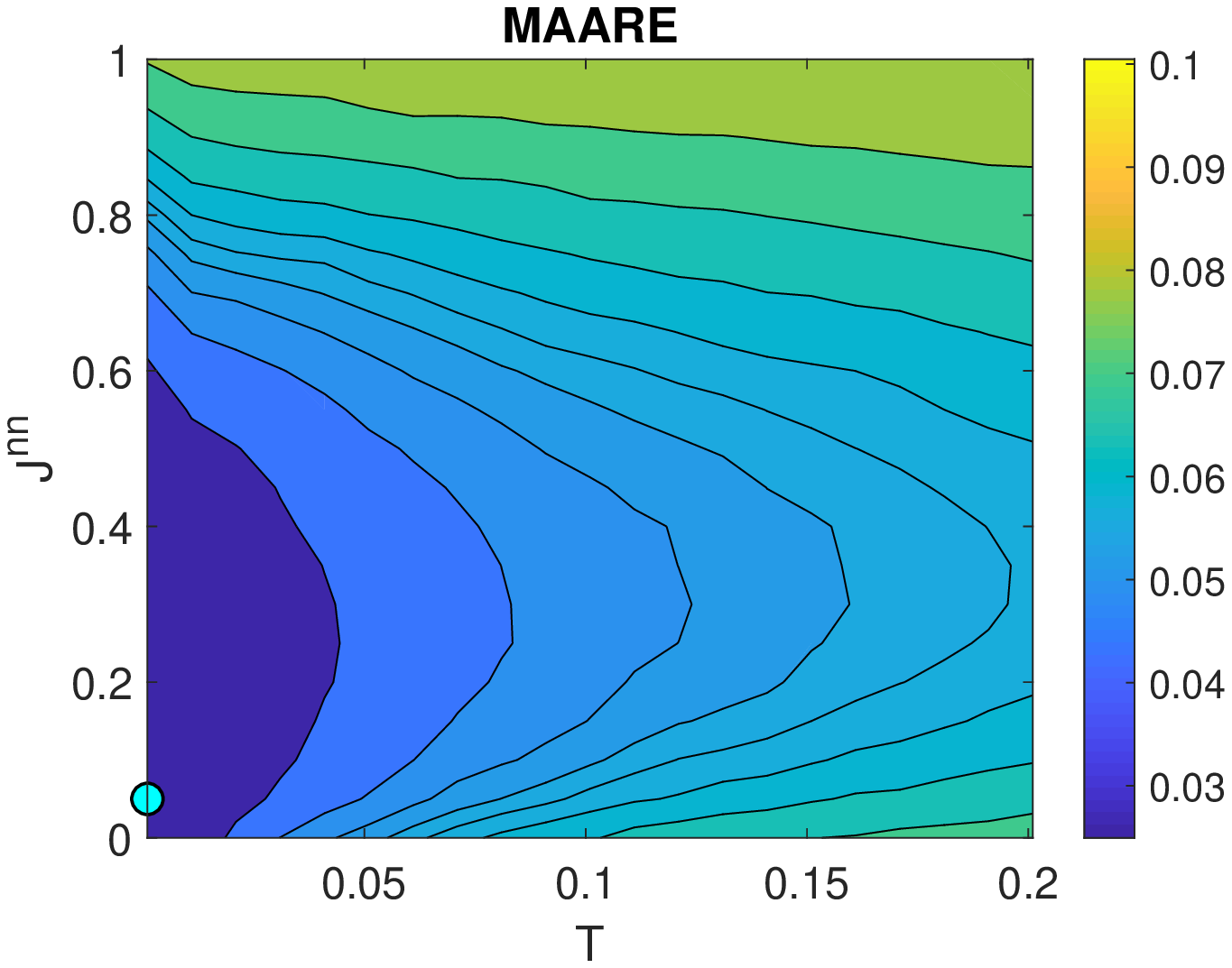}\label{fig:ani_b2_p033_maare}}
\subfigure{\includegraphics[scale=0.43,clip]{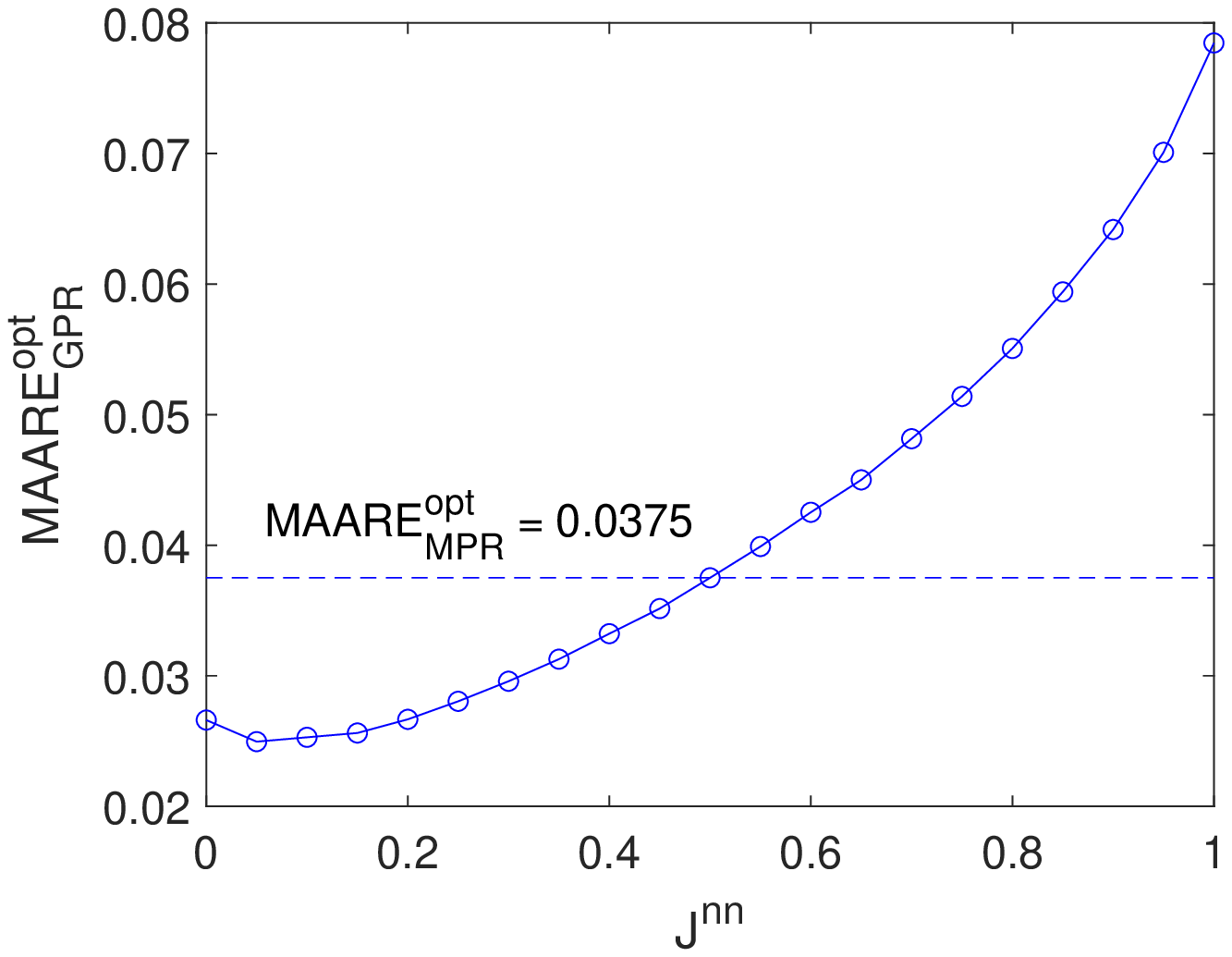}\label{fig:ani_b2_p033_maare_opt}}\\ \vspace{-3mm}
\subfigure{\includegraphics[scale=0.43,clip]{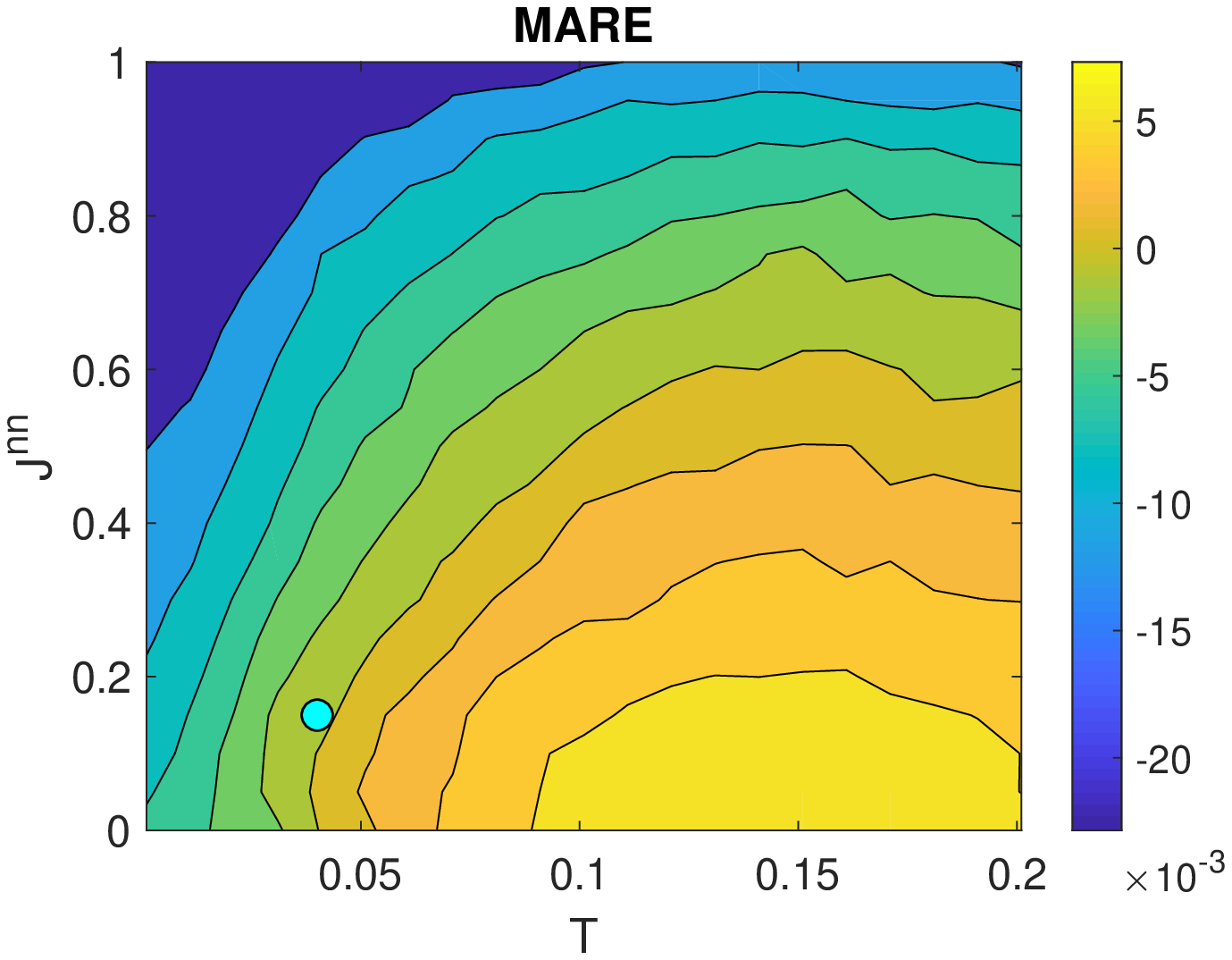}\label{fig:ani_b2_p033_mare}}
\subfigure{\includegraphics[scale=0.43,clip]{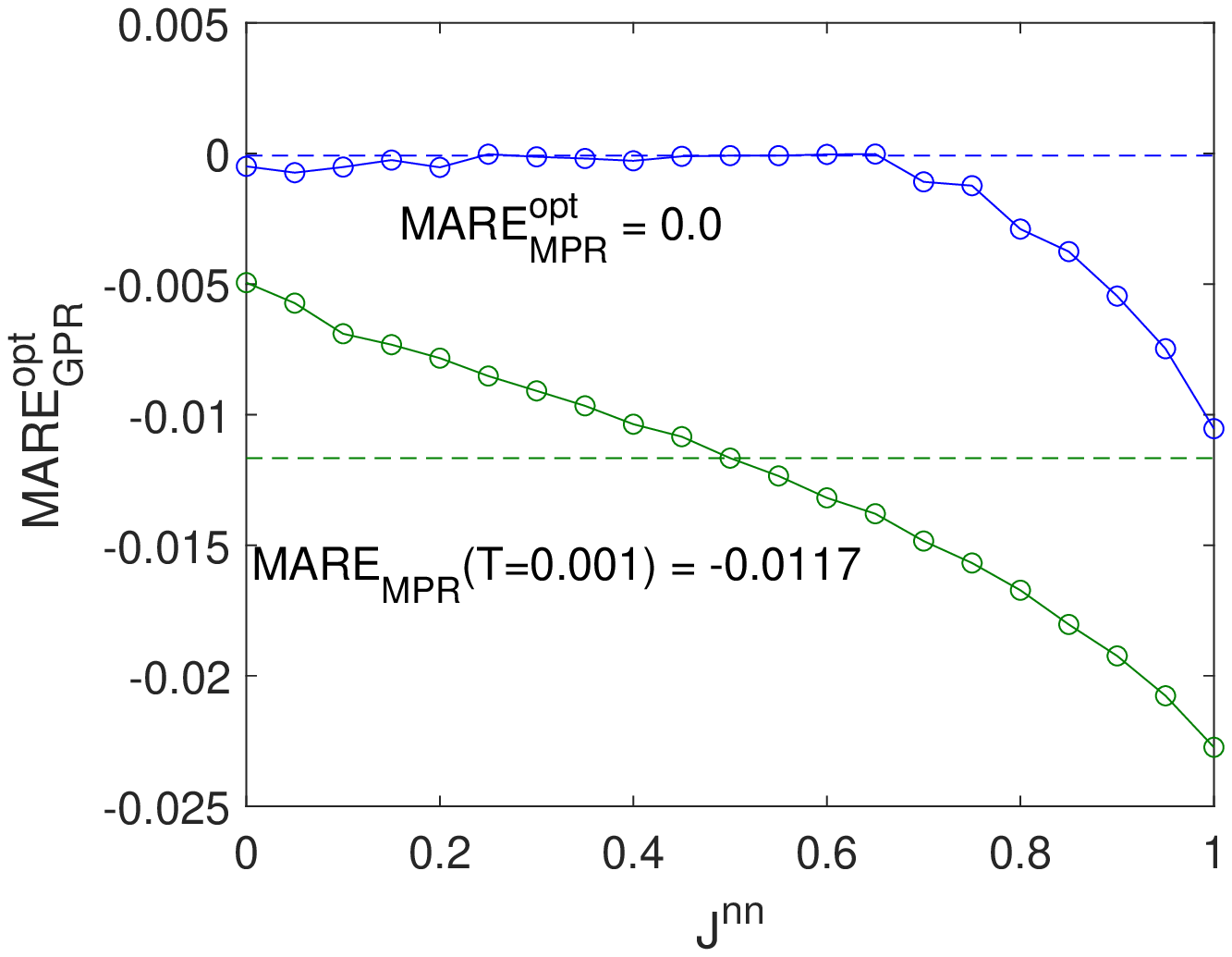}\label{fig:ani_b2_p033_mare_opt}}\\ \vspace{-3mm}
\subfigure{\includegraphics[scale=0.43,clip]{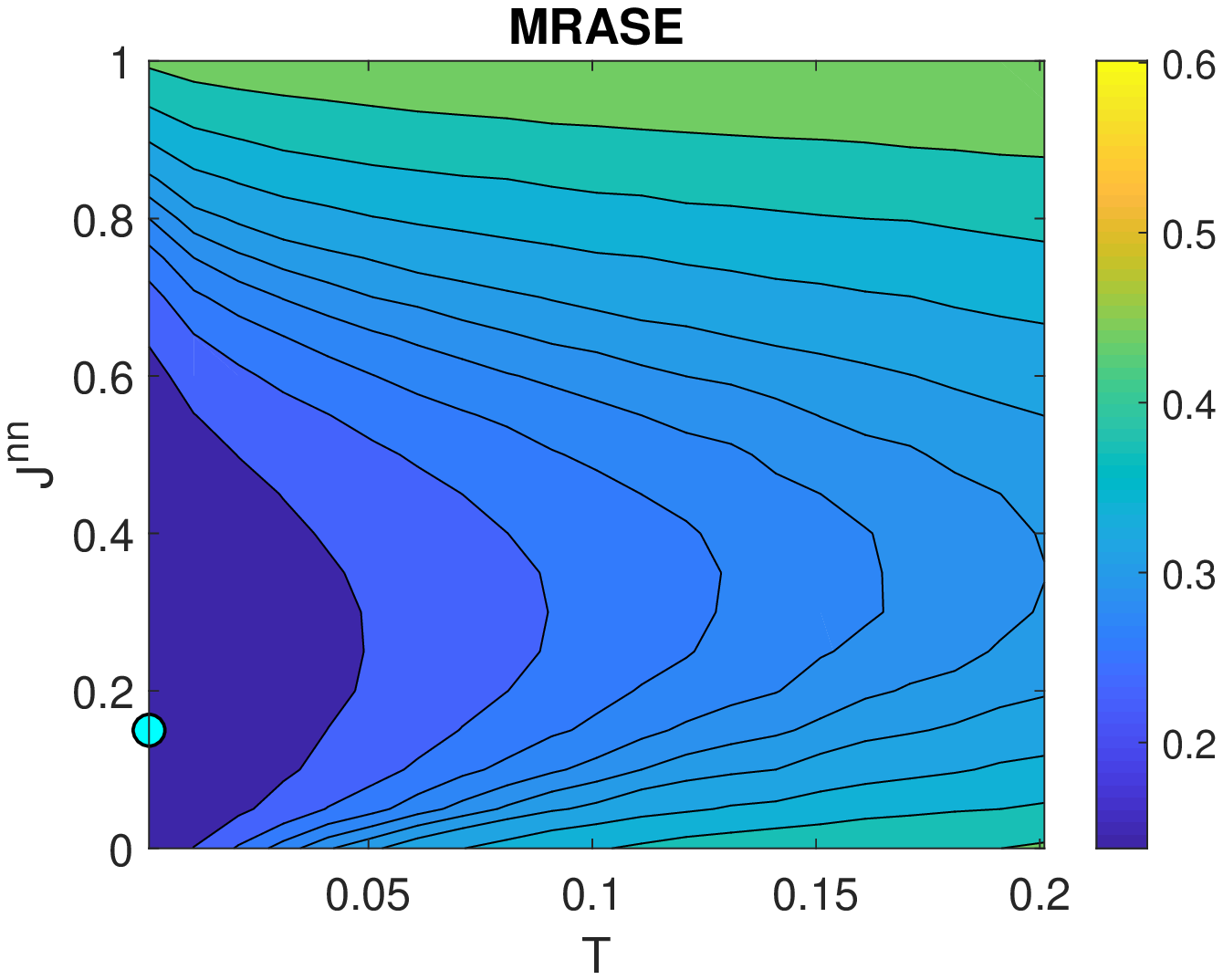}\label{fig:ani_b2_p033_mrase}}
\subfigure{\includegraphics[scale=0.43,clip]{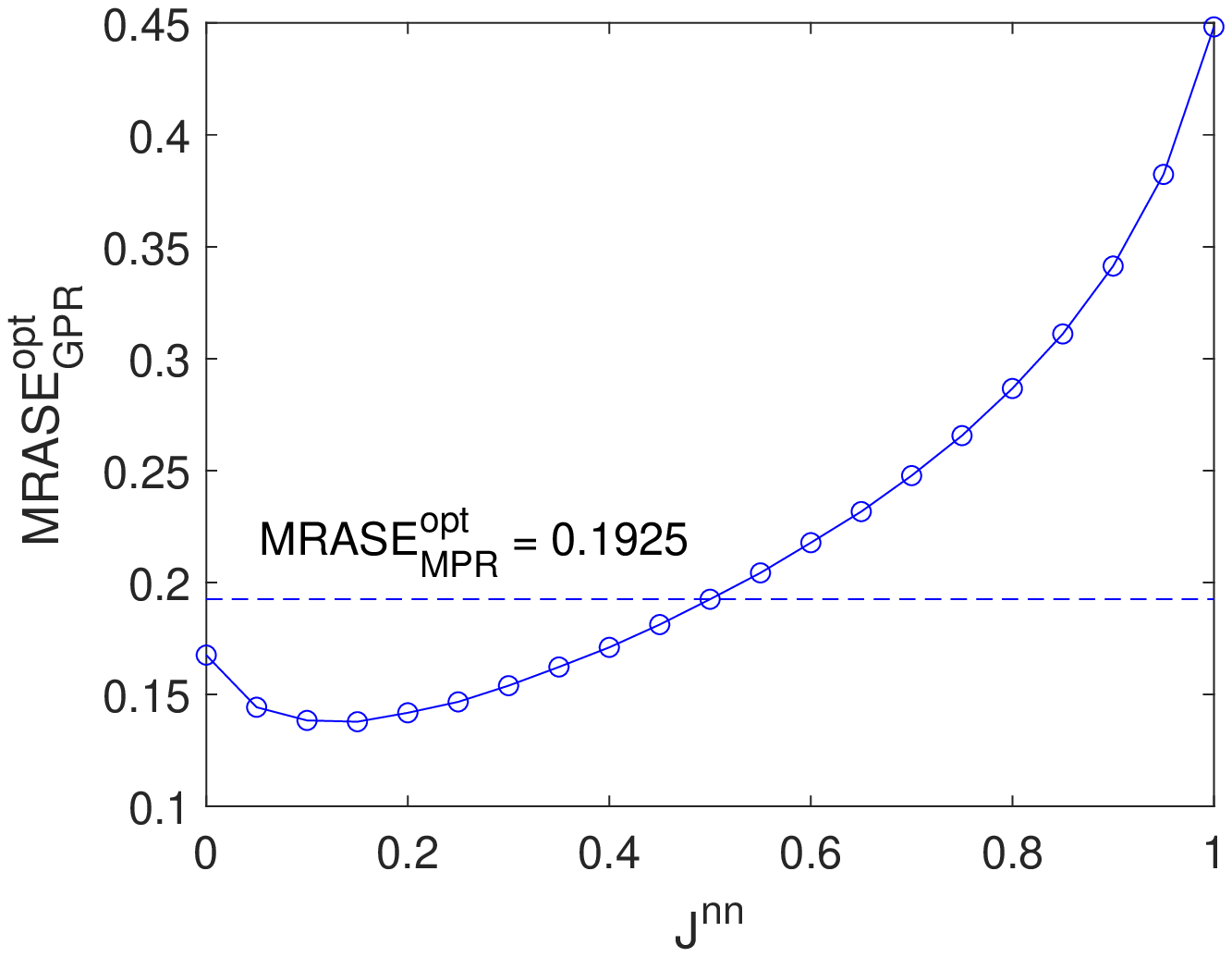}\label{fig:ani_b2_p033_mrase_opt}} \vspace{-3mm}
\caption{Left column: contour plots of the validation measures in the $T-\Jnn$ parameter plane;  the cyan circles mark the optimal values. Right column: validation measures as functions of  $\Jnn$ at optimal (blue curves) and $T=0.001$ (green curve) temperatures. The dashed lines mark the optimal values obtained by means of the MPR model. $S=100$ samples are generated from the Gaussian random field with $\xi_1=4,\xi_2=2, \nu = 2.5$. The percentage of missing data is $p=33\%$.}
\label{fig:errors_ani}
\end{figure}

To test the effect of introducing direction-dependent nn interactions $\Jnn_x$ and $\Jnn_y$ we consider the anisotropic data with normal distribution $Z \sim N(m = 5, \sigma = 2)$, the correlation lengths $\xi_{1}=4$ and $\xi_{2}=2$ along the $x$ and $y$ axes, respectively, and the smoothness parameter $\nu =2.5$. Missing data were generated by random thinning with $p=33\%$ and the predictions were obtained by using the GPR model at various values of the parameters $T$ and $\Jnn$, with the remaining model parameters kept fixed: $n=1$, $\alpha = \infty$, $\Jfn=0$, and $K=0$. The respective validation measures are presented in Fig.~\ref{fig:errors_ani} as contour plots in the $T-\Jnn$ parameter plane (left column). The cyan circles show the optimal values, corresponding to the minimal errors.

It is worth noticing that except MARE all the optimal values correspond to the lowest simulated temperature $T=0.001$. This can be attributed to the fact that  lower temperatures generate smoother realizations~\citep{mz-dth15}. Since the present data are rather smooth, the best prediction performance is achieved at the lowest temperature. From all the figures it is apparent that the respective measures are asymmetric with respect to the the axis $\Jnn=0.5$, corresponding to the isotropic case $\Jnn_x=\Jnn_y$. In particular, better prediction performance can be observed for $\Jnn<0.5$ with the optimal values at around $\Jnn \approx 0.1$. This corresponds to a strongly anisotropic interaction with the intensities $\Jnn_x \approx 0.9$ and $\Jnn_y \approx 0.1$ along the $x$ and $y$ axes, respectively. The behavior of MARE is specific in the sense that the values in the $T-\Jnn$ plane can be negative or positive and thus the optimal values corresponding to zero can be found along the isoline crossing the area $[T_{\min},T_{\max}]\times [\Jnn_{\min},\Jnn_{\max}] \approx [0.03,0.2] \times [0,0.6]$.

The panels in the right column compare the respective measures obtainable by the GPR method for different $\Jnn$ with those obtained by the MPR method, i.e., the case of $\Jnn=0.5$. All the presented GPR (symbols) and MPR (dashed lines) values correspond to temperatures at which they are optimal. In the specific case of MARE we also show the values corresponding to $T=0.001$ at which all the remaining validation measures show optimal performance (green curve)~\footnote{There can be only one temperature corresponding to overall optimum of all the prediction measures as a set, which in this case would be $T \approx 0.001$.}. From all the figures it is evident that by allowing anisotropic nn interactions the MPR prediction performance can be substantially improved. For example, close to the optimal parameter values $(T_{\mathrm{opt}},\Jnn_{\mathrm{opt}}) \approx (0.001,0.1)$ the MAAE, MARE, MAARE, and MRASE errors can be decreased by $32\%, 33\%, 41\%$ and $28\%$, respectively.

\subsection{Effect of further-neighbor interaction}
\label{ssec:fn_exchange}

\begin{figure}[t!]
\centering \vspace{-10mm}
\subfigure{\includegraphics[scale=0.43,clip]{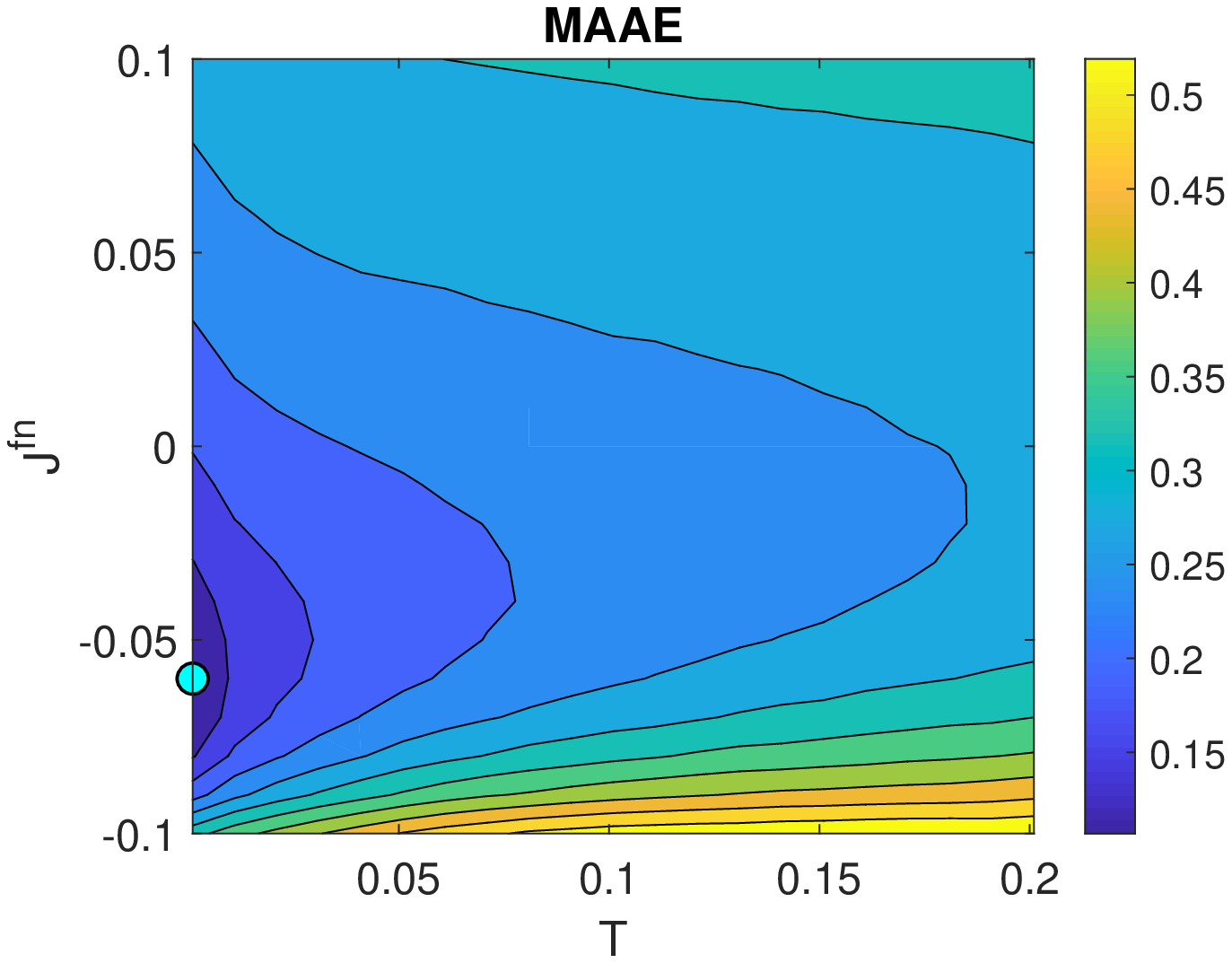}\label{fig:jnnn_p033_maae}}
\subfigure{\includegraphics[scale=0.43,clip]{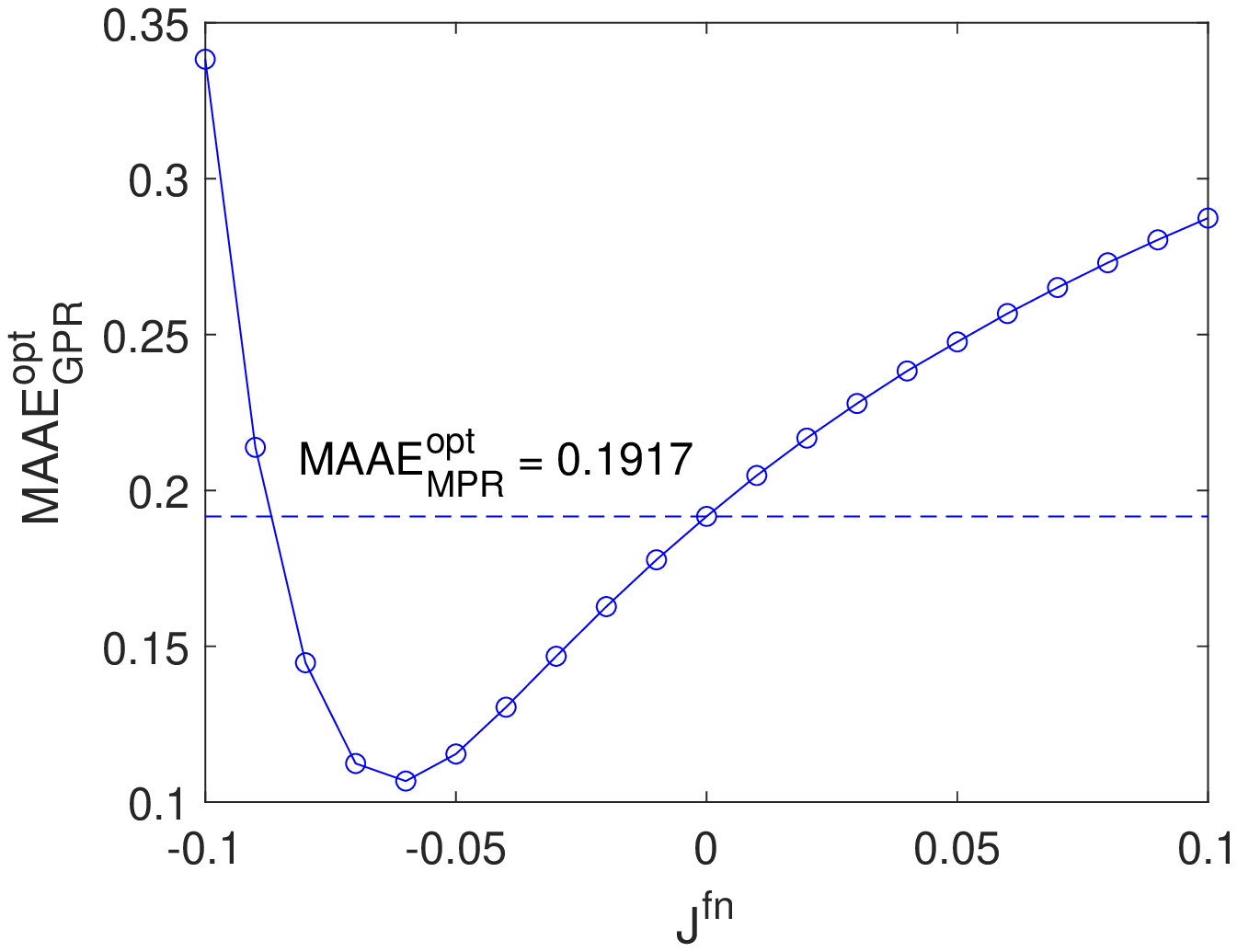}\label{fig:jnnn_p033_maae_opt}}\\ \vspace{-3mm}
\subfigure{\includegraphics[scale=0.43,clip]{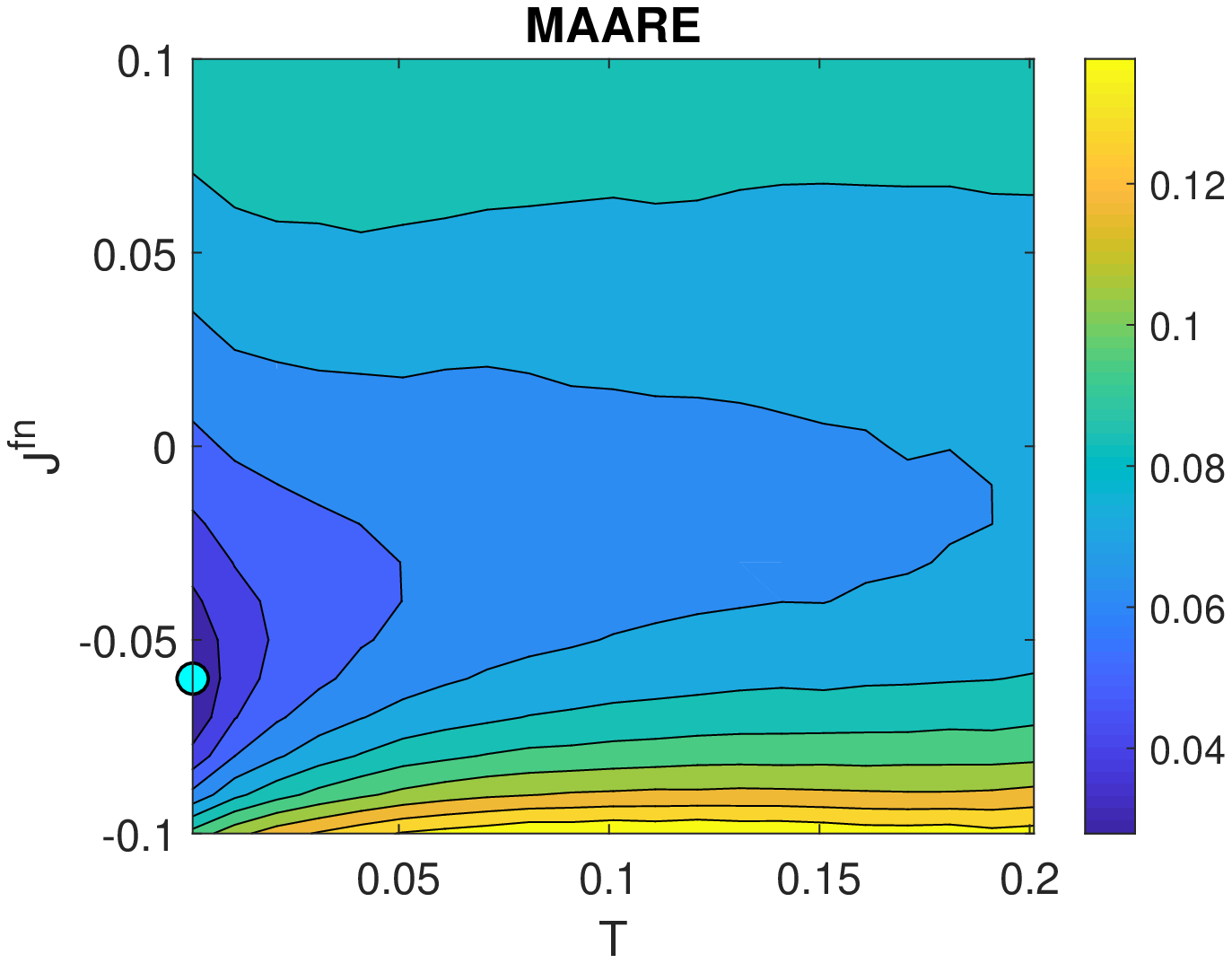}\label{fig:jnnn_p033_maare}}
\subfigure{\includegraphics[scale=0.43,clip]{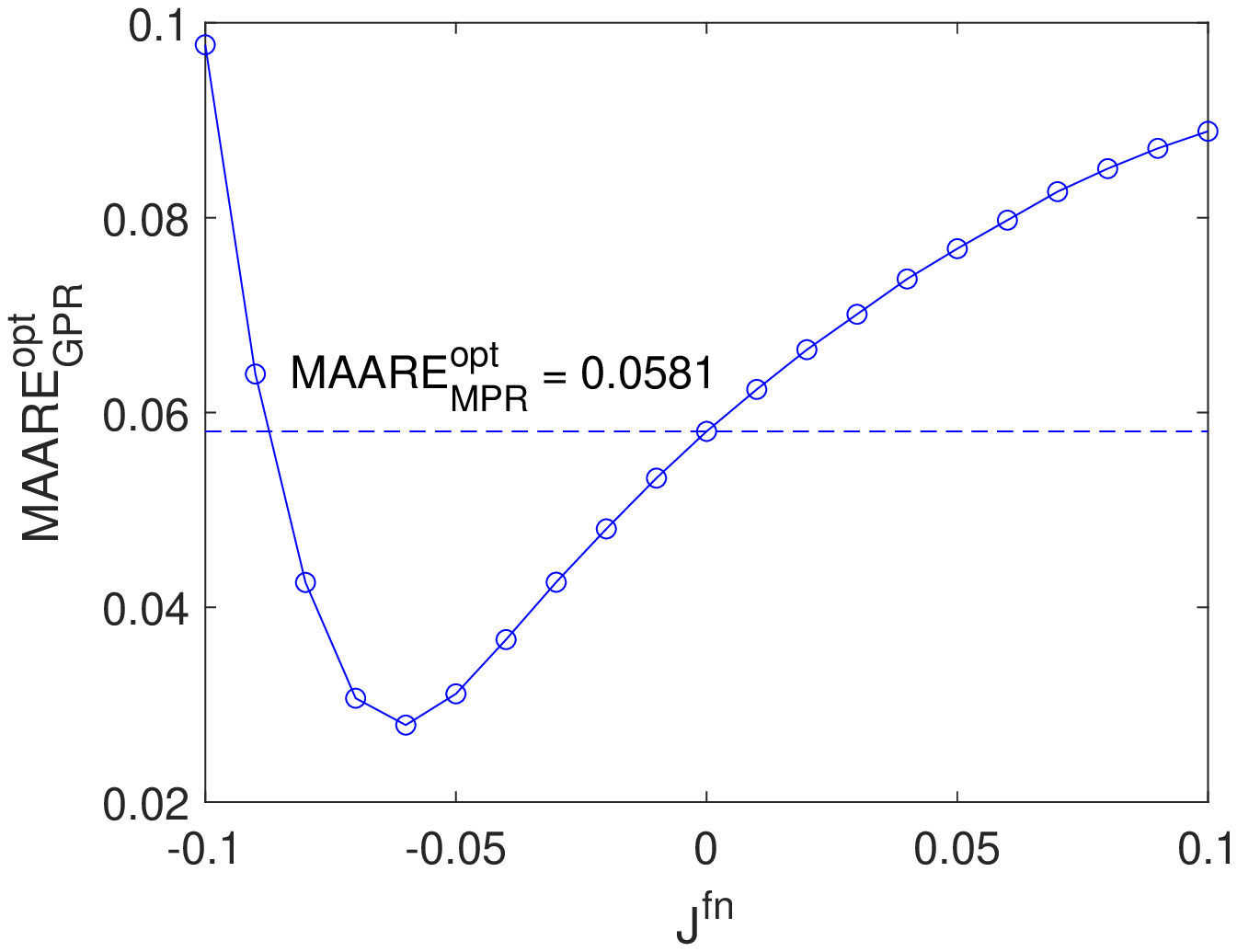}\label{fig:jnnn_p033_maare_opt}}\\ \vspace{-3mm}
\subfigure{\includegraphics[scale=0.43,clip]{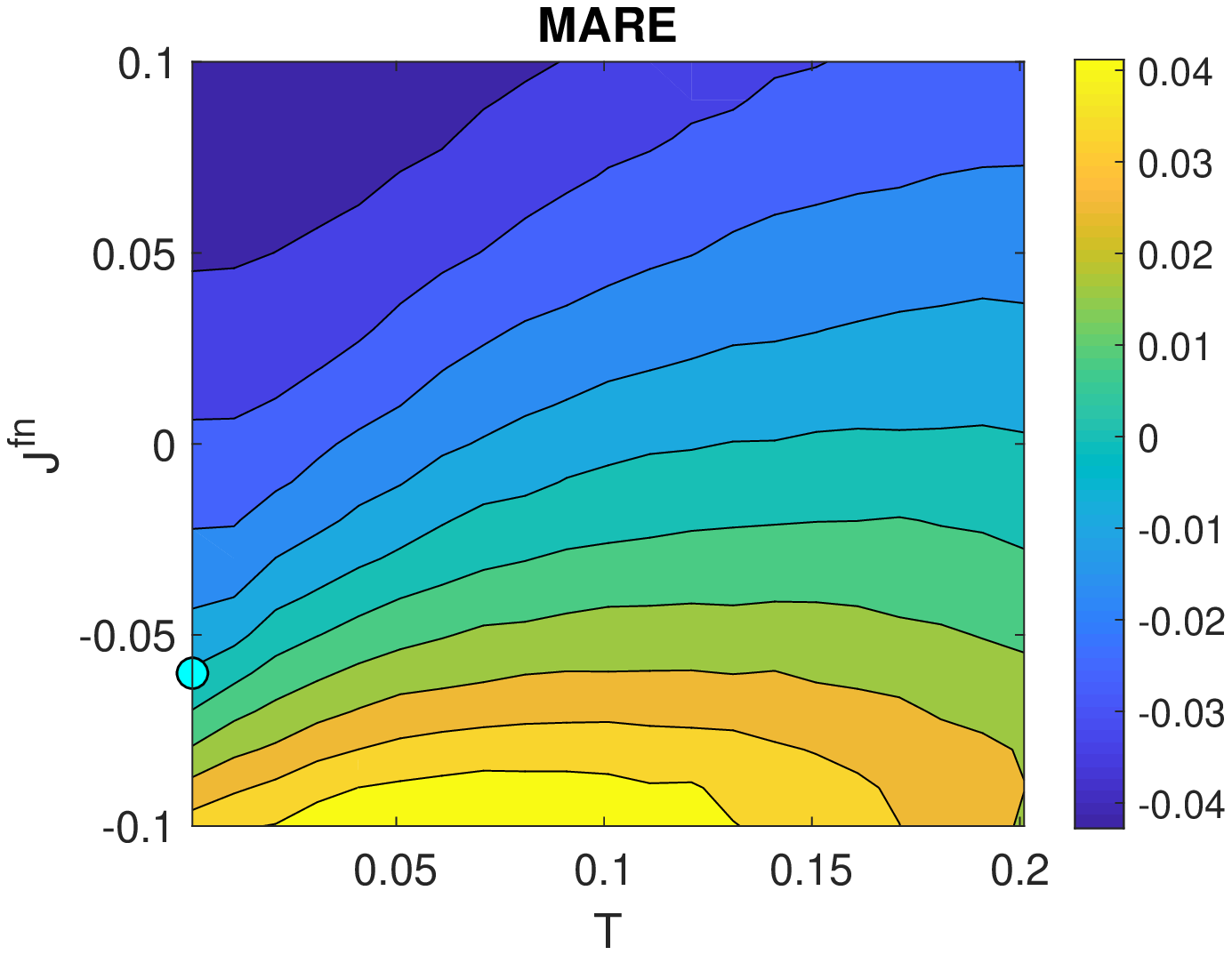}\label{fig:jnnn_p033_mare}}
\subfigure{\includegraphics[scale=0.43,clip]{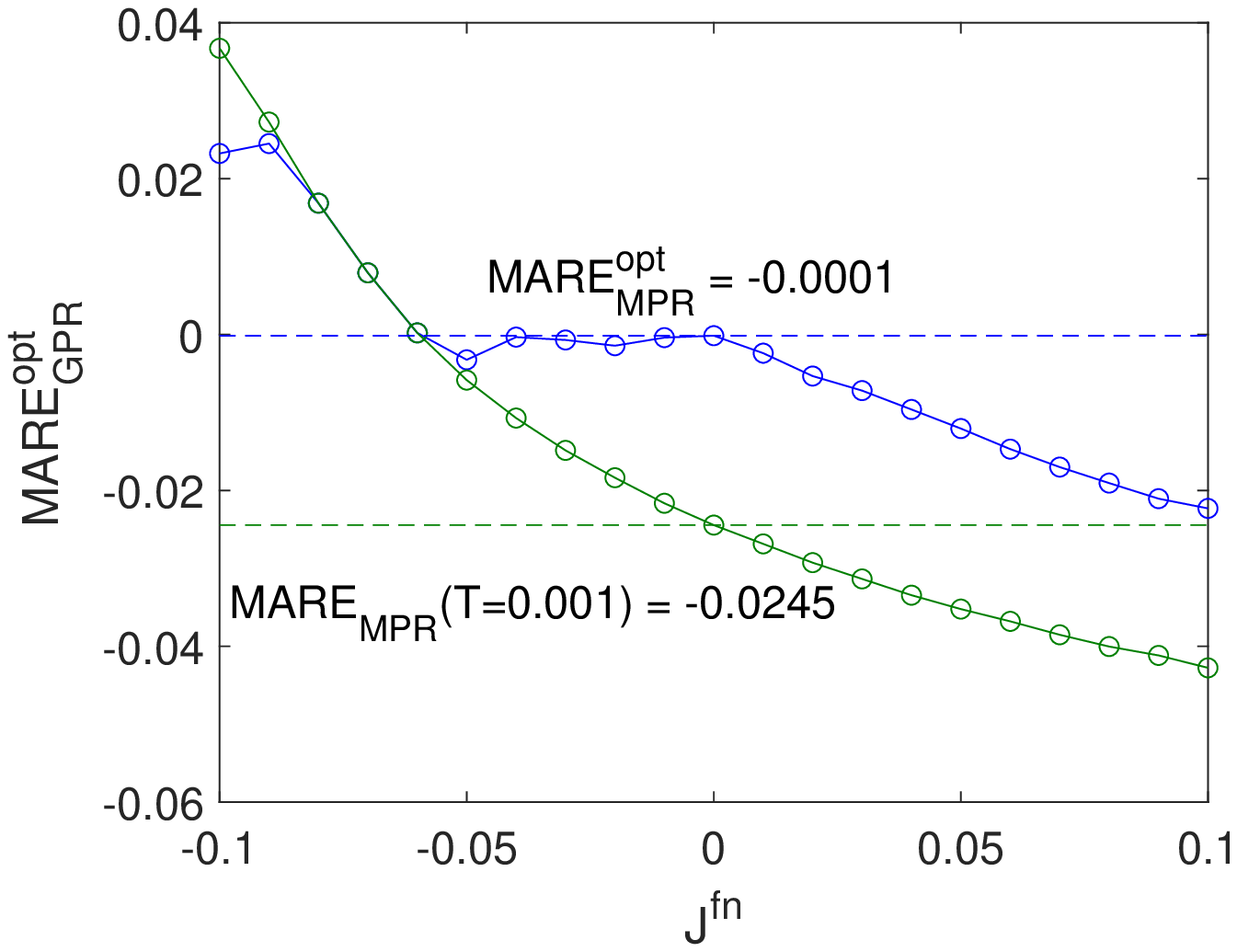}\label{fig:jnnn_p033_mare_opt}}\\ \vspace{-3mm}
\subfigure{\includegraphics[scale=0.43,clip]{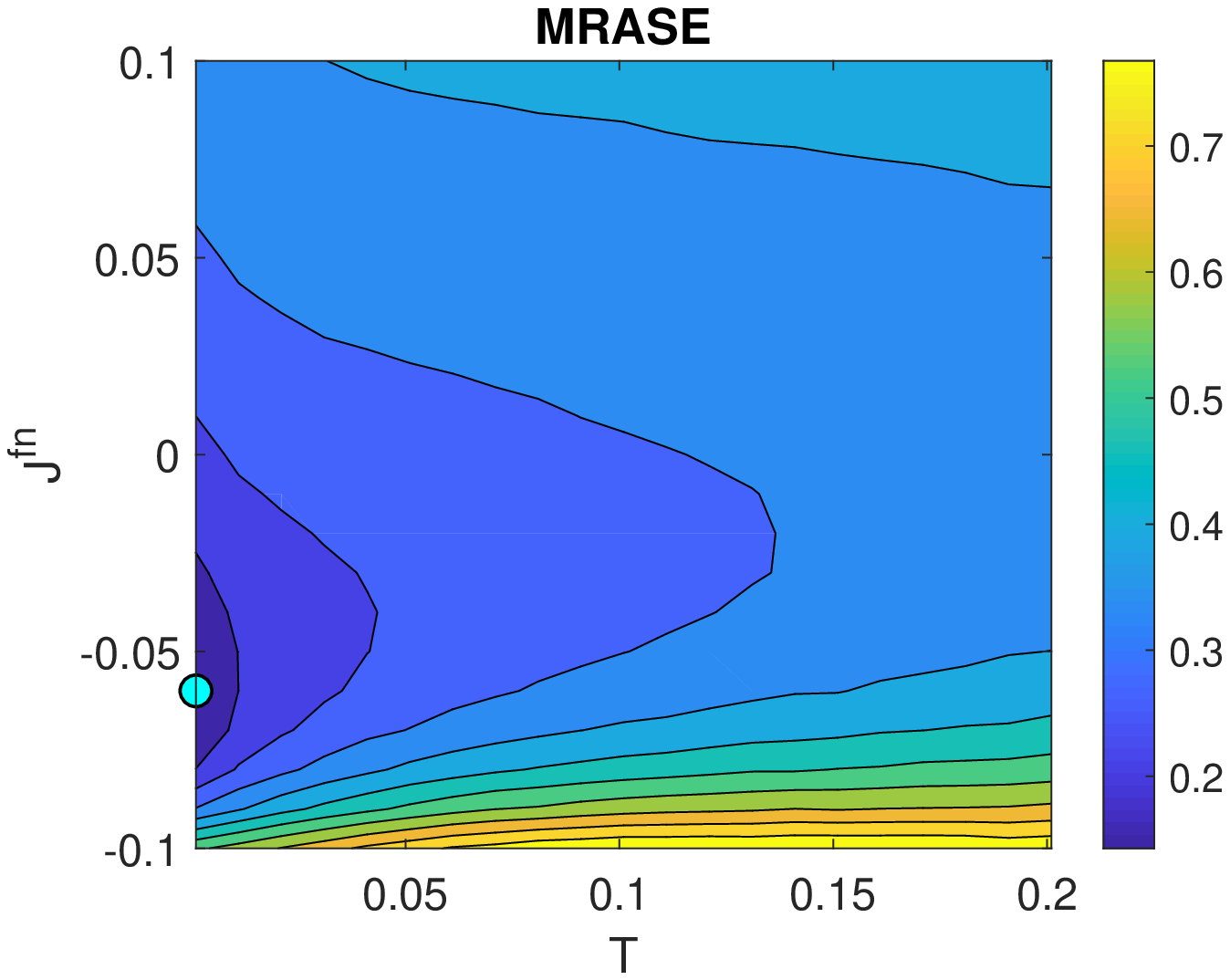}\label{fig:jnnn_p033_mrase}}
\subfigure{\includegraphics[scale=0.43,clip]{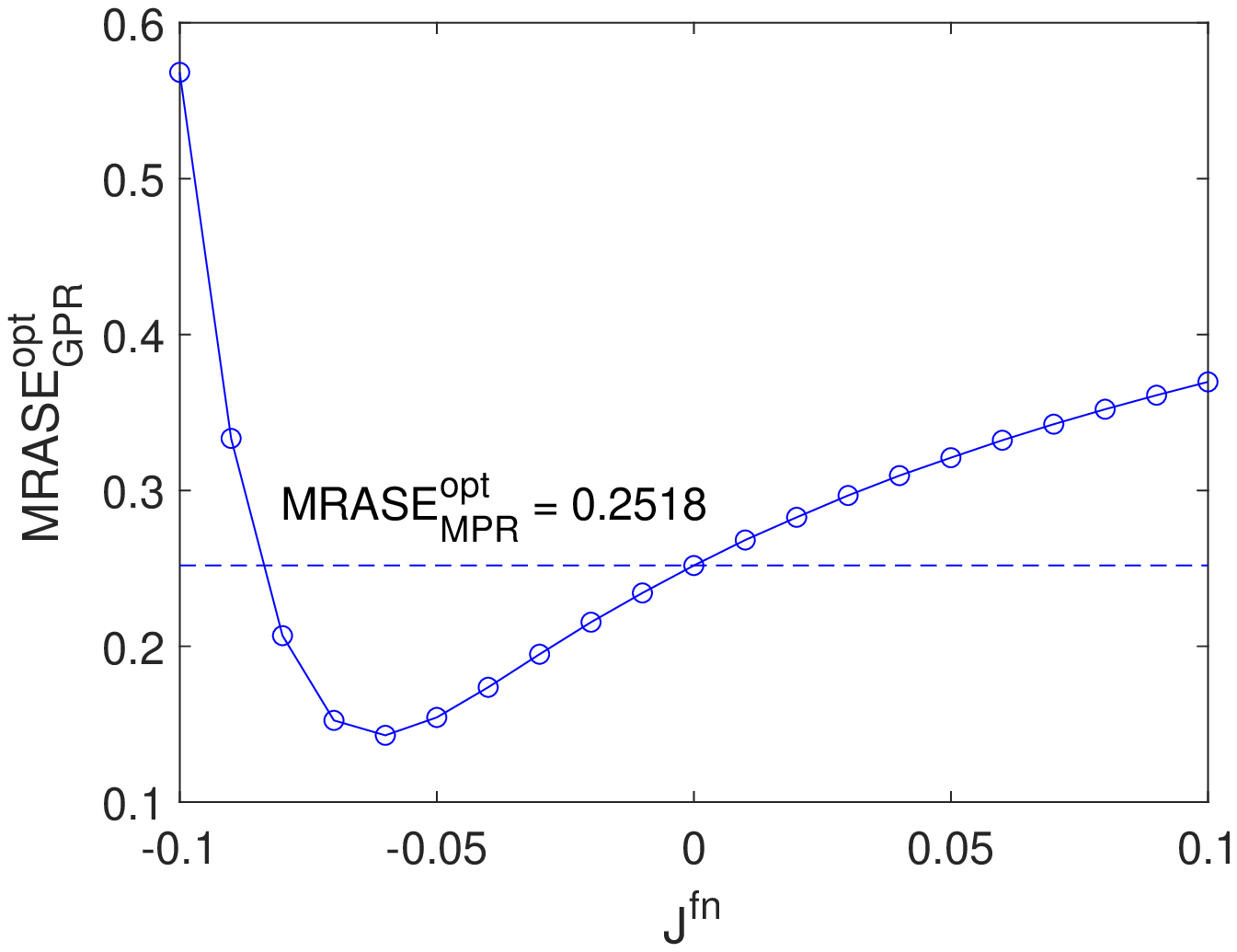}\label{fig:jnnn_p033_mrase_opt}} \vspace{-3mm}
\caption{Left column: contour plots of the validation measures in the $T-\Jfn$ parameter plane;  cyan circles correspond to the optimal values. Right column: validation measures as functions of $\Jfn$ at optimal (blue curves) and $T=0.001$ (green curve) temperatures. The dashed lines mark the optimal values obtained by means of the MPR model. $S=100$ samples are generated from the Gaussian random field with $\xi_1=\xi_2=2$, $\nu = 2.5$. The percentage of missing data is $p=33\%$.}
\label{fig:errors_jnnn}
\end{figure}

Below, we study the effect of further-neighbor interaction, $\Jfn$. We expect that its inclusion can help better model spatial variability at more distant lags and, thus, also better capture the data smoothness. Therefore, in our test we chose the data with Gaussian distribution $Z \sim N(m = 5, \sigma = 2)$ and the WM($\xi_1=\xi_2=2,\nu = 2.5$) covariance, which show relatively smooth spatial variation. In the left column of Fig.~\ref{fig:errors_jnnn} we present the calculated validation measures in the $T-\Jfn$ parameter planes. Missing data are generated by $p=33\%$ random thinning. The contour plots show variations of the respective measures corresponding to different values of the parameters $T$ and $\Jfn$, while all the remaining model parameters are kept fixed: $n=1$, $\alpha = \infty$, $\Jnn=0.5$, and $K=0$.

For the same reason as in the above case of the anisotropic data, the optimal values marked by the cyan circles again all except MARE correspond to the lowest simulated temperature $T=0.001$, as one would expect for smooth data. Nevertheless, the temperature alone is not sufficient to entirely control the smoothness and the fn interaction can serve as an additional parameter that adds some more flexibility. One can observe that the inclusion of the further-neighbor interaction can improve the prediction performance. In each instance the optimal values have the same coordinates $(T_{\mathrm{opt}},\Jfn_{\mathrm{opt}})=(0.001,-0.06)$, which means that the MPR best performance can be further improved by including the ``antiferromagnetic'' (negative) fn interaction with the strength $\Jfn=-0.06$. Moreover, the resulting values, shown in the right column of Fig.~\ref{fig:errors_jnnn}, indicate that the improvement can be quite substantial. In particular, the MAAE, MAARE, MARE, and MRASE errors respectively dropped by about $44\%, 52\%, 100\%$, and $43\%$. We note that, similar to the anisotropic data above, MARE shows optimal (zero) values along the isoline in this case crossing the area $[T_{\min},T_{\max}]\times [\Jfn_{\min},\Jfn_{\max}] \approx [0.001,0.2] \times [-0.06,0]$ and, therefore, $(T_{\mathrm{opt}},\Jfn_{\mathrm{opt}})=(0.001,-0.06)$ can be considered optimal also for this measure.

\subsection{Effect of nonlinearity controlling parameters}
\label{ssec:nonlinearity}

Further, we analyze the effect of the parameters $n$ and $\alpha$, which come from inclusion of higher-order couplings and control the shape of the potential function. Therefore, we expect that their inclusion gives more flexibility by adjusting the shape of the potential function with regard to the data distribution. For this purpose, in our test we chose the data with lognormal distribution $\log Z \sim N(m = 5, \sigma = 2)$ and the WM($\xi_1=\xi_2=2,\nu = 0.25$) covariance, which has a highly skewed non-Gaussian distribution. Missing data are generated by $p=33\%$ random thinning. Since now we have two coupled parameters, following the examination of their effects on the potential function in Fig.~\ref{fig:potential}, we will study their individual effects by varying only one parameter and fixing the other to some value.  The values of the remaining model parameters are also kept fixed: $\Jnn=0.5$, $\Jfn=0$, and $K=0$. 

\begin{figure}[t!]
\centering \vspace{-10mm}
\subfigure{\includegraphics[scale=0.43,clip]{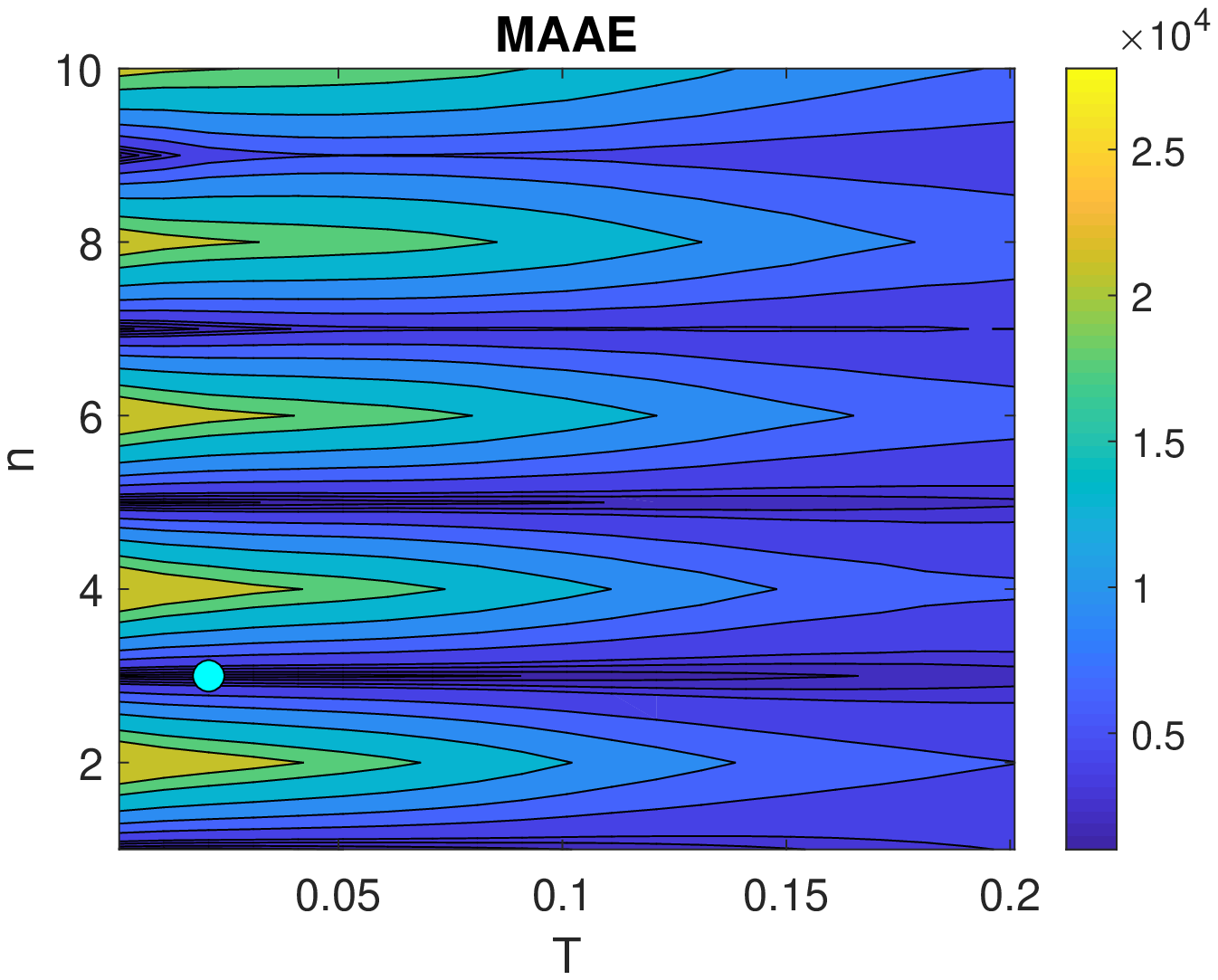}\label{fig:p_p033_maae}}
\subfigure{\includegraphics[scale=0.43,clip]{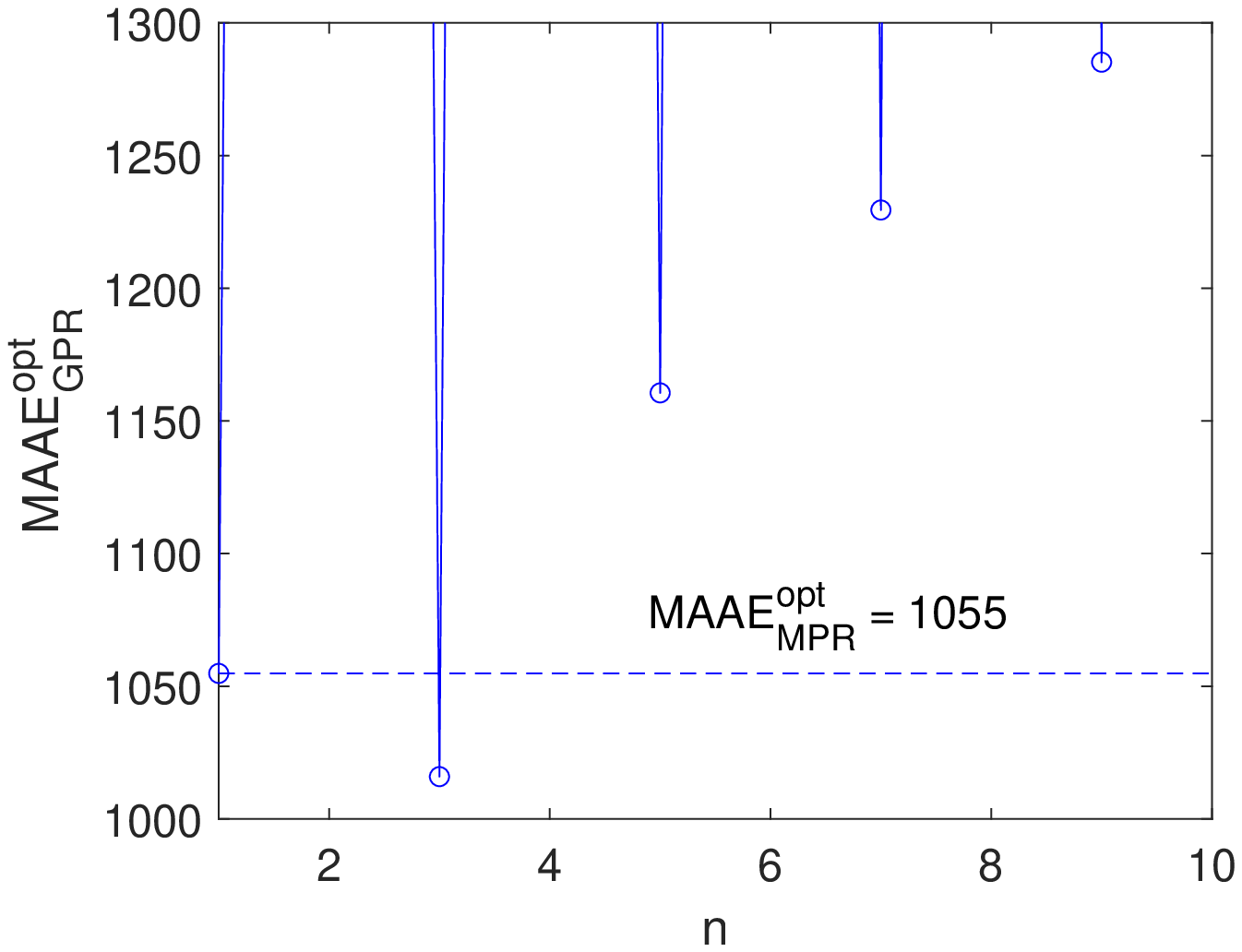}\label{fig:p_p033_maae_opt}}\\ \vspace{-3mm}
\subfigure{\includegraphics[scale=0.43,clip]{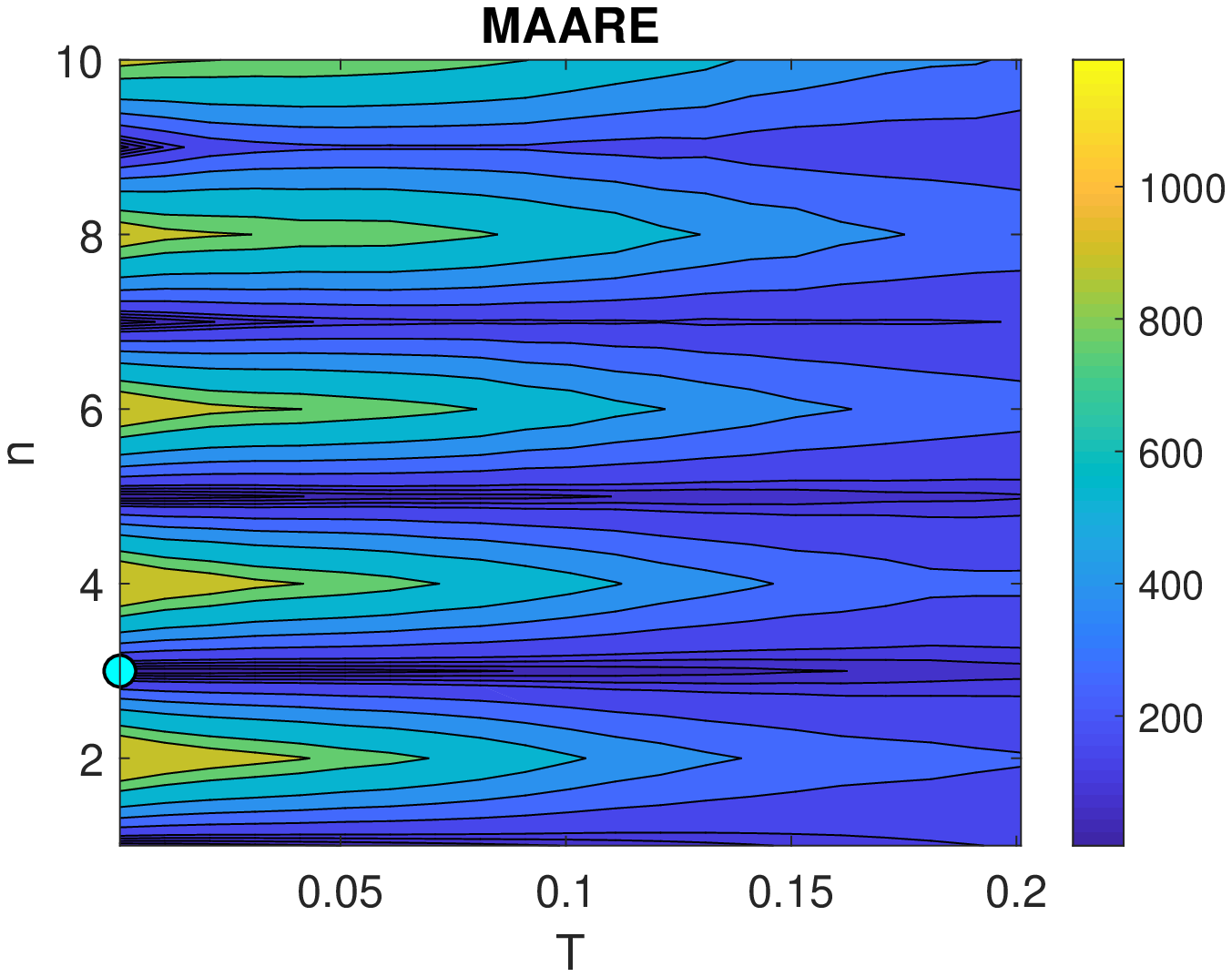}\label{fig:p_p033_maare}}
\subfigure{\includegraphics[scale=0.43,clip]{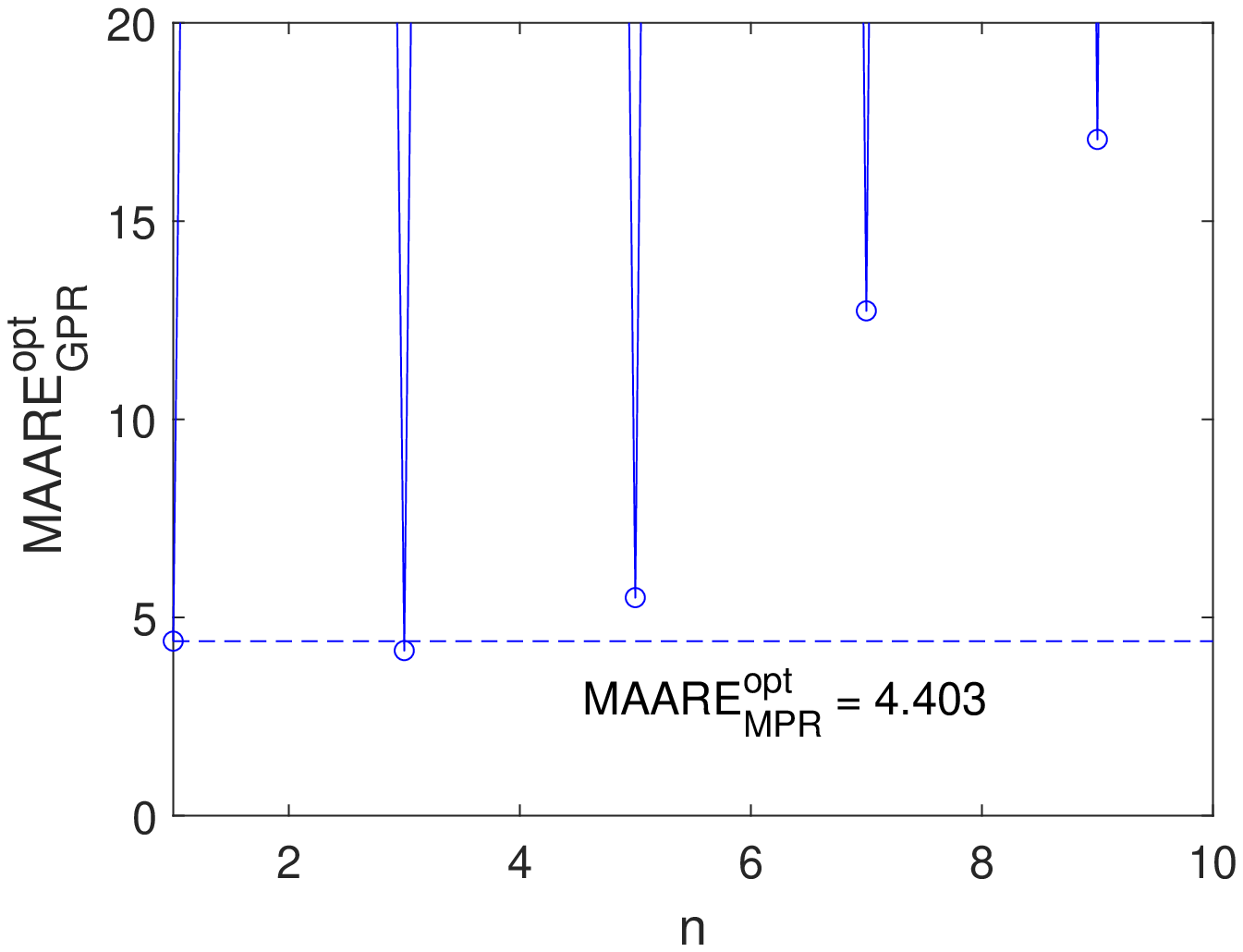}\label{fig:p_p033_maare_opt}}\\ \vspace{-3mm}
\subfigure{\includegraphics[scale=0.43,clip]{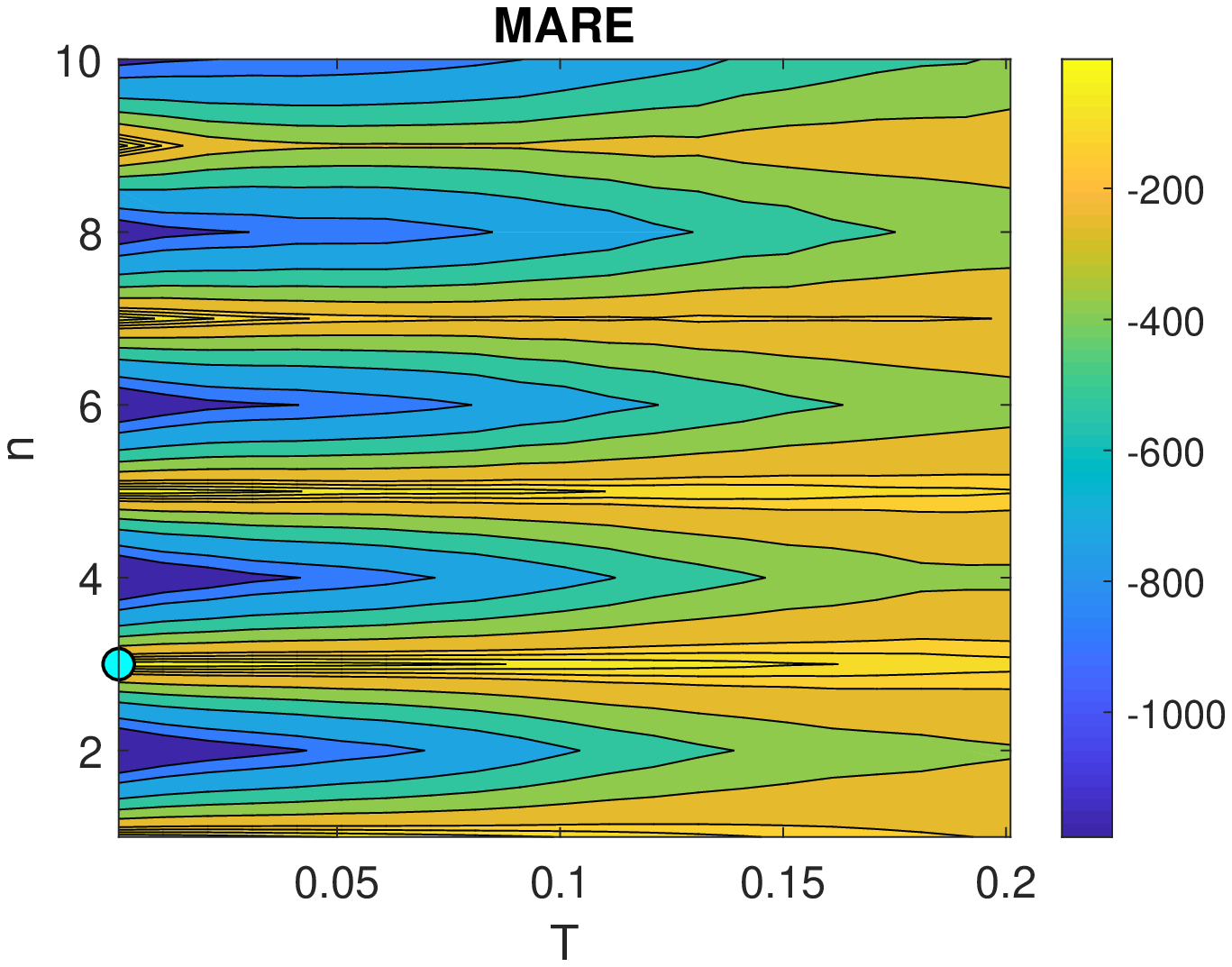}\label{fig:p_p033_mare}}
\subfigure{\includegraphics[scale=0.43,clip]{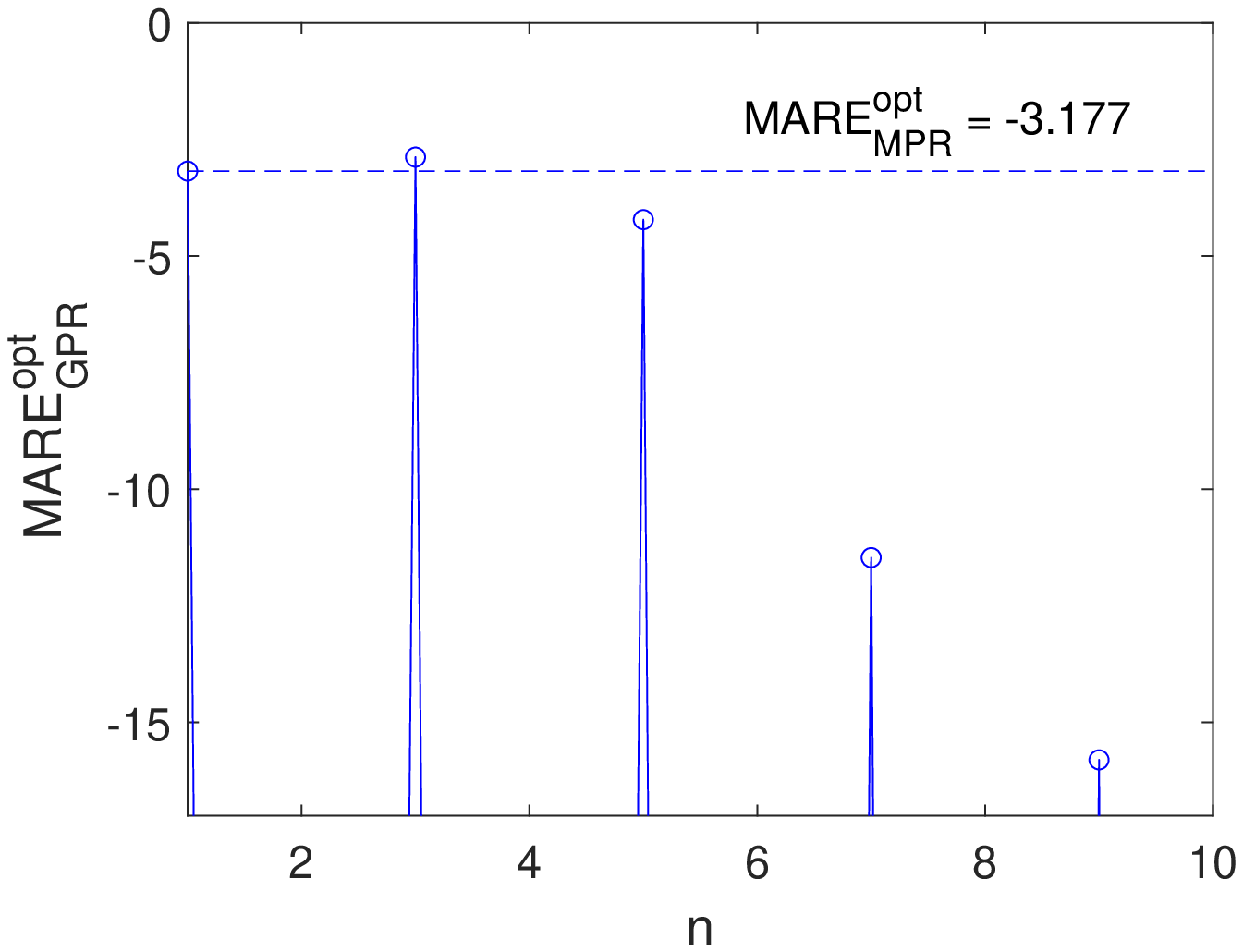}\label{fig:p_p033_mare_opt}}\\ \vspace{-3mm}
\subfigure{\includegraphics[scale=0.43,clip]{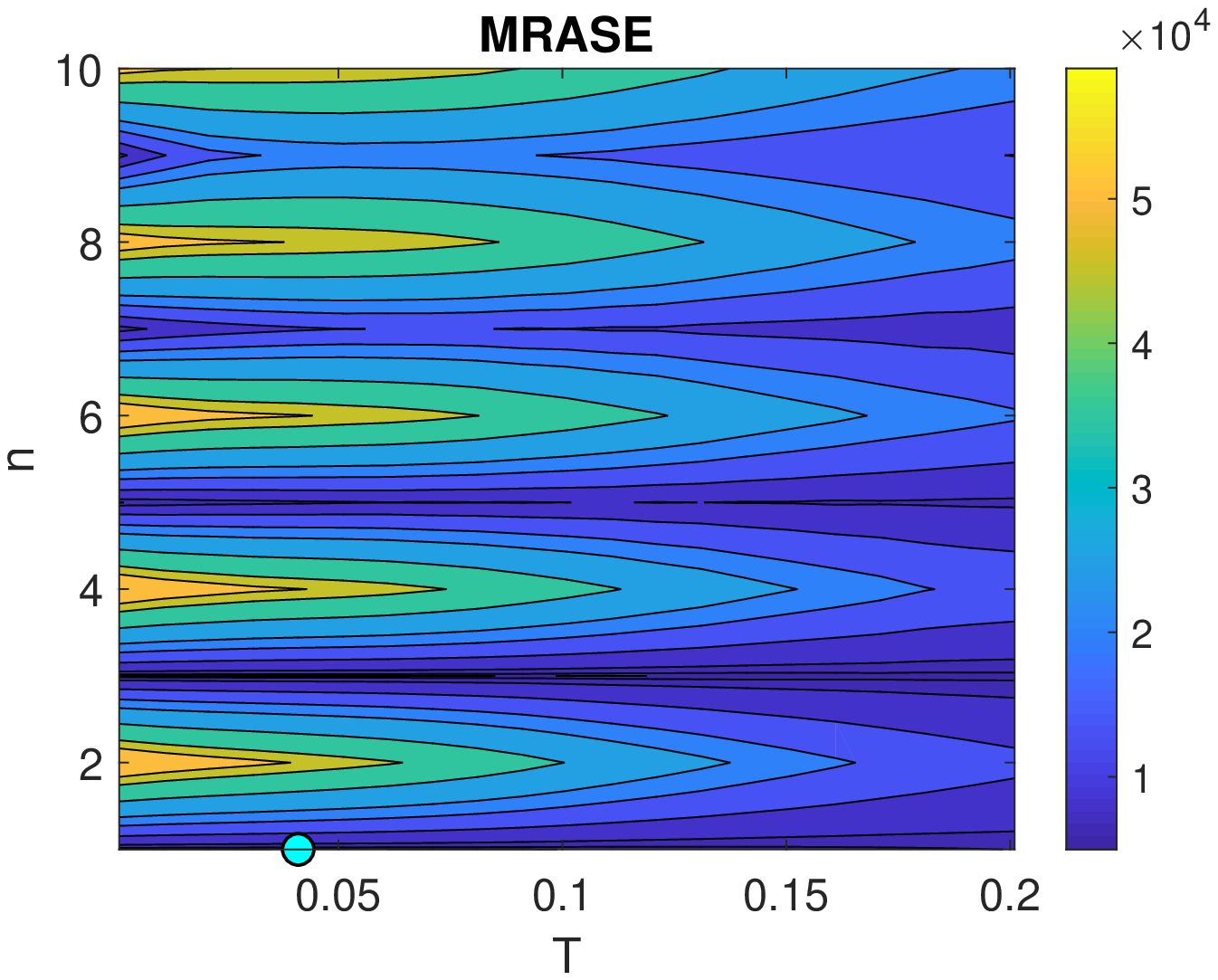}\label{fig:p_p033_mrase}}
\subfigure{\includegraphics[scale=0.43,clip]{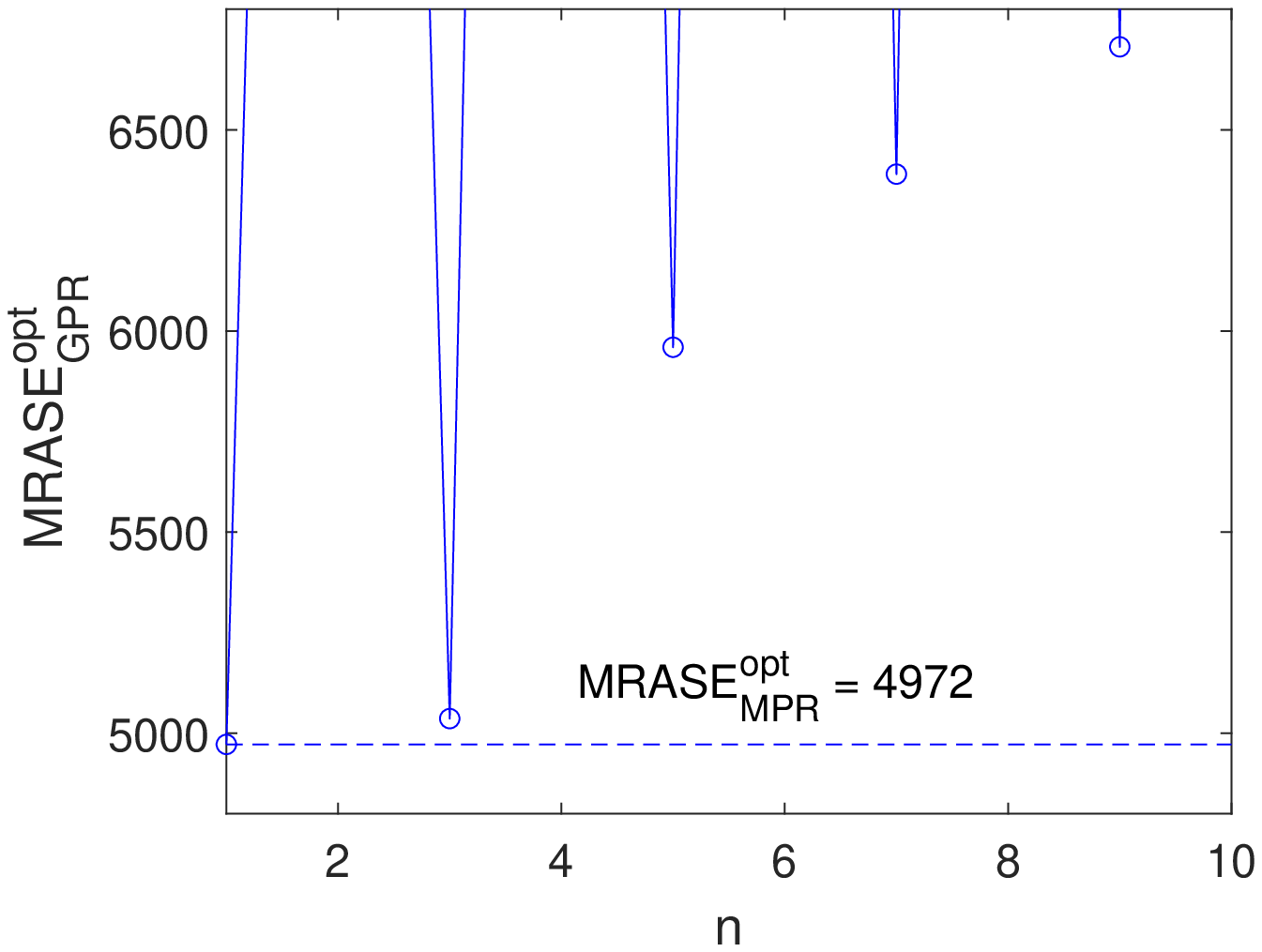}\label{fig:p_p033_mrase_opt}} \vspace{-3mm}
\caption{Left column: contour plots of the validation measures in the $T-n$ parameter plane for $\alpha=1.01$;  cyan circles mark the optimal values. Right column: validation measures as functions of the parameter $n$ at optimal temperature. The dashed lines mark the optimal values obtained by means of the MPR model. $S=100$ samples are generated from the lognormal random field with $\xi_1=\xi_2=2$, $\nu = 0.25$. The percentage of missing data is $p=33\%$.}
\label{fig:errors_n}
\end{figure}

In the left column of Fig.~\ref{fig:errors_n} we first show the calculated validation measures in the $T-n$ parameter plane for the fixed $\alpha=1.01$. The sensitivity of the shape of the potential function to whether $n$ is odd or even (see Fig.~\ref{fig:H_ij_phi-n}) is reflected in the oscillating landscapes: the valleys (MAAE, MAARE, MARE, and MRASE) and ridges (MARE) correspond to regions of better prediction performance obtained for odd values of $n$. On the other hand, the errors dramatically increase for even $n$. In particular, the optimal performance seems to be obtained for $n=3$, except MRASE which is the lowest for $n=1$. Nevertheless, as one can see in the right column of Fig.~\ref{fig:errors_n}, compared to the MPR model, the improvement is not substantial for the chosen value of $\alpha=1.01$. Namely, the MAAE, MAARE, MARE, and MRASE errors decrease by about $4\%, 5\%$, $9\%$, and $0\%$, respectively.

\begin{figure}[t!]
\centering \vspace{-10mm}
\subfigure{\includegraphics[scale=0.43,clip]{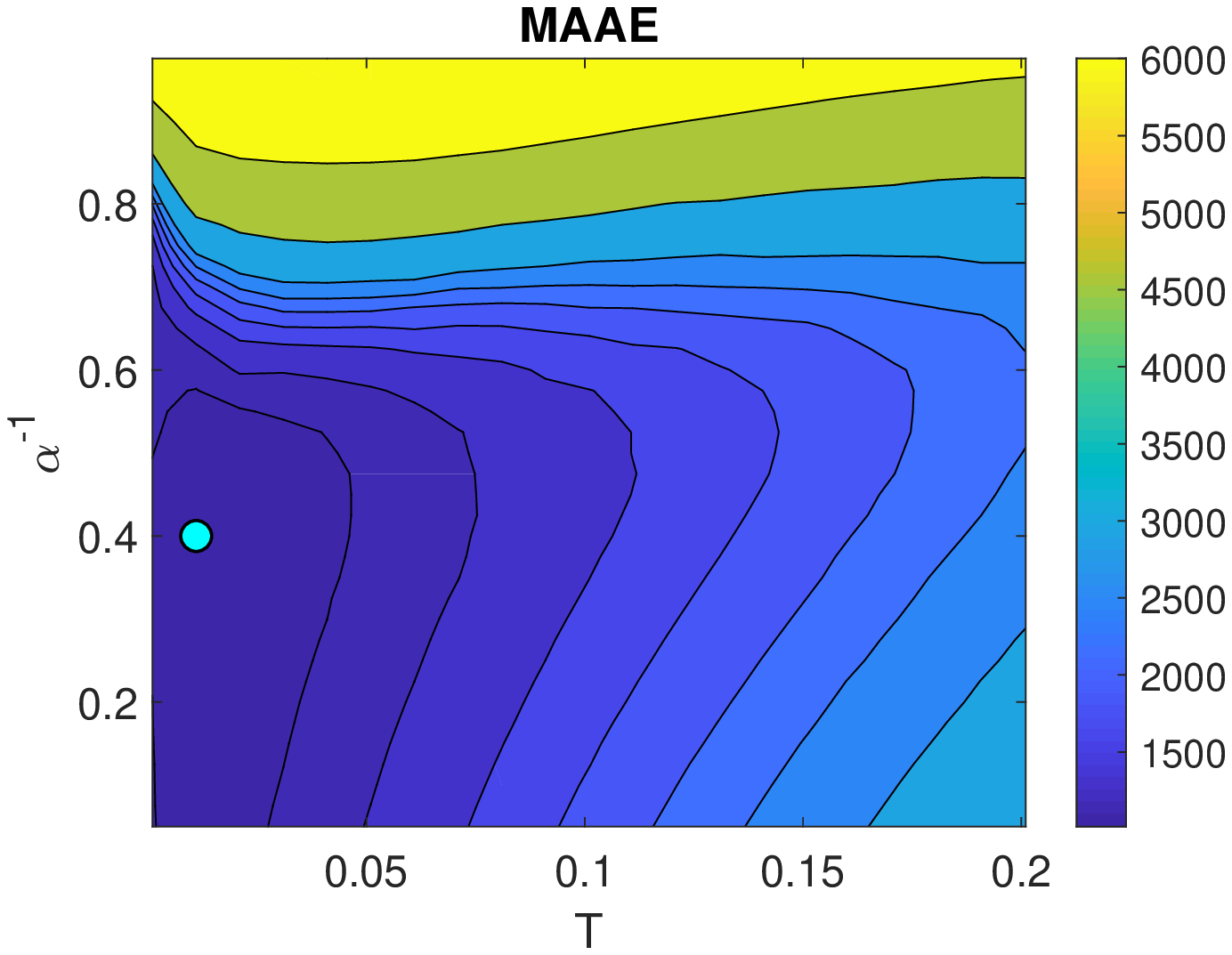}\label{fig:alp_p033_maae}}
\subfigure{\includegraphics[scale=0.43,clip]{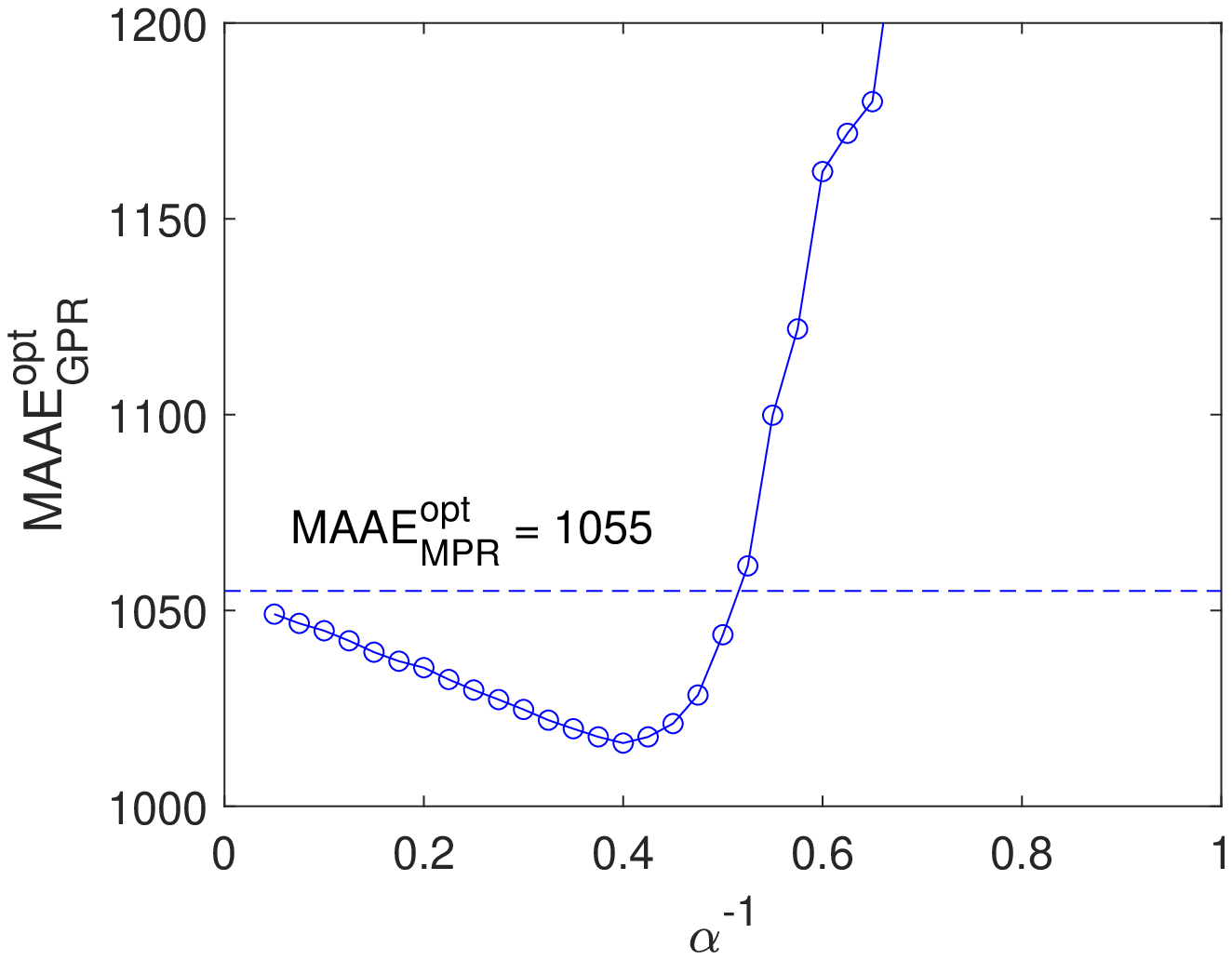}\label{fig:alp_p033_maae_opt}}\\ \vspace{-3mm}
\subfigure{\includegraphics[scale=0.43,clip]{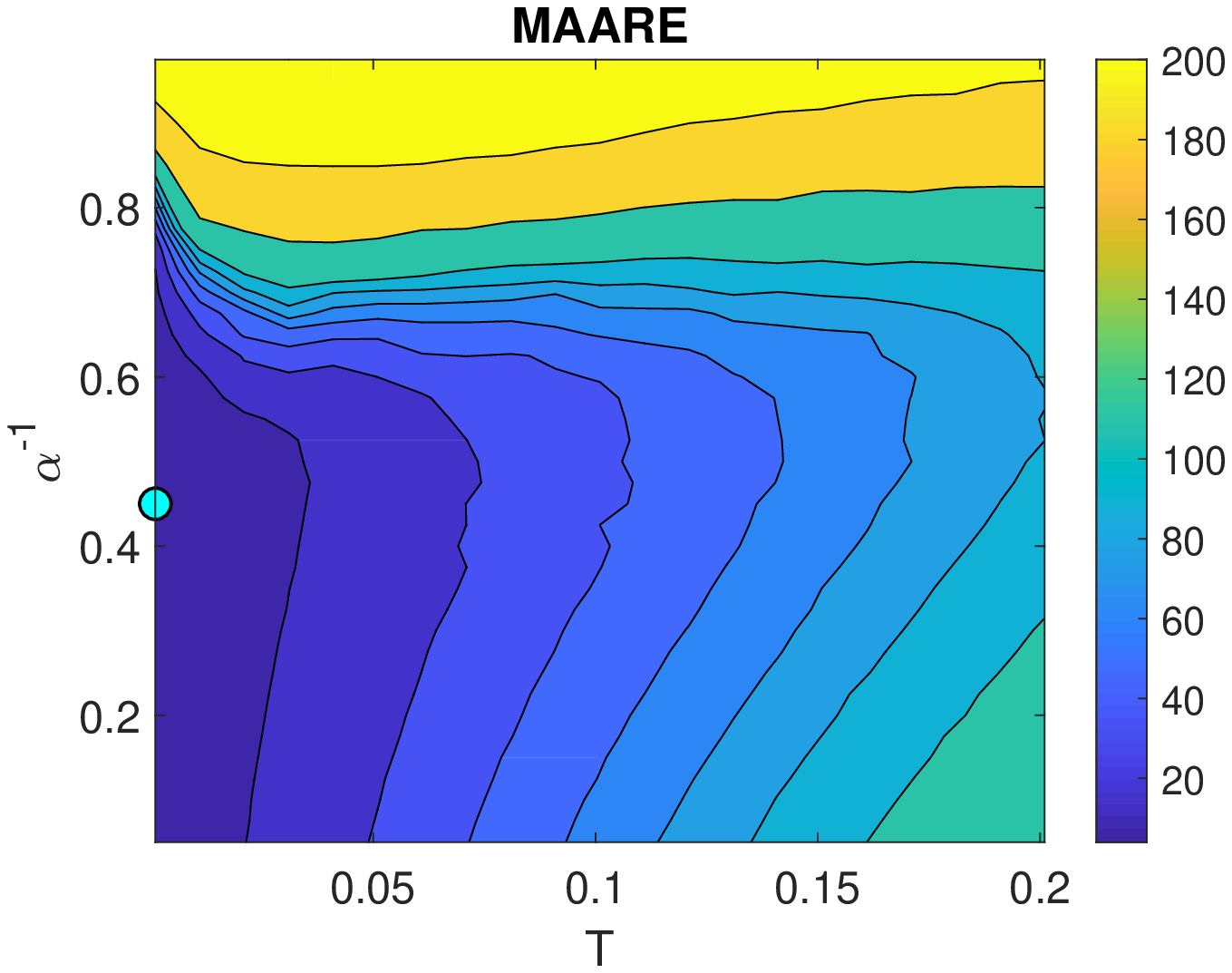}\label{fig:alp_p033_maare}}
\subfigure{\includegraphics[scale=0.43,clip]{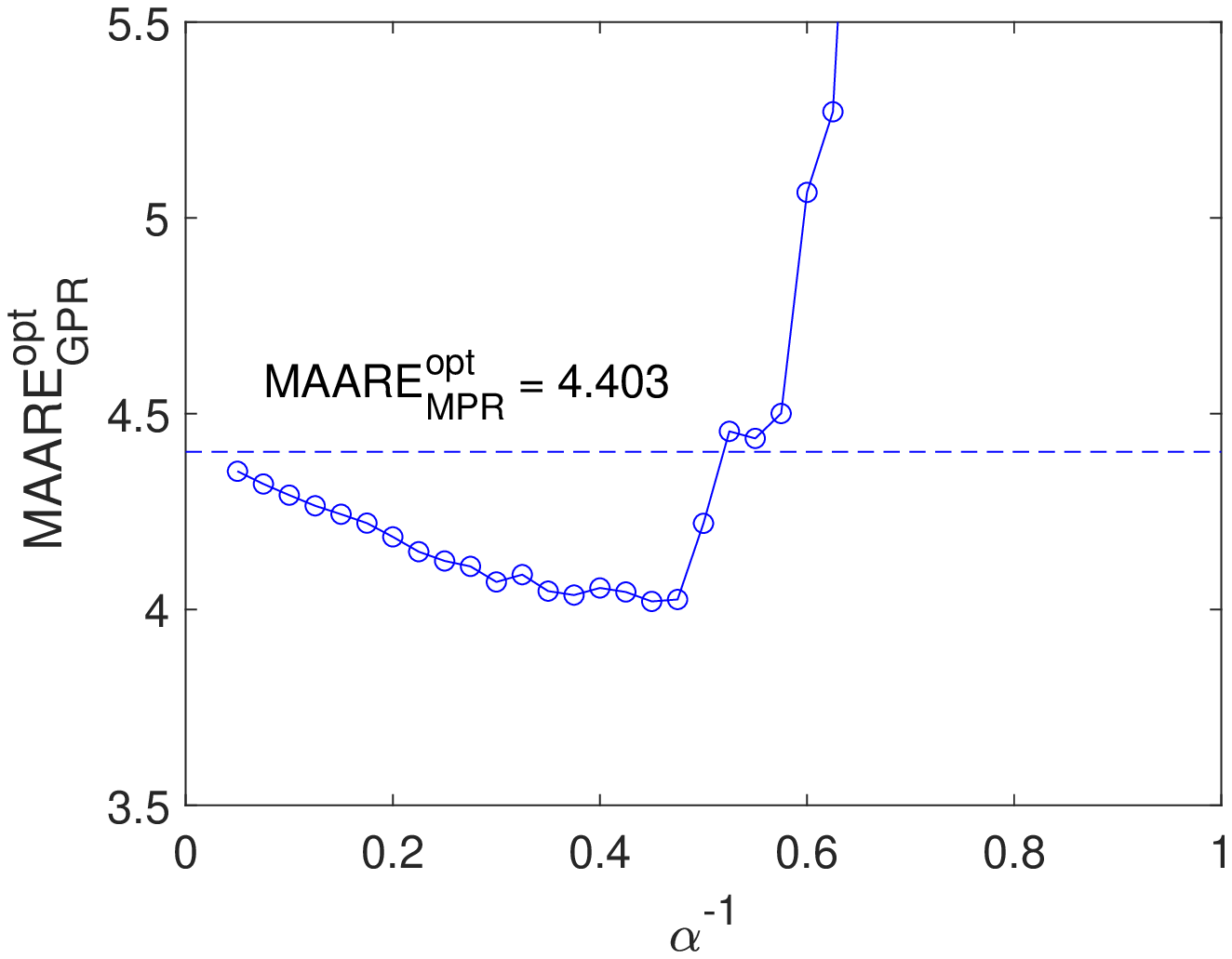}\label{fig:alp_p033_maare_opt}}\\ \vspace{-3mm}
\subfigure{\includegraphics[scale=0.43,clip]{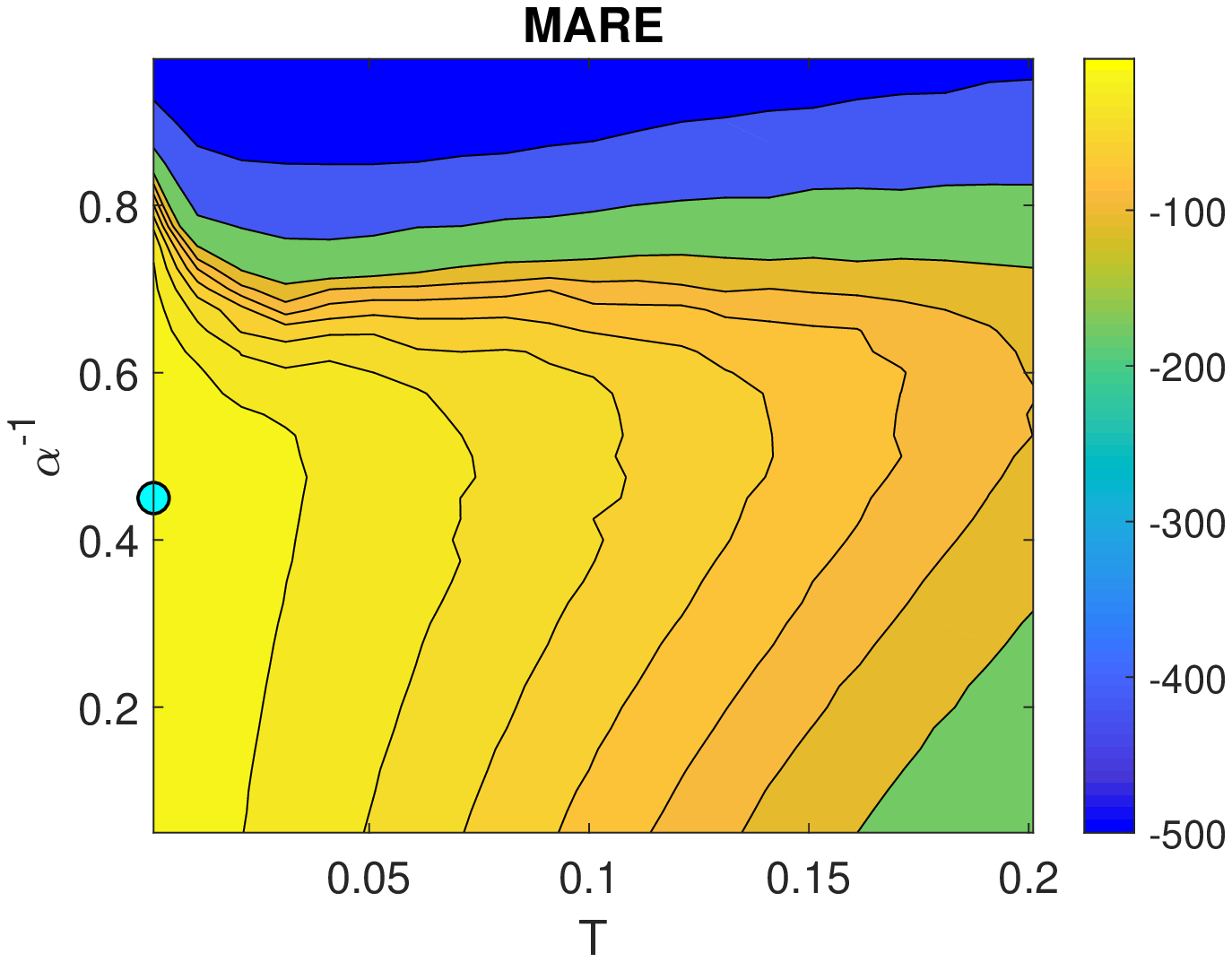}\label{fig:alp_p033_mare}}
\subfigure{\includegraphics[scale=0.43,clip]{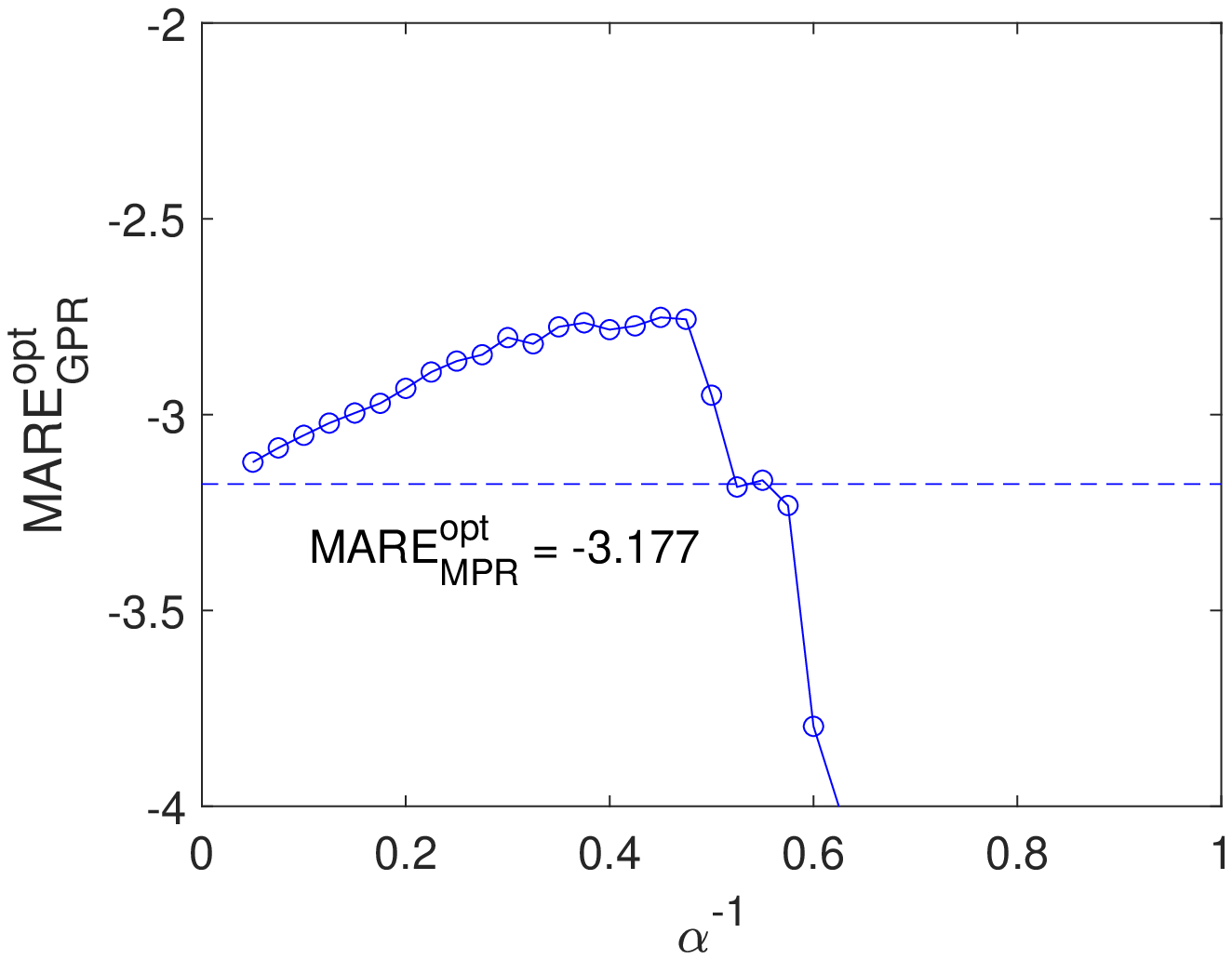}\label{fig:alp_p033_mare_opt}}\\ \vspace{-3mm}
\subfigure{\includegraphics[scale=0.43,clip]{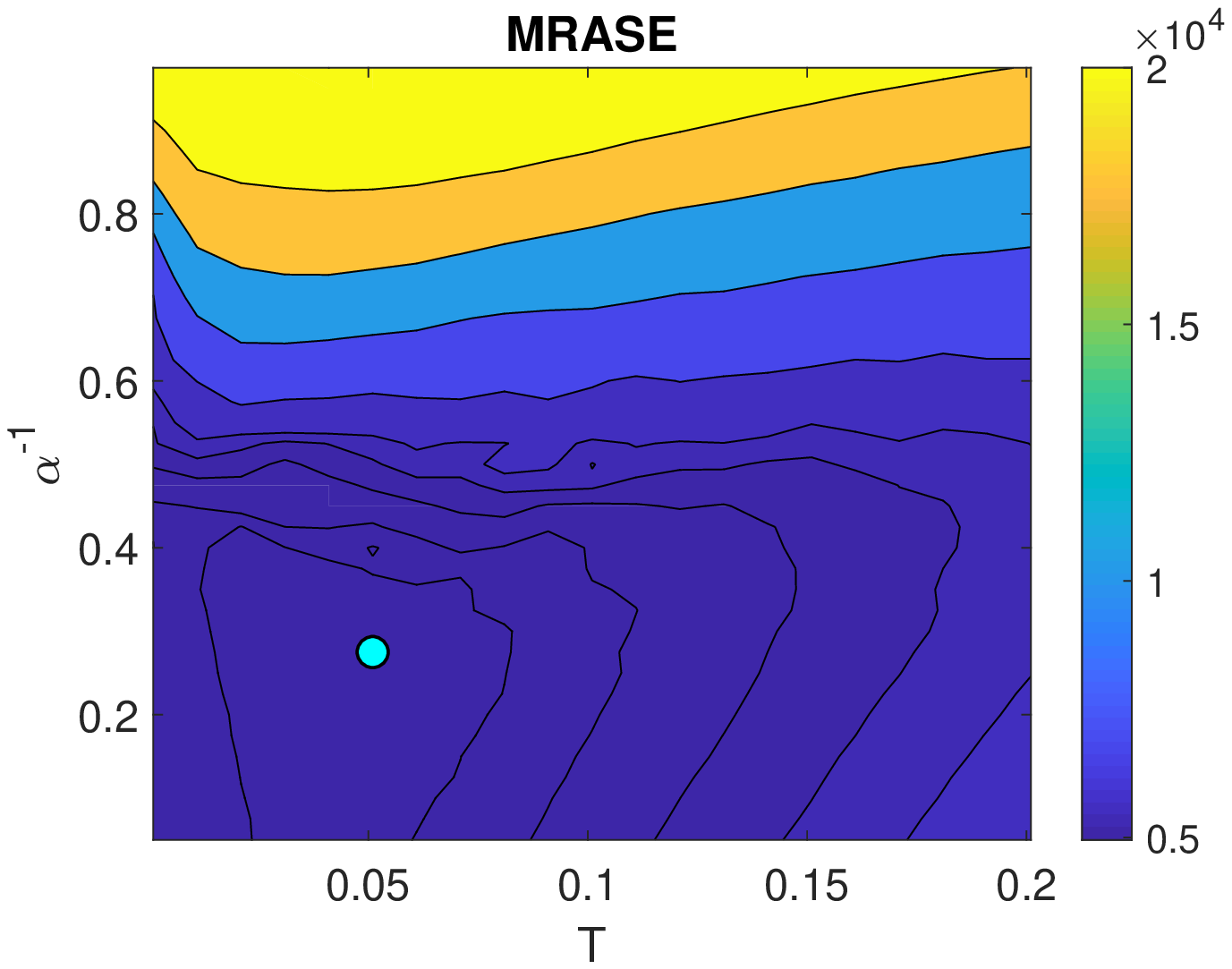}\label{fig:alp_p033_mrase}}
\subfigure{\includegraphics[scale=0.43,clip]{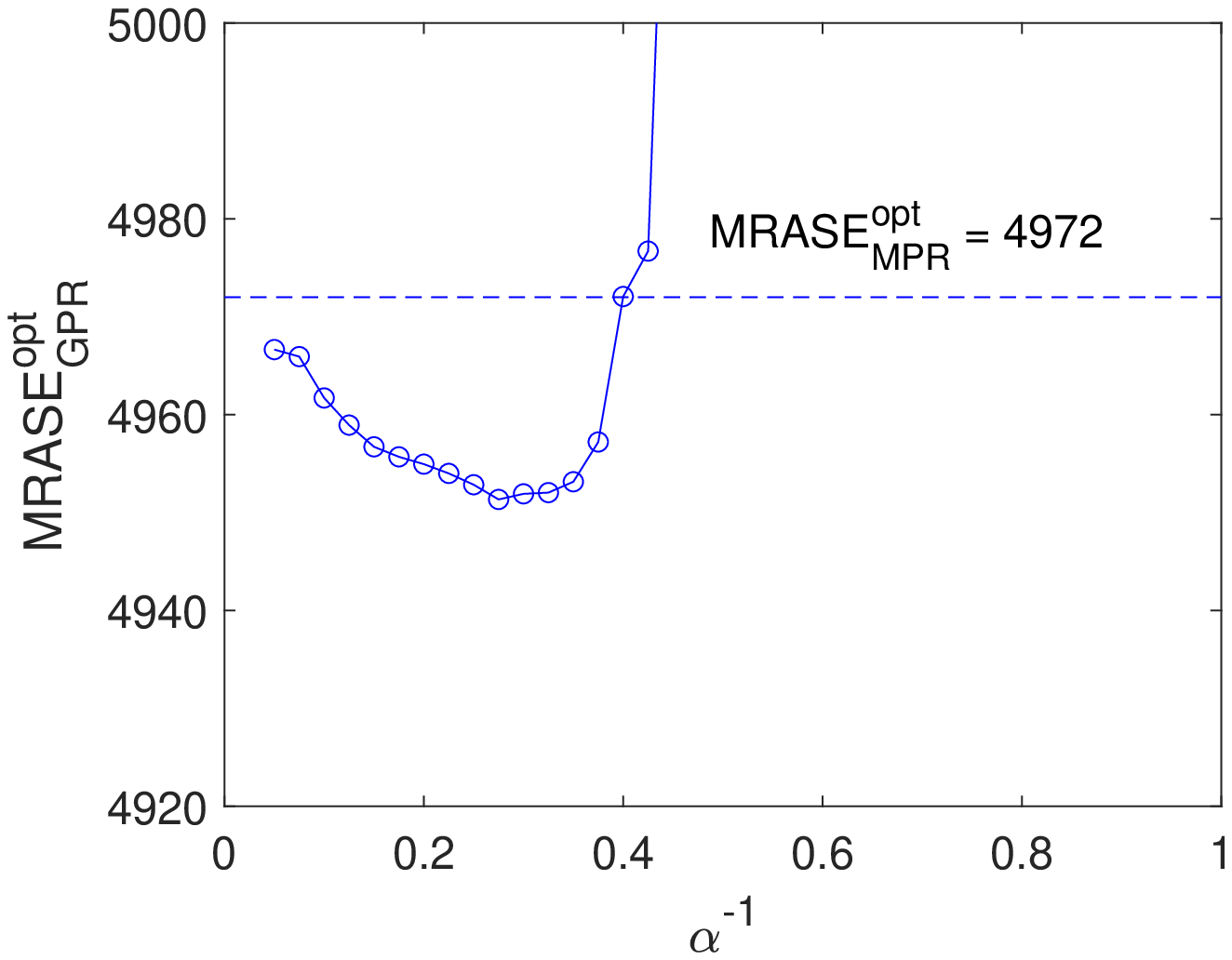}\label{fig:alp_p033_mrase_opt}} \vspace{-3mm}
\caption{Left column: contour plots of the validation measures in the $T-\alpha^{-1}$ parameter plane for $n \to \infty$; the cyan circles mark the optimal values. Right column: validation measures as functions of  $\alpha$ at optimal temperature. The dashed lines mark the optimal values obtained by means of the MPR model. $S=100$ samples are generated from the lognormal random field with $\xi_1=\xi_2=2$, $\nu = 0.25$. The percentage of missing data is $p=33\%$.}
\label{fig:errors_alp}
\end{figure}

In the left column of Fig.~\ref{fig:errors_alp} we present the obtained results in the $T-\alpha^{-1}$ parameter plane for $n \to \infty$. Let us recall that the MPR results correspond to the case of $\alpha \to \infty$ or $\alpha^{-1} = 0$, which is not displayed. It is clear, however, that better performance is obtained for $\alpha^{-1} > 0$. In particular, the optimal value is close to $\alpha^{-1} \approx 0.4$. It is worthwhile noticing the similarity of the potential functions for $n =3$ and $\alpha =1.01$ with the one for $\alpha^{-1} \approx 0.4$ and $n \to \infty$ (see the green curves for $n=3$ in Fig.~\ref{fig:H_ij_phi-n} and for $\alpha=2$ in Fig.~\ref{fig:H_ij_phi-alp}). Unlike in the previous cases, looking at different measures there is somewhat larger scatter of the optimal temperatures, ranging from $T=0.001$ for MAARE and MARE to $T=0.05$ for MRASE. Again, compared to the MPR values, the improvement achieved by adjusting of the parameter $\alpha$ (for $n \to \infty$) is rather moderate. As shown in the right column of Fig.~\ref{fig:errors_alp}, the MAAE, MAARE, MARE and MRASE errors can be reduced by about $4\%, 9\%$, and $13\%$ and less than $1\%$, respectively. Nevertheless, we would like to remark that the presented results do not represent the limits of the effects achievable by inclusion of the parameters $n$ and $\alpha$. The full benefits would be obtained by optimizing both parameters simultaneously, instead of optimizing only one while arbitrarily fixing the value of the other.

\subsection{Effect of the external field}
\label{ssec:field}

Considering our expectations about the ability of this extension to bring more control on the data distribution and smoothness, the test was performed on the data with lognormal distribution $\log Z \sim N(m = 5, \sigma = 2)$ and the WM($\xi_1=\xi_2=2,\nu = 2.5$) covariance, i.e., rather smoothly varying data with highly skewed distribution. Furthermore, we simulated contiguous blocks of missing data, which makes it more difficult to reproduce the distribution than in the case of randomly missing data due to the absence of conditioning data inside the blocks. In particular, the missing data are generated by random removal of a square data block with side length $L_B=20$. Below we present validation measures in the $T-K$ and $T-K'$ parameter planes and the remaining parameters take the following fixed values: $n=1$, $\alpha = \infty$, $\Jnn=0.5$, and $\Jfn=0$.

\begin{figure}[t!]
\centering \vspace{-10mm}
\subfigure{\includegraphics[scale=0.43,clip]{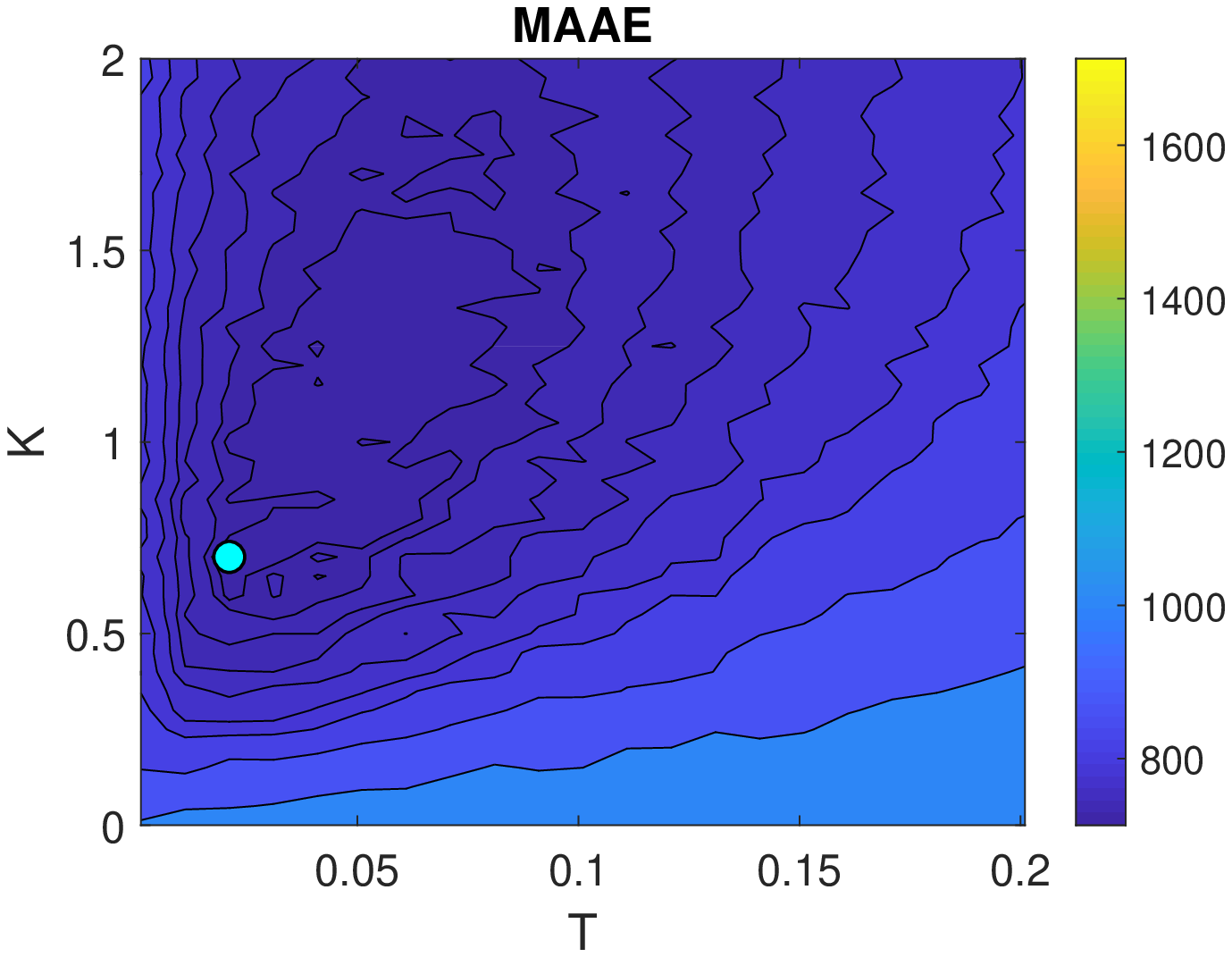}\label{fig:h_blk20_maae}}
\subfigure{\includegraphics[scale=0.43,clip]{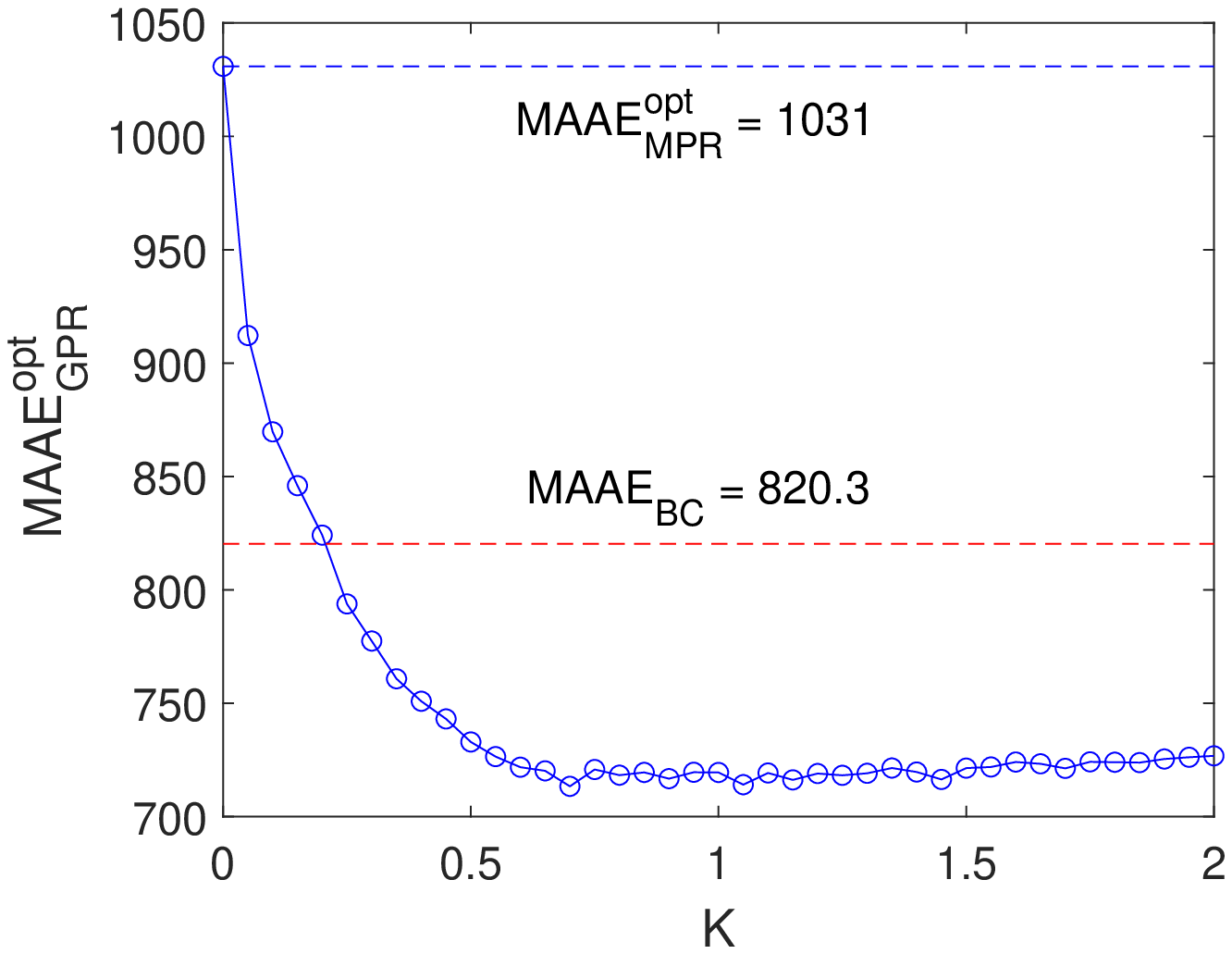}\label{fig:h_blk20_maae_opt}}\\ \vspace{-3mm}
\subfigure{\includegraphics[scale=0.43,clip]{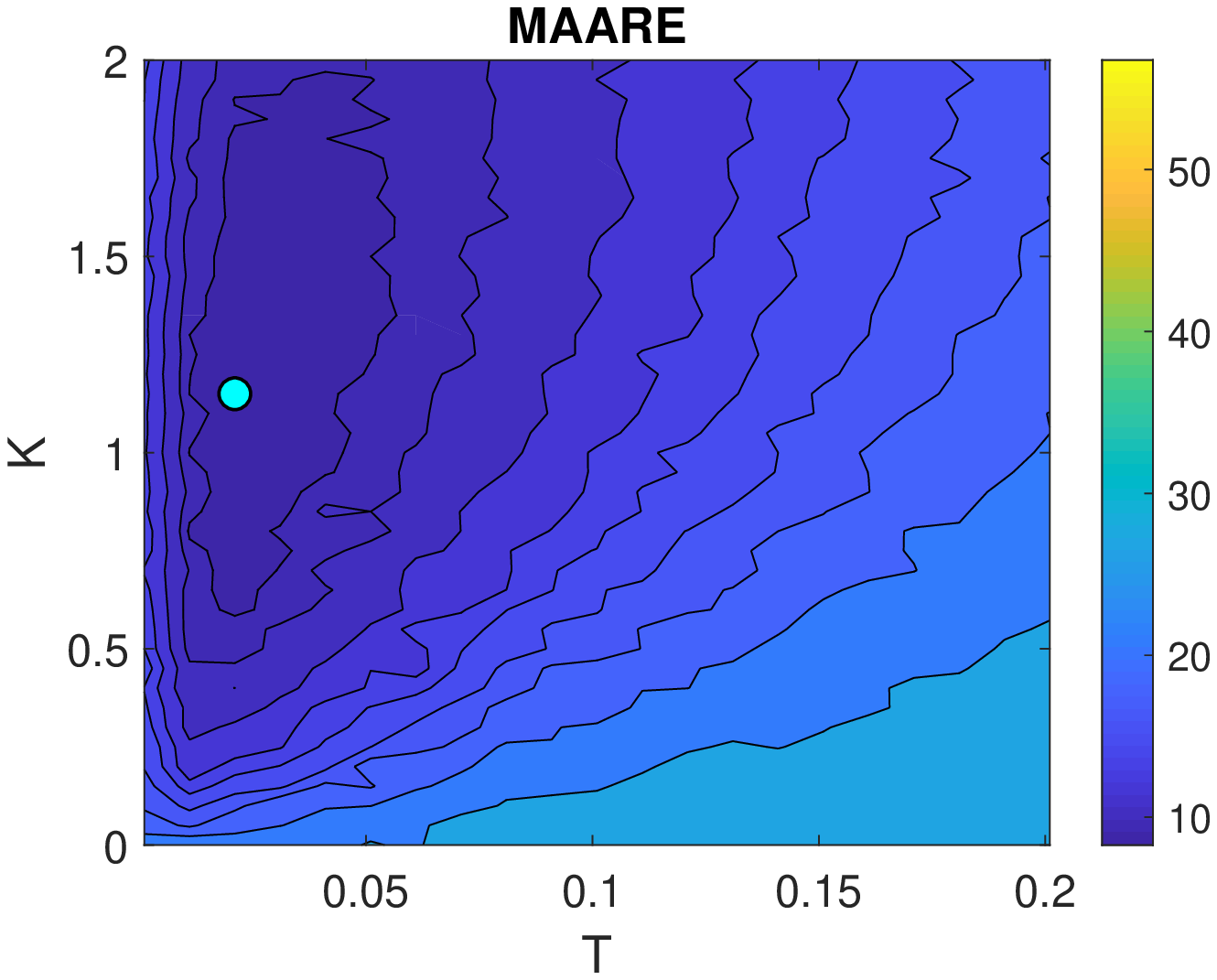}\label{fig:h_blk20_maare}}
\subfigure{\includegraphics[scale=0.43,clip]{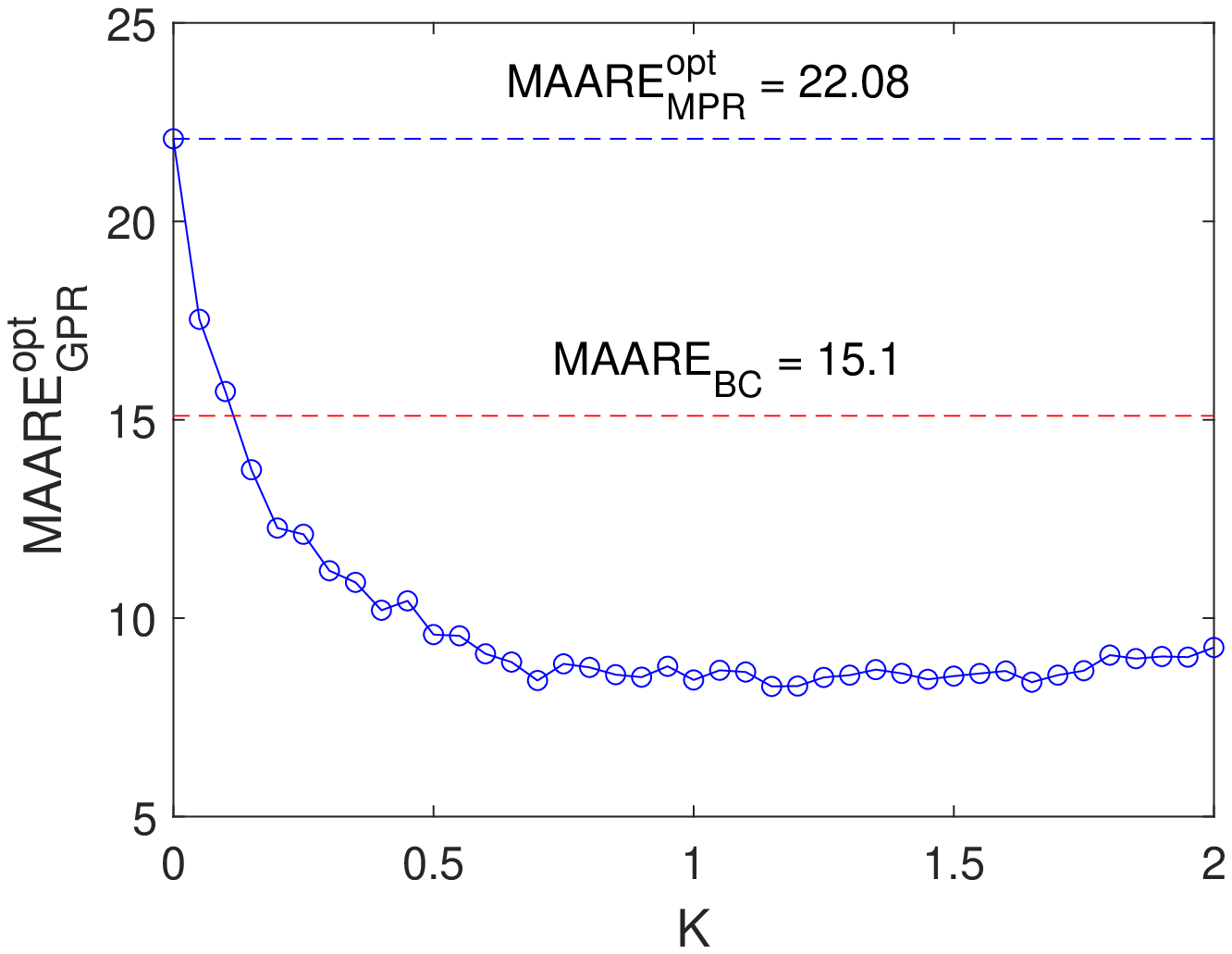}\label{fig:h_blk20_maare_opt}}\\ \vspace{-3mm}
\subfigure{\includegraphics[scale=0.43,clip]{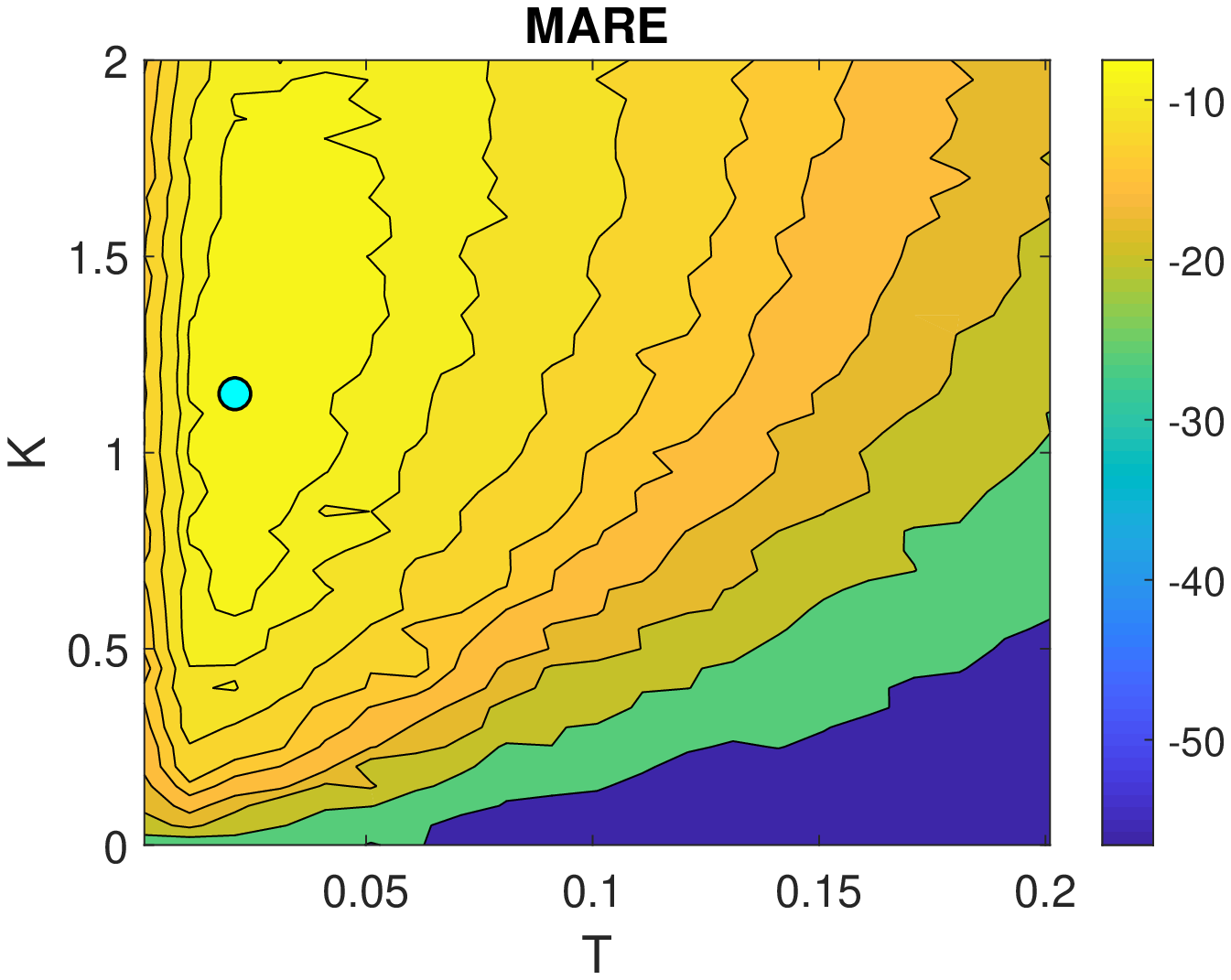}\label{fig:h_blk20_mare}}
\subfigure{\includegraphics[scale=0.43,clip]{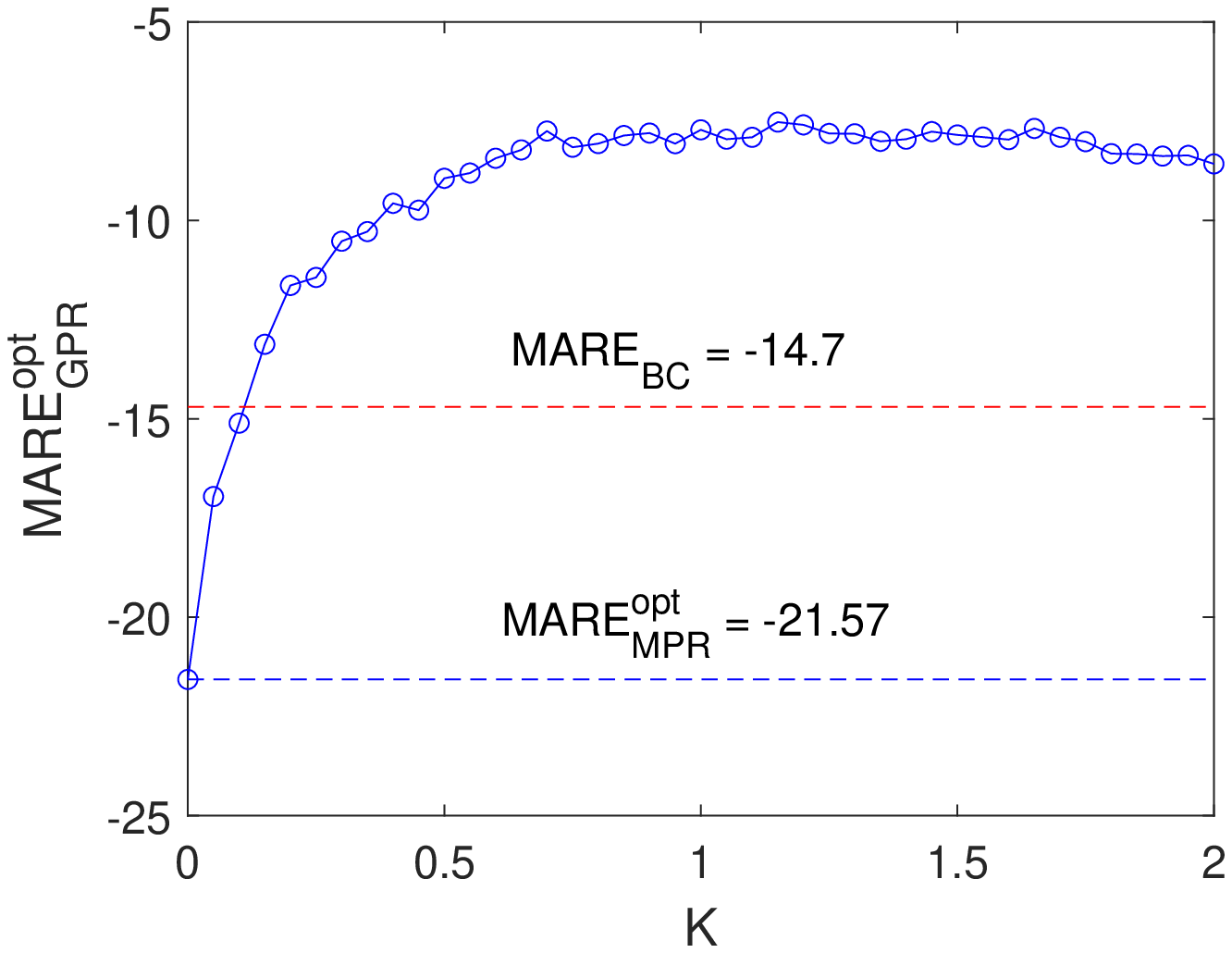}\label{fig:h_blk20_mare_opt}}\\ \vspace{-3mm}
\subfigure{\includegraphics[scale=0.43,clip]{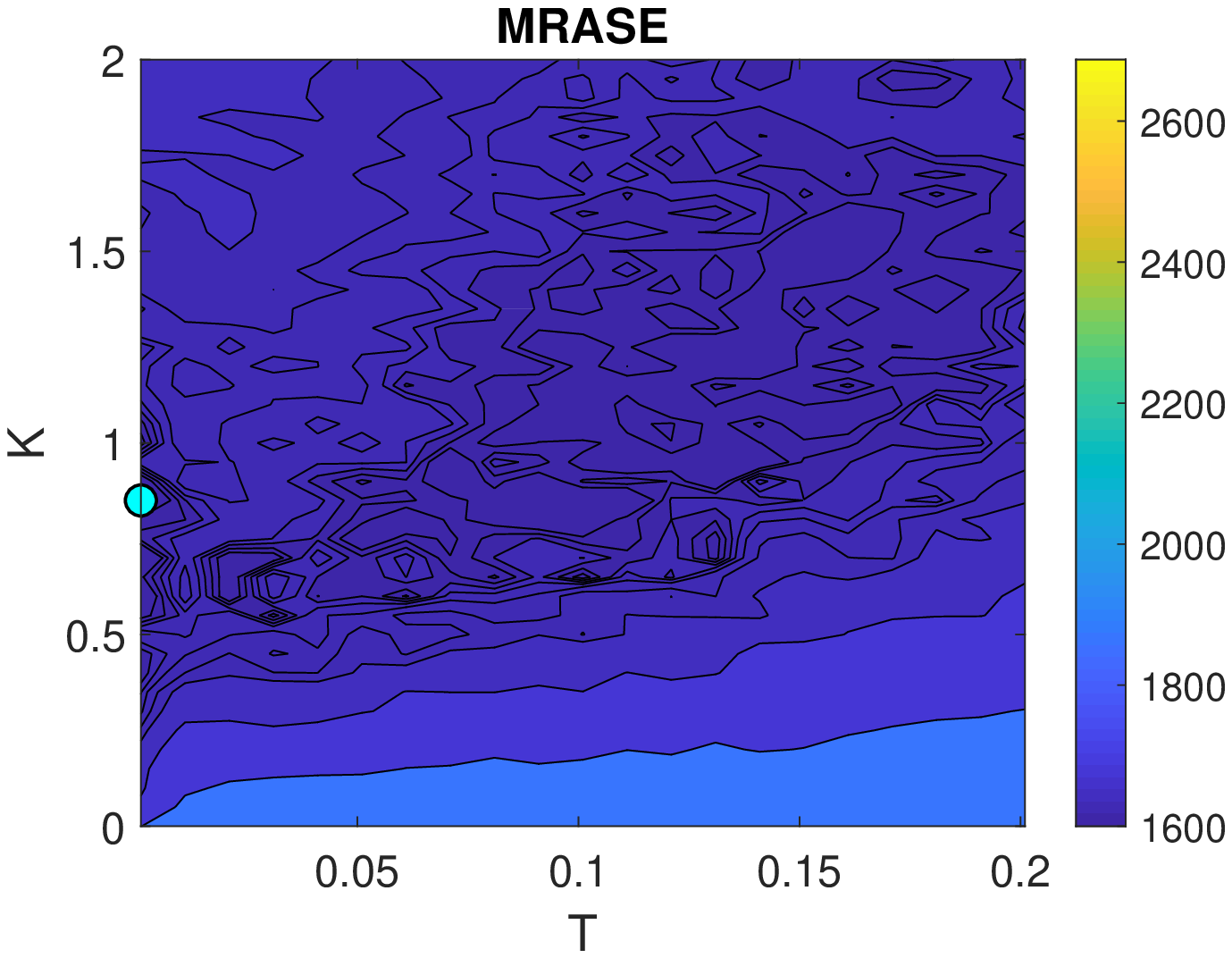}\label{fig:h_blk20_mrase}}
\subfigure{\includegraphics[scale=0.43,clip]{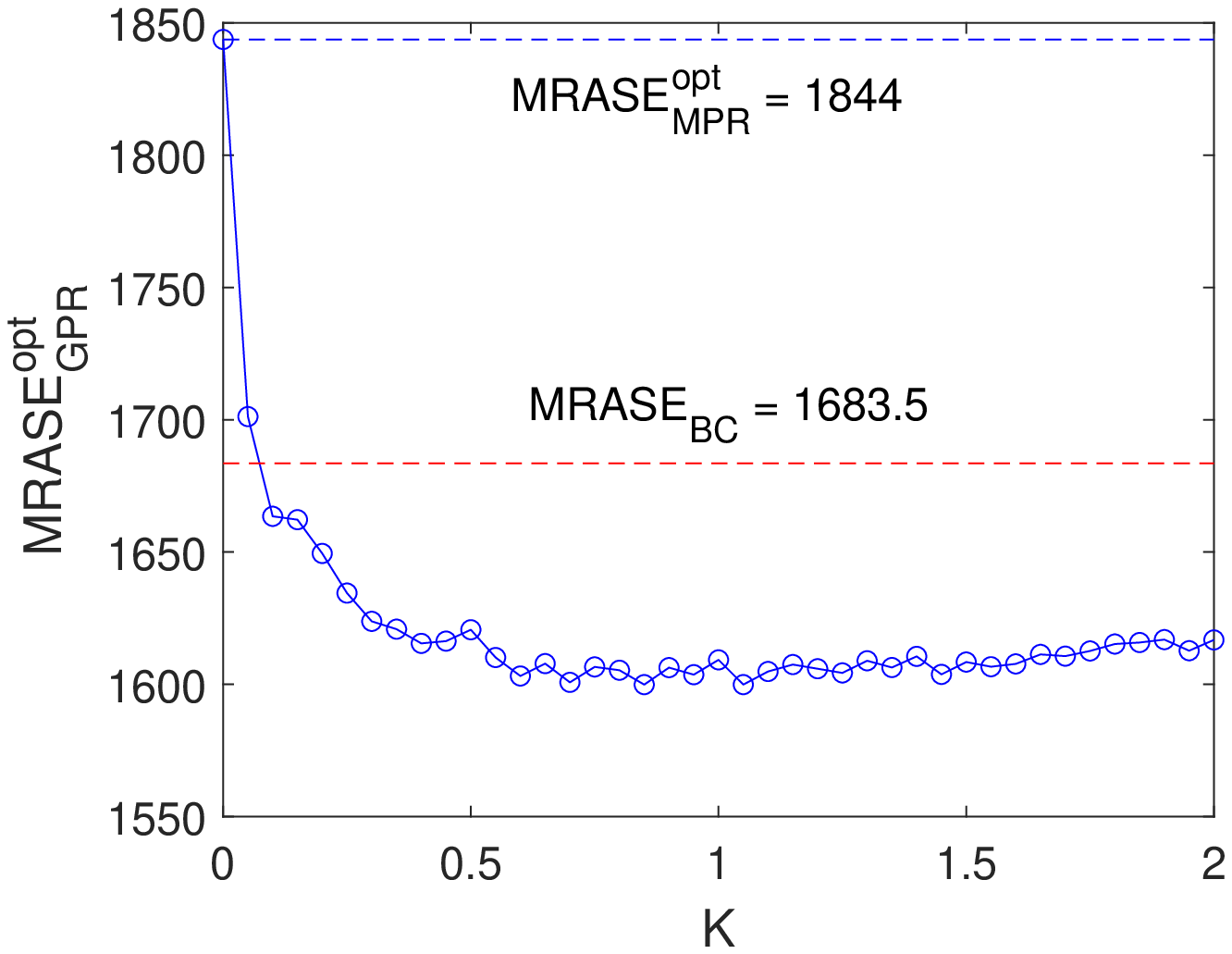}\label{fig:h_blk20_mrase_opt}} \vspace{-3mm}
\caption{Left column: contour plots of the validation measures in the $T-K$ parameter plane;  the cyan circles mark the optimal values. Right column: validation measures as functions of the parameter $K$ at optimal temperatures. The blue and red dashed lines mark the optimal values obtained by the MPR model and BC interpolation, respectively. $S=100$ samples are generated from the lognormal random field with $\xi_1=\xi_2=2$, $\nu = 2.5$. A missing box with $L_B=20$ is used.}
\label{fig:errors_K}
\end{figure}

Let us first consider effects of the bias field parameter $K$. The left column panels in Fig.~\ref{fig:errors_K} show the calculated validation measures in the $T-K$ parameter plane. For all the measures in zero field, i.e. by the MPR method, the optimal values are achieved at the lowest temperature $T=0.001$, as it could be expected and as it has been observed above for smooth data. However, by applying the bias field the low error areas are shifted to finite fields and somewhat increased temperatures.

In the right column of Fig.~\ref{fig:errors_K} we present variations of the validation measures as functions of the bias field parameter $K$ at optimal temperatures and compare them with those obtained by the MPR method (dashed blue lines) also at optimal temperatures. In this case, it is also interesting to include the results obtained by the bicubic (BC) interpolation (dashed red lines), which was employed as the bias field $h_i$. We note that the latter would result from the GPR method in the limit of  $K \to \infty$, when the interaction terms of the GPR Hamiltonian become negligibly small compared to the bias field term. One can see that all the measures obtained by the MPR method are considerably inferior to those from the BC interpolation. However, by inclusion of the bias field with even very small parameter $K$ all the validation measures dramatically improve. The presented figures demonstrate that the optimal prediction performance is achieved at moderately small values of the bias field parameter, which is significantly superior to both the MPR and BC methods. In particular, the MAAE, MAARE, MARE and MRASE errors obtained by the GPR method at $K=K_{\mathrm{opt}}>0$ are smaller than those by the BC method ($K \to \infty$) by about $13\%, 45\%$, and $49\%$ and $5\%$, respectively, and those by the MPR method ($K=0$) by about $31\%, 62\%$, and $65\%$ and $13\%$, respectively. This implies a synergic effect of the interaction and field terms which results in the prediction performance that cannot be achieved by either of the individual terms.

\begin{figure}[t!]
\centering \vspace{-10mm}
\subfigure{\includegraphics[scale=0.43,clip]{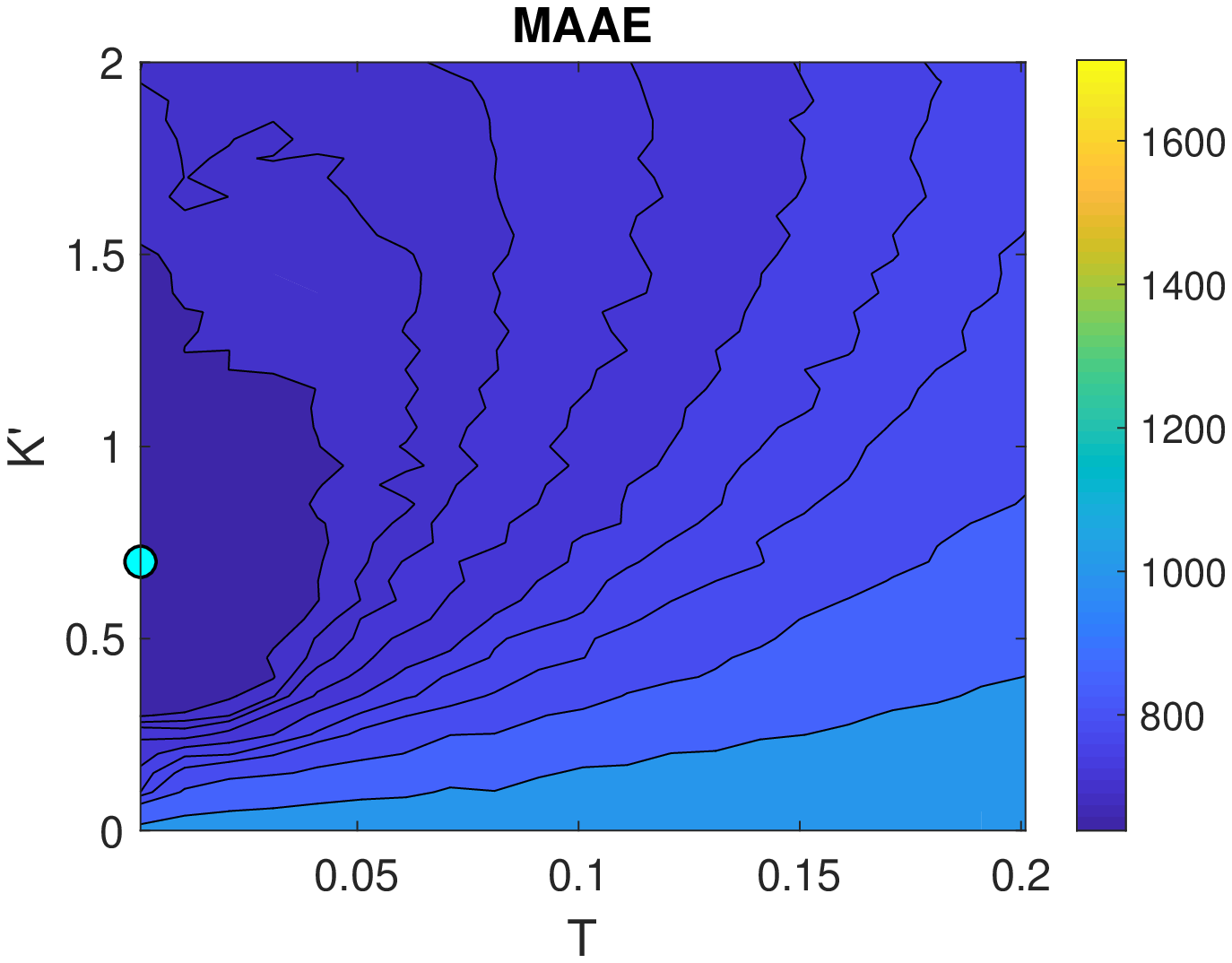}\label{fig:h0_blk20_maae}}
\subfigure{\includegraphics[scale=0.43,clip]{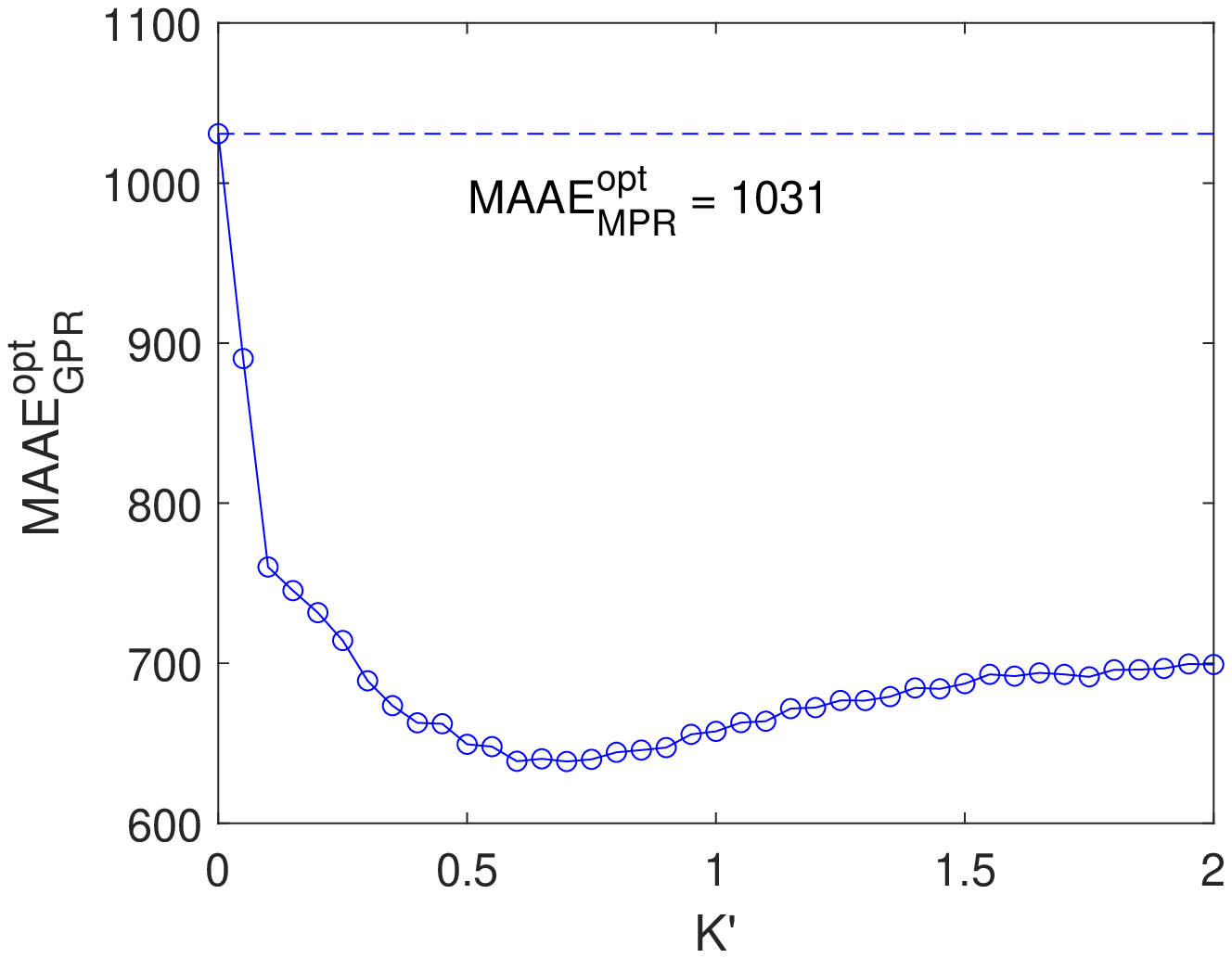}\label{fig:h0_blk20_maae_opt}}\\ \vspace{-3mm}
\subfigure{\includegraphics[scale=0.43,clip]{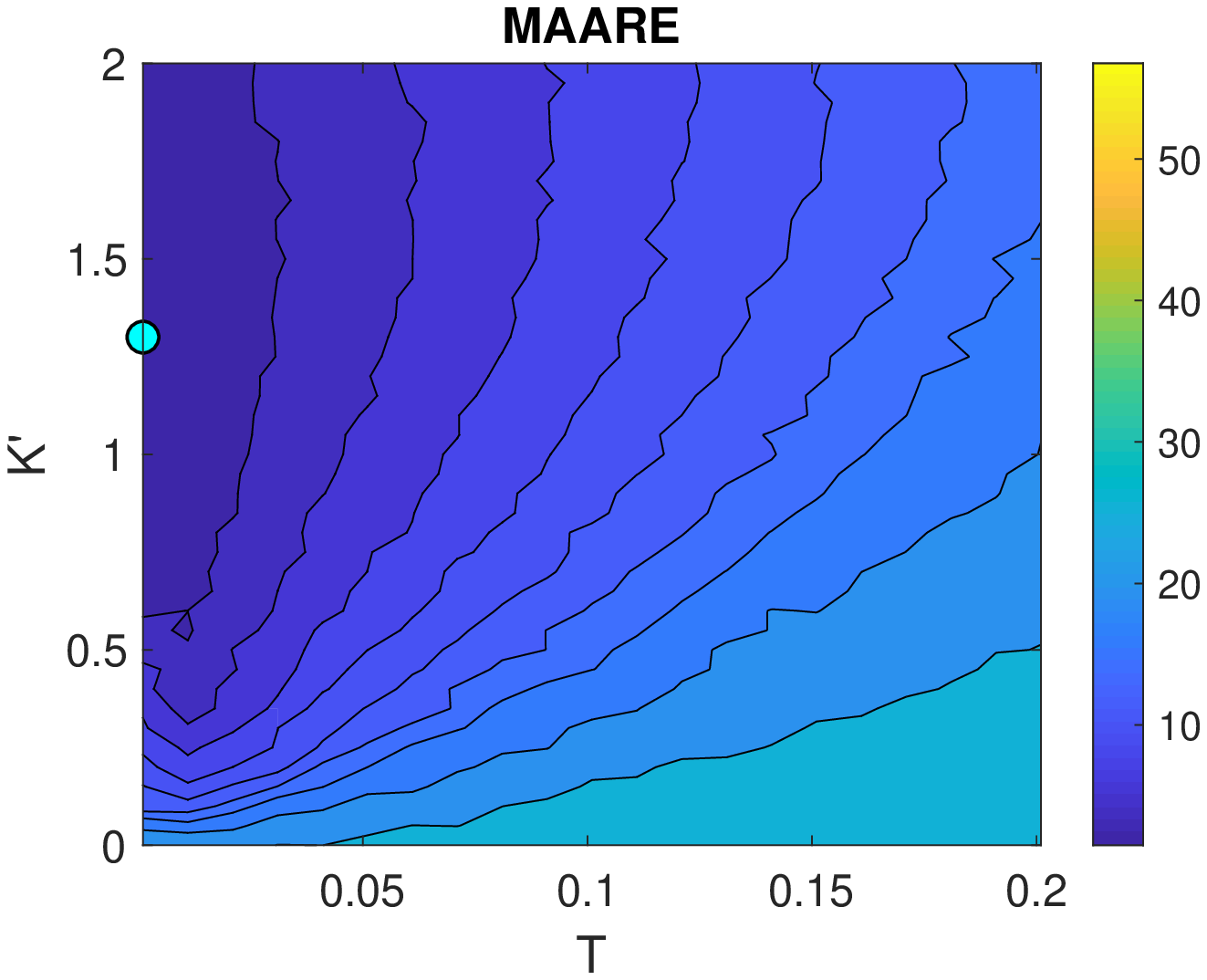}\label{fig:h0_blk20_maare}}
\subfigure{\includegraphics[scale=0.43,clip]{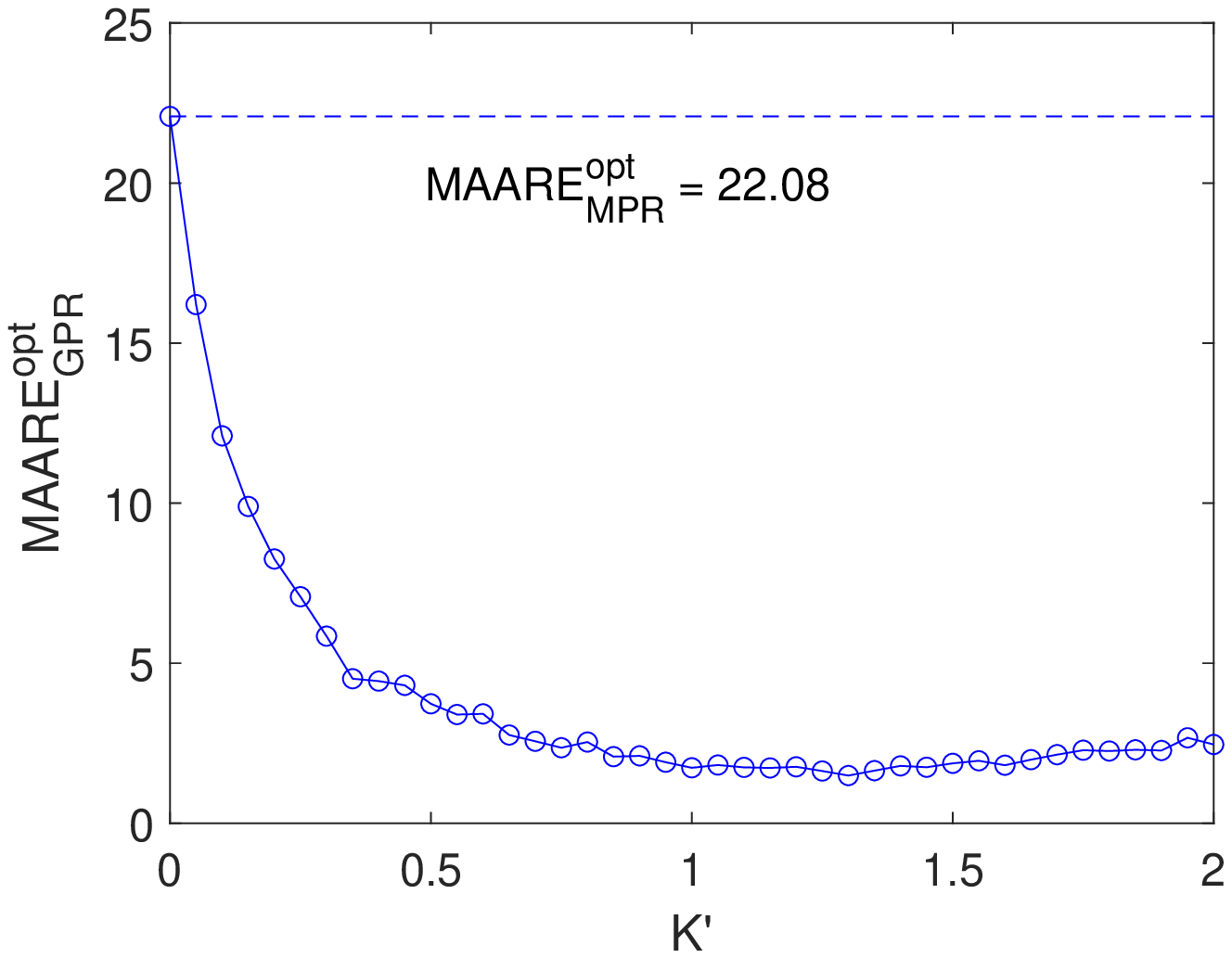}\label{fig:h0_blk20_maare_opt}}\\ \vspace{-3mm}
\subfigure{\includegraphics[scale=0.43,clip]{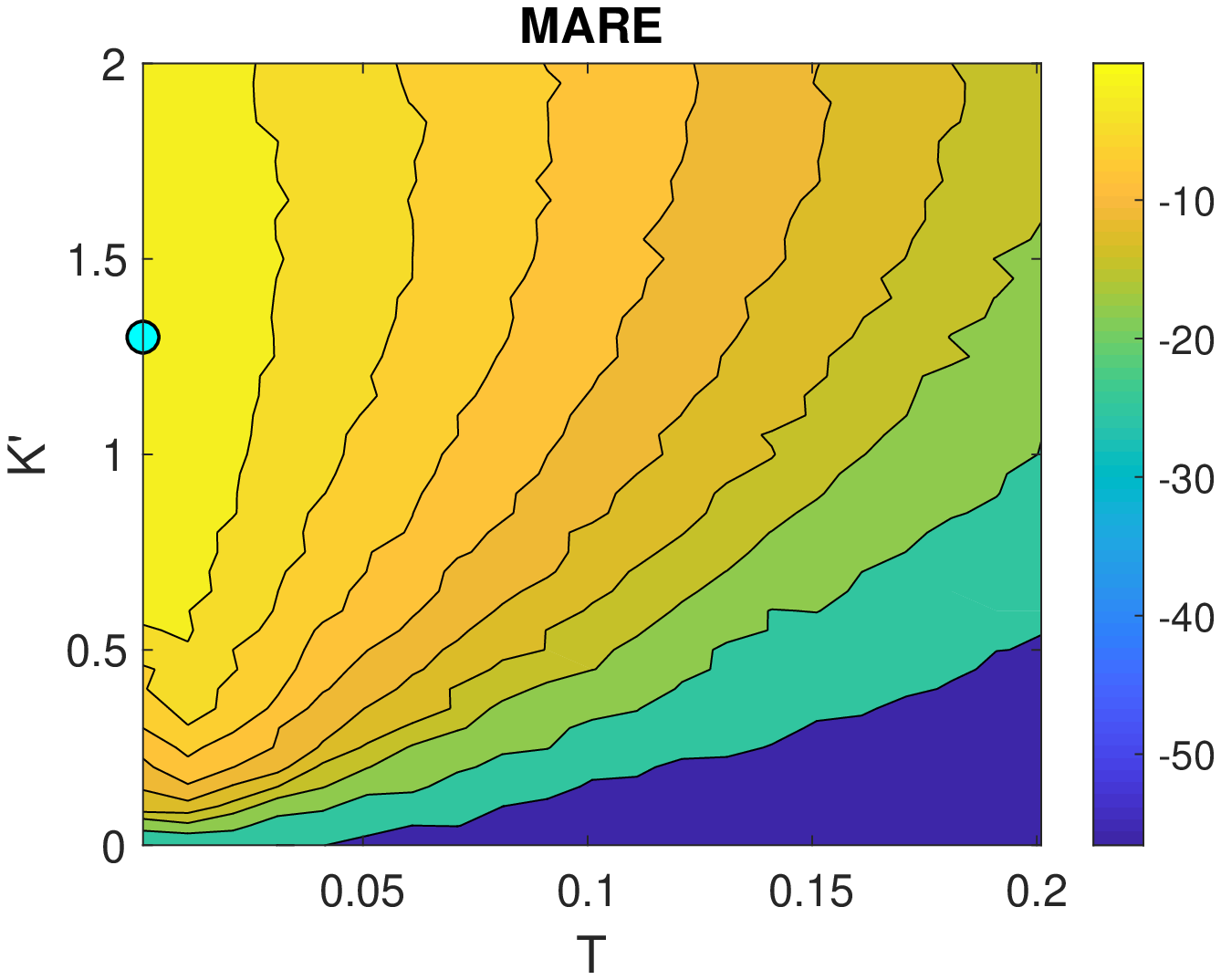}\label{fig:h0_blk20_mare}}
\subfigure{\includegraphics[scale=0.43,clip]{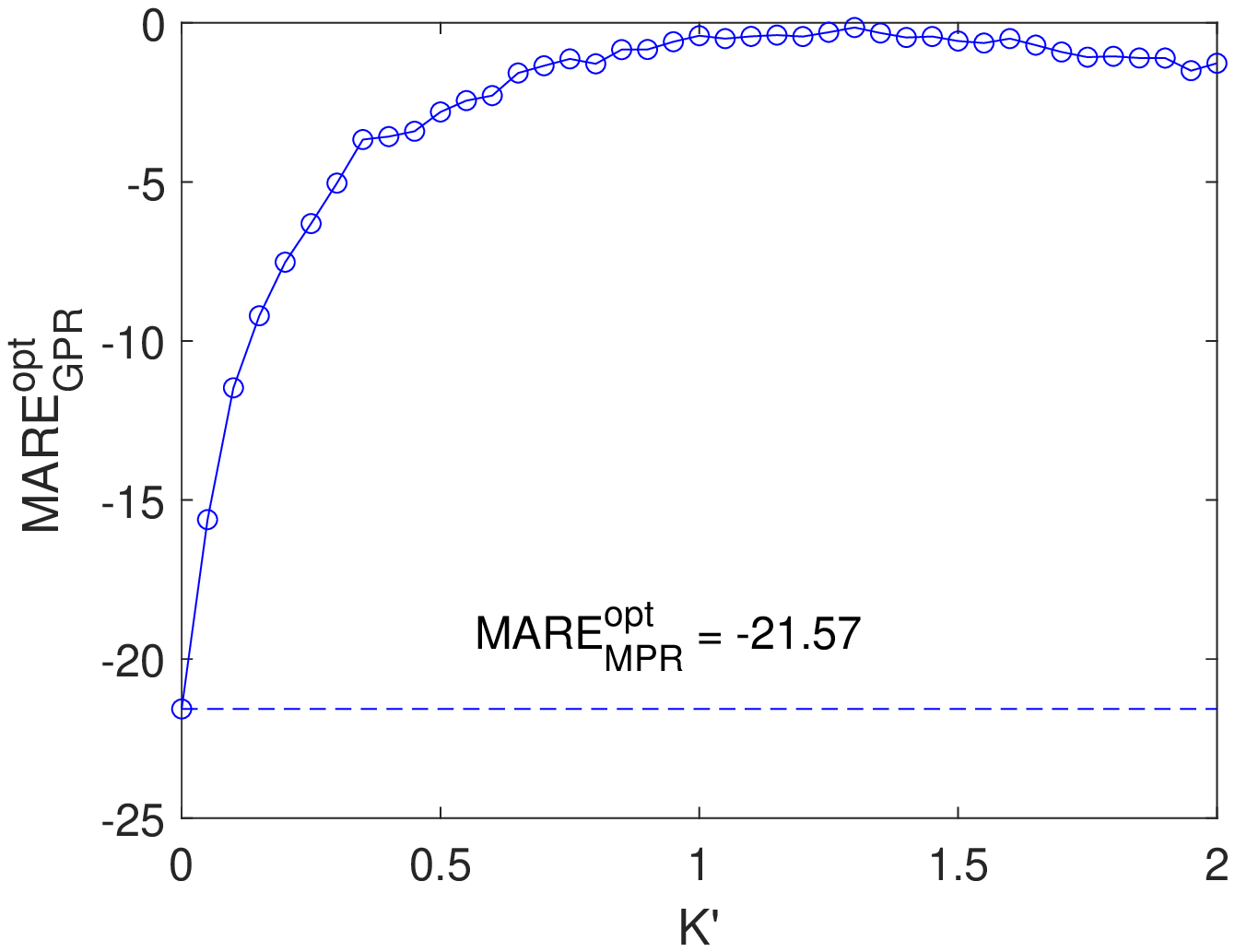}\label{fig:h0_blk20_mare_opt}}\\ \vspace{-3mm}
\subfigure{\includegraphics[scale=0.43,clip]{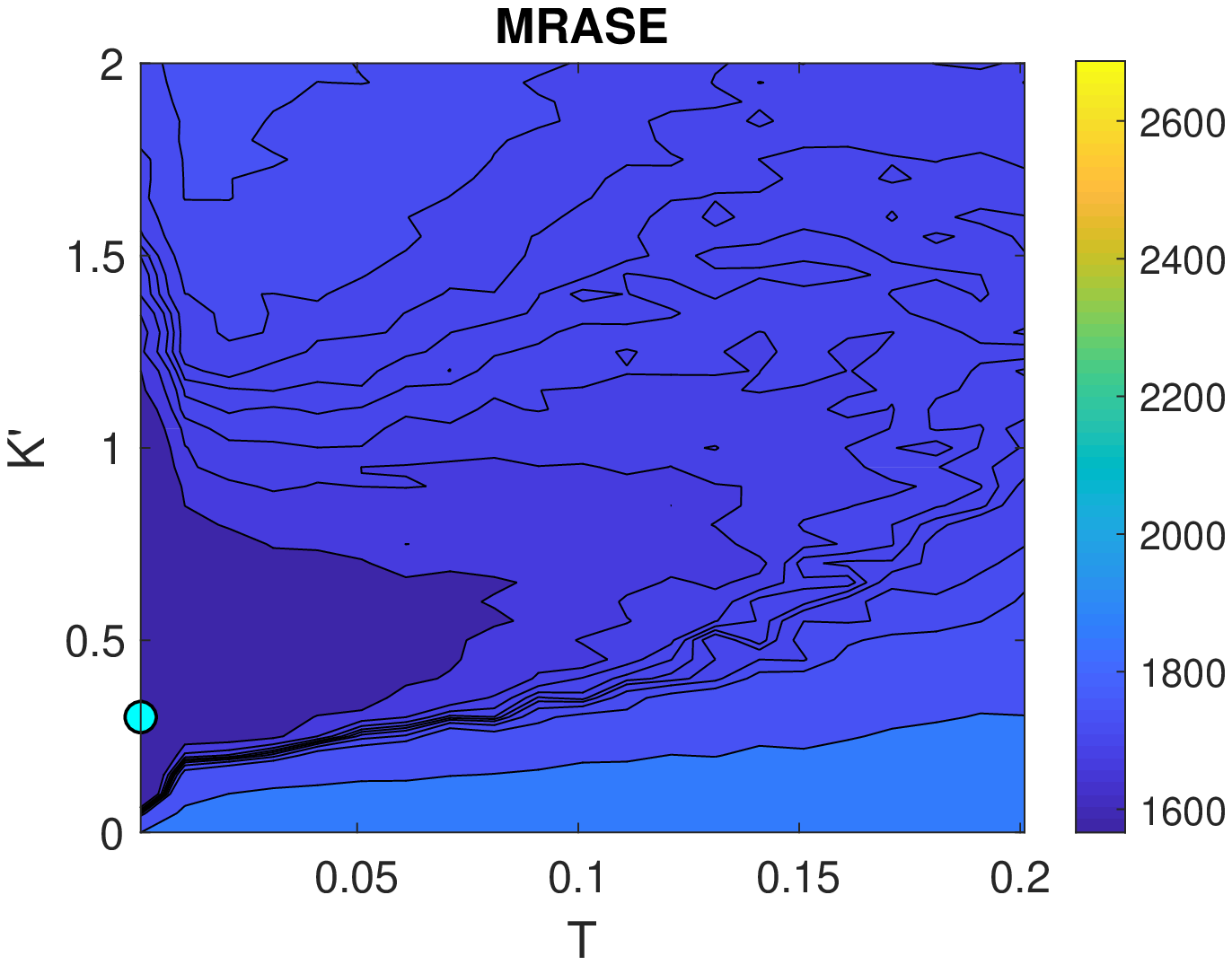}\label{fig:h0_blk20_mrase}}
\subfigure{\includegraphics[scale=0.43,clip]{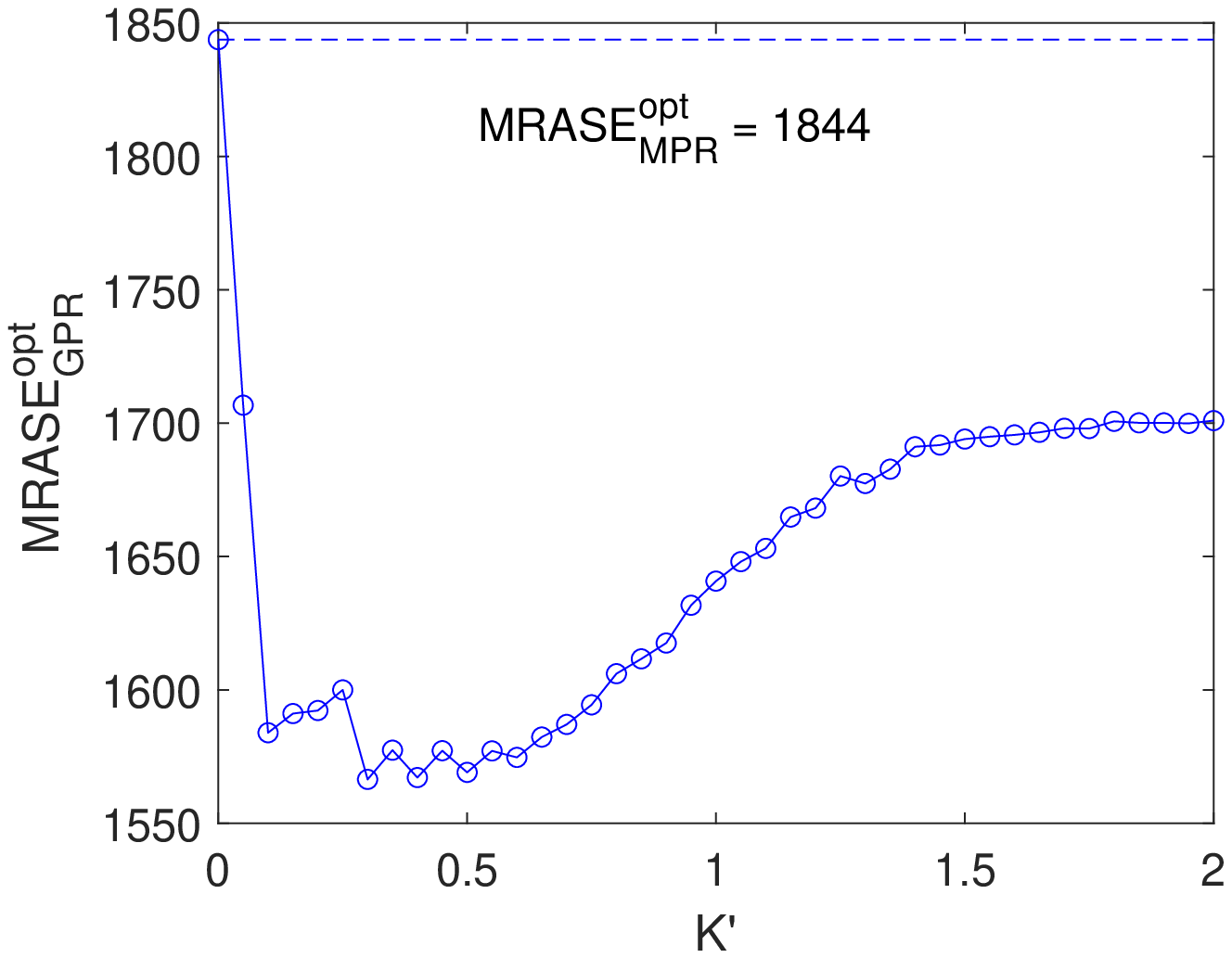}\label{fig:h0_blk20_mrase_opt}} \vspace{-3mm}
\caption{Left column: contour plots of the validation measures in the $T-K'$ parameter plane;  the cyan circles mark the optimal values. Right column: validation measures as functions of the parameter $K'$ at optimal temperatures. The dashed lines mark the optimal values obtained by the MPR model. $S=100$ samples are generated from the lognormal random field with $\xi_1=\xi_2=2$, $\nu = 2.5$. A missing box with side length $L_B=20$ is used.}
\label{fig:errors_K'}
\end{figure}

Finally, in the case of the uniform external ``magnetic field'' the validation measures in $T-K'$ parameter plane are presented in Fig.~\ref{fig:errors_K'}. One can notice that, in contrast to the non-uniform bias field presented above, the optimal prediction performance is achieved at the lowest temperature $T=0.001$ not only for $K'=0$ but also for $K'>0$. Thus, all the validation measures, shown in the right column of Fig.~\ref{fig:errors_K'} as functions of the external ``magnetic field'' parameter $K'$, correspond to the optimal temperature $T=0.001$. As in the the non-uniform bias field, the optimal prediction performance is achieved at similar values of the field parameter, i.e., $K'_{\mathrm{opt}} \approx K_{\mathrm{opt}}$, however, the corresponding errors are even much lower. Namely, the MAAE, MAARE, MARE and MRASE errors obtained by the GPR method at $K'_{\mathrm{opt}}$ are smaller than those at $K_{\mathrm{opt}}$ by about $10\%, 82\%$, and $98\%$ and $2\%$, respectively, and those by the MPR method ($K'=0$) by about $38\%, 93\%$, and $99\%$ and $15\%$, respectively. We note that $K'_{\mathrm{opt}}$ differ for different validation measures and, thus, if we defined some $K'_{gopt}$ that would globally optimize all the prediction errors than the above listed improvements would be generally smaller. 

\section{Conclusions and Further Research}
\label{sec:conclusions}
We have introduced a Gibbs Markov random field based on the generalized planar rotator (GPR) model. The latter was obtained by extending the earlier introduced modified planar rotator (MPR) model~\citep{mz-dth18} by inclusion of several additional terms in the Hamiltonian. In particular, in addition to the bilinear interaction term between nearest neighbors on the grid, which we generalized to distinguish interactions along vertical and horizontal axes, we considered up to infinite number of higher-order harmonics with exponentially vanishing interaction strength, further-neighbor interaction, and two types of  external bias field. This generalization added to the temperature $T$ (the only parameter of the MPR model) five more parameters: $n$ - the number of higher-order terms, $\alpha$ - the parameter controlling decay rate of higher-order interactions,  $\Jnn$ - the exchange anisotropy parameter, $\Jfn$ - the further-neighbor interaction, and $K$ or $K'$ - the external bias field parameter.

Using empirical tests, we have demonstrated benefits of the respective terms in the GPR model by showing that their inclusion can improve prediction performance. The tests were performed on synthetic data with both randomly missing data and contiguous block missing data. The results were presented for the random thinning $p=33\%$ and missing blocks of the size $L_B=20$ but similar results were also obtained for other values of $p$ and $L_B$. Potential of the respective terms to decrease prediction errors ranged from relatively small (inclusion of $n$ and $\alpha$) up to substantial (inclusion of $\Jnn$, $\Jfn$, $K$ and $K'$) but it strongly depended on the character of the data. Since in our tests we intentionally selected data sets for which we anticipated that the given term can improve prediction performance, the demonstrated effects could be smaller on different data sets. On the other hand, we always tested impact of only one selected parameter by finding its optimal value, while switching off the remaining ones. Therefore, we assume that simultaneous optimization of all the parameters could lead to further improvement and even better prediction performance than the one presented in our tests. 

The optimal parameter set could be obtained by one of the available parameter estimation methods. Since the present method targets efficient prediction of massive data it is desirable to employ also a computationally efficient parameter estimation method. In the case of the one-parameter MPR model the temperature (or the reduced parameter $J/T$) was efficiently estimated using the so called specific energy matching method. In the GPR model such an approach could be extended to differentiate between ``temperatures'' in different directions, $\Jnn_x/T$ and $\Jnn_y/T$, and thus perform inference of the parameter $\Jnn$. The parameter $\Jfn$ (or $\Jfn/T$) could be estimated analogically as $J/T$ by the specific energy matching method with the specific energy calculated between further instead of nearest neighbors. Furthermore, the specific energy matching method for estimation of the exchange interaction parameter(s) can be extended to the (specific) magnetization matching method for estimation of the external ``magnetic field'' parameter $K'$. Then, in the simplest case of the model with only the isotropic nn interaction and the field $K'$, this approach would assume knowledge of the specific energy and magnetization surfaces in the $T-K'$ parameter plane. These can be obtained from unconditional MC simulations of the GPR model with the $J$ and $K'$ terms. However, if simultaneous estimation of all the model parameters is targeted such an approach becomes more involved and computationally intractable. We leave finding an efficient way of the GPR parameters' inference for the future considerations.

\begin{acknowledgments}
This work was supported by the Scientific Grant Agency of Ministry of Education of Slovak Republic (Grant No. 1/0531/19). We also acknowledge support for a short visit by M.~\v{Z}. at the Technical University of Crete from the Hellenic Ministry of Education - Department of Inter-University Relations, the State Scholarships Foundation of Greece and the Slovak Republic's Ministry of Education through the Bilateral Programme of Educational Exchanges between Greece and Slovakia.
\end{acknowledgments}

\bibliographystyle{apsrev}
\bibliography{bibliography}

\end{document}